\documentclass[11pt,graphicx,amsmath]{article}
\usepackage{amsmath}
\usepackage{graphicx}
\usepackage{bm}
\usepackage[dvips]{color}
\usepackage{amssymb}
\usepackage{amsfonts}
\usepackage{comment}
\usepackage{cite}
\usepackage{todonotes}
\usepackage{caption}
\usepackage{subcaption}
\usepackage{CJKutf8}

\usepackage[utf8]{inputenc}

\def\be{\begin{equation}}
\def\ee{\end{equation}}
\def\ba{\begin{eqnarray}}
\def\ea{\end{eqnarray}}
\def\nn{\nonumber}
\newcommand{\blue}[1]{{\color{blue} #1}}

\def\bl#1\el{\begin{align}#1\end{align}}

\title{ The Conserved  Effective Stress Tensor of Gravitational Wave }

\author{\small
             Yang  Zhang \thanks{yzh@ustc.edu.cn} , \,
           Xuan Ye   \thanks{yyyyy@ustc.edu.cn}   ,
            \\
 \small  Department of  Astronomy,
         CAS Key Laboratory for Researches in Galaxies and Cosmology, \\
 \small  School of Astronomy and Space Sciences, \\
 \small  University of Science and Technology of China, Hefei, Anhui, 230026, China \\
 }

 \date{}

\def\be{\begin{equation}}
\def\ee{\end{equation}}
\def\ba{\begin{eqnarray}}
\def\ea{\end{eqnarray}}
\def\nn{\nonumber}
\def\bl#1\el{\begin{align}#1\end{align}}

\begin{document}
\begin{CJK}{UTF8}{gbsn}
\maketitle

\allowdisplaybreaks

\abstract

\large

{We present a detailed study of the effective stress tensor of gravitational wave (GW)
as the source for the background Einstein equation
and examine three candidates in literature.
The second order perturbed Einstein tensor $G^{(2)}_{\mu\nu}$,
up to a coefficient, proposed  by Brill, Hartle, and Isaacson,
has long been known to be covariantly nonconserved
with respect to the background spacetime.
We observe that $G^{(2)}_{\mu\nu}$
is not a true tensor on the background spacetime.
More importantly,  we find that,
by expressing $G^{(2)}_{\mu\nu}$ in terms of the perturbed Hilbert-Einstein actions,
  the nonconserved part of $G^{(2)}_{\mu\nu}$
is actually canceled out by the perturbed fluid stress tensors
in the back-reaction equation,
or is vanishing in absence of fluid.
The remaining part of  $G^{(2)}_{\mu\nu}$
is just the conserved effective stress tensor $\tau_{\mu\nu}$
proposed by Ford and Parker.
As the main result,  we derive $\tau_{\mu\nu}$ for a general curved spacetime
by varying the GW action
and show its conservation using the equation of GW.
The stress tensor $T_{\text{MT}}^{\mu\nu}$ proposed  by MacCallum and Taub
was based on an action  $J_2$.
We derive $T_{\text{MT}}^{\mu\nu}$ and find that it is nonconserved,
and that $J_2$ does not give the correct GW equation
in presence of matter.
The difficulty with $J_2$ is due to a background Ricci tensor term,
which should be also canceled out by the fluid term
or vanishing in absence of fluid.
We also demonstrate these three candidates in a flat Robertson-Walker spacetime.
The conserved $\tau_{\mu\nu}$ has a positive energy density spectrum,
and  is adequate for the  back-reaction  in a perturbation scheme,
while the two nonconserved stress tensors
have a negative  spectrum at long wavelengths and are unphysical.
}

\newpage

\section{Introduction}

The stress tensor as the source of the Einstein equation
 determines  the background spacetime.
For ordinary matter contents,
such as  fluids \cite{MTW,Weinberg1972},
scalar fields \cite{ZhangYeWang2020,ZhangWangYe2020,ZhangYe2022PointSpl},
vector fields \cite{ZhangYe2022,YeZhang2024,ZhangYe2025}, etc,
the corresponding stress tensor is well defined.
But the stress tensor of gravitational field itself is not so straightforward.
Landau and Lifshitz \cite{LandauLifshitz1975} put the Einstein equation
into a generalized d'Alembert wave equation of the metric
with a stress pseudo-tensor of gravitational field as the source.
By the equivalence principle,
the gravitational field can be made vanish locally
by coordinate transformation \cite{Weinberg1972,MTW}, and so is the pseudo-tensor.
Ref. \cite{BabakGrischuk1999} further developed
the Landau and Lifshitz  formulation into a covariant version.

In the approach of the effective stress tensor
\cite{BrillHartle1964,Isaacson1968a,Isaacson1968b},
the metric of spacetime is divided
into the background and perturbation parts,
and one studies the back-reaction of the metric perturbation
upon the background spacetime.
The effective stress tensor serves as
the source for the background Einstein equation,
unlike  the Landau and Lifshitz approach.
When a true stress tensor of the metric perturbation can be constructed,
the tensor will not be made vanish by gauge transformations of
background metric \cite{Isaacson1968a}.
In general, the scalar metric perturbation is gauge dependent
\cite{Matarrese1998,WangZhang2017},
so is its associated effective stress tensor
\cite{AbramoPRD1997,Unruh1998,Cho2020,IshibashiWald2006}.
In contrast, the GW  as the tensor metric perturbation
is invariant under certain coordinate transformations
\cite{WangZhang2017,WangZhang2018,WangZhang2019,WangZhang2024},
in this case a gauge-invariant  effective stress tensor of GW
can be defined.

In literature  three candidates were proposed
for the effective stress tensor of GW.
The back-reaction of GW upon the background spacetime
was first investigated by Brill and Hartle \cite{BrillHartle1964}
in the context of gravitational geons,
the effective stress tensor of GW was defined by
$T^{\text{eff}}_{\mu\nu} \equiv \frac{-1}{8\pi G} G^{(2)}_{\mu\nu}$.
Isaacson \cite{Isaacson1968a,Isaacson1968b}
further developed the formulation for a general vacuum case.
This  was adopted to study the expanding Universe
by  Refs.\cite{AbramoPRD1997,AbramoPRD1999,BrandTaka2018}.
  $T^{\text{eff}}_{\mu\nu}$ is not covariantly conserved
with respect to the background spacetime,
and its associated energy density spectrum is negative at long wavelengths.
These difficulties of $T^{\text{eff}}_{\mu\nu}$ were discussed
in Refs.\cite{Giovannini2006,SuZhang2012,Giovannini2019prd,Giovannini2020}.

MacCallum and Taub \cite{MacCallum1973} divided the second order perturbed
Hilbert-Einstein action into two parts, $J_1+J_2$,
where $J_1$ consists of the $\bar R_{\mu\nu}$ terms,
and $J_2$ represents a curved-spacetime generalization of
the Fierz-Pauli action of spin-2 massless field.
They suggested an effective stress tensor $T_{\text{MT}}^{\mu\nu}$
by variation of $J_2$,
without giving its explicit expression.
Isaacson  \cite{Isaacson1968a} first mentioned $J_2$,
but rejected it in search for an effective stress tensor of GW.
Refs.\cite{stein2011,Deepen2022,Butcher2009} also used the action $J_2$.
However, as we notice,
$J_2$ does not give the correct equation of GW in presence of matter.

As a more realistic model,
Ford and  Parker \cite{FordParker1977} studied GW in the presence of fluid,
and gave the second order perturbed action $I_{\text{FP}}$
which has contributions from GW and fluid.
For a fRW spacetime, by canceling  the background Ricci tensor term
by the fluid term,
they gave the action of GW  which is equal to that of
a pair of minimally-coupling massless scalar fields,
and the equation of GW which is that first derived by Lifshitz \cite{Lifshitz1946}.
They also obtained the conserved effective stress tensor $\tau_{\mu\nu}$ in a fRW spacetime,
but did not give the expression $\tau_{\mu\nu}$ for a general curved spacetime.

Given $T^{\text{eff}}_{\mu\nu}$, $T_{\text{MT}}^{\mu\nu}$ and $\tau_{\mu\nu}$,
which is pertinent for the back-reaction?
So far the issue has not been sufficiently investigated in literature.
In this paper we shall examine the main properties (or difficulties)
of the candidates,
trace the origin of their differences,
and find  the adequate candidate and
its corresponding back-reaction scheme.

For a general curved spacetime,
we shall derive  the expression $\tau_{\mu\nu}$ from the GW action   $I_{gw}$,
and show its   conservation,  as implied by the GW equation.
Next, we shall show that $G^{(2)}_{\mu\nu}$ is
not a true tensor on the background spacetime.
By writing $G^{(2)}_{\mu\nu}$ in terms of perturbed actions,
we find that the nonconserved and nontensorial part of $G^{(2)}_{\mu\nu}$
comes from the background Ricci tensor terms,
which is canceled out by the fluid terms in presence of fluid,
or is just vanishing  in absence of fluid.
After this correction, $G^{(2)}_{\mu\nu}$  will reduce to
the conserved $\tau_{\mu\nu}$.
We shall also derive  $T_{\text{MT}}^{\mu\nu}$ from $J_2$
and show that it is not conserved.
The difficulty of $J_2$ is due to a  background Ricci tensor term
which is actually canceled out by a corresponding fluid term.
In a fRW spacetime, we shall demonstrate that
$\tau_{\mu\nu}$ has a positive energy density spectrum,
while $T^{\text{eff}}_{\mu\nu}$ and $T_{\text{MT}}^{\mu\nu}$
have a negative spectrum.

The paper is organized as follows.

In sect. \ref{Sect1GWeq},  we review the model of GW in presence of fluid,
and give the GW action $I_{gw}$ and list the equation of GW.

In sect.  \ref{action}, we expand the Hilbert-Einstein action
and the fluid action up to the second order.
In particular, we show how the Ricci tensor terms are canceled
by the corresponding fluid terms,
and that only  $I_{gw}$  remains.
We also point out the  difference between  $J_{2}$ and $I_{gw}$.

In sect. \ref{sect4FP},
we present the conserved  $\tau_{\mu\nu}$ following from $I_{gw}$,
and the nonconserved   $T^{\text{MT}}_{\mu\nu}$ following from $J_2$.

In sect. \ref{effectivest},  we show that
$G^{(2) \mu\nu}$ is not  a true tensor on the background spacetime,
and that the four divergence  $G^{(2) \mu\nu}\, _{|\nu} \ne 0$ in general.

In sect. \ref{sectionAnal}, we rewrite
$G^{(2)}_{\mu\nu}$ in terms of the perturbed actions,
and show that  its nonconserved  part $X_{\mu\nu}$
is canceled out by the fluid terms in the back-reaction,
leaving only  $\tau_{\mu\nu}$.

In sect. \ref{sectionfordpark},
we demonstrate $T^{\text{eff}}_{\mu\nu}$, $T_{\text{MT}}^{\mu\nu}$,  $\tau_{\mu\nu}$
in a flat RW spacetime.

In sect. \ref{Disscusson},  we give the conclusion and discussions.

In Appendix. A, we list the perturbation quantities up to the second order.

In Appendix \ref{Conditions}, we discuss the TT gauge conditions of GW
and the gauge invariance.

In Appendix \ref{Appvaaction}, we list
the  detailed calculation of the variation $\delta I_{gw}$
with respect to the background metric to give $\tau_{\mu\nu}$.

In Appendix  \ref{j2append} we list
the  detailed calculation of $\delta J_2$ to give $T_{\text{MT}}^{\mu\nu}$.

In Appendix \ref{G3delta},
we   prove  that the varied action of the nonconserved part
is canceled by the fluid term.

In Appendix \ref{usefulquantites}, we list some quantities in fRW spacetimes.

In this paper
we use the units $16\pi G=\hbar=c=1$ and the metric signature $(+,-,-,-)$.

\section{The System of GW and Fluid}\label{Sect1GWeq}

We describe briefly the model of GW in presence of fluid.
The Einstein equation $G_{\mu\nu}=8 \pi G T_{\mu\nu}$
 can be written as (in the unit $16\pi G=1$)
\bl
R_{\mu\nu}= \frac12 \big(T_{\mu\nu}-\frac12g_{\mu\nu}T \big),\label{ese}
\el
where $g_{\mu\nu}$ is the metric, $R_{\mu\nu}$ is the Ricci tensor,
$T_{\mu\nu}$ is the stress tensor of a perfect fluid given by
\bl
T_{\mu\nu} =(\rho+p)u_{\mu}u_{\nu}-pg_{\mu\nu},
\label{strfld}
\el
where $\rho$ and $p$ are energy density and pressure of the fluid,
$u_{\mu}$ is the four-velocity of fluid
with the normalization $u^{\mu}u_{\mu}=1$,
and the trace $T=T^{\lambda}_{~~ \lambda}=\rho-3p$.
The stress tensor of fluid \eqref{strfld}
can describe a class of models of matter-energy content.
For instance, $p=0$ gives  the dust model,
 $p=\frac13 \rho$ gives the radiation model,
and $\rho=\Lambda=-p$ gives the cosmological constant model.

Consider  the metric perturbation
\bl \label{expandg}
g_{\mu\nu}=\gamma_{\mu\nu} + h_{\mu\nu},
\el
where $\gamma_{\mu\nu}$ is the background metric,
and  $h_{\mu\nu}$ is a small metric perturbation
which generally  contains  the scalar, vector and tensorial modes
 \cite{WangZhang2017,WangZhang2018,WangZhang2019}.
Note that we do not separate a second order metric perturbations from
the expression \eqref{expandg},
and the treatment here differs from
Refs.\cite{Matarrese1998,WangZhang2017,WangZhang2018,WangZhang2019,WangZhang2024},
in which the second order perturbations of metric were calculated.

The 0th order Einstein equation is
\bl
\bar G_{\mu\nu} = \frac12 \bar T_{\mu\nu},
\label{beisteineq}
\el
or,  written as
\bl
\bar{R}_{\mu\nu} & = \frac12   (\bar{\rho}+\bar{p})\bar{u}_{\mu}  \bar{u}_{\nu}
-\frac14 \gamma_{\mu\nu}(\bar\rho-\bar{p})   ,
\label{ei1st}
\el
where a bar denotes the background quantity.
The trace of  \eqref{ei1st} is
\bl
\bar{R}=\frac12(-\bar{\rho}+3\bar{p}) .
\label{bkR}
\el
Given a fluid model, the background equation \eqref{beisteineq} will determine
the background metric  $\gamma_{\mu\nu}$.

The 1st order perturbed Einstein equation is
\bl
 R^{(1)}_{\mu\nu} = \frac12  \big(T^{(1)}_{\mu\nu}
 -\frac12 h_{\mu\nu} \bar{T} -\frac12 \gamma_{\mu\nu} T^{(1)} \big) \, ,
 \label{compact1st}
\el
where $R_{\mu\nu}^{(1)}$ denotes the first order perturbed Ricci tensor,
and $T_{\mu\nu}^{(1)}$ is the first order stress tensor.
(See Appendix \ref{AppendixA}.)
Eq.\eqref{compact1st} determines the metric perturbation $h_{\mu\nu}$.
At the linear level,  GW is not coupled with the matter perturbations
\cite{Lifshitz1946,Matarrese1998,WangZhang2017,WangZhang2018,WangZhang2019,WangZhang2024,Giovannini2020}.
Therefore,   in this paper for simplicity,
the matter perturbation can be set to zero,
$\delta\rho=\delta p=\delta u^{\mu}=0$,
as in Ford-Parker's treatment of GW  \cite{FordParker1977}.
We consider GW that satisfies
the following three conditions
\bl
& \bar{u}_{\mu}h^{\mu\nu}=0,
\label{coordcondition}
\\
& h^{\mu\nu}_{~~|\nu}=0, \label{transvcondition}
\\
& h = 0 , \label{conditionsforGW}
\el
where $h=h^{\mu}_{~\mu}$ is the trace of perturbation,
 $|_{\nu}$ represents the covariant derivative with respect to the background spacetime.
The condition \eqref{coordcondition} is the coordinate condition
which specifies that the GW is perpendicular to the four velocity of the fluid.
The conditions \eqref{transvcondition} and \eqref{conditionsforGW}
are  the  transverse  and traceless (TT)   gauge condition.
We shall refer to the set of the conditions
\eqref{coordcondition}  \eqref{transvcondition} \eqref{conditionsforGW}
as the GW conditions.
In appendix \ref{Conditions} we show that
for a wide class of spacetimes where a shearless four-vector can be defined,
the conditions \eqref{coordcondition} \eqref{transvcondition} \eqref{conditionsforGW}
can be consistently imposed and constitute eight independent constraints.
This class
includes the fRW spacetimes, the Schwarzschild spacetime, the Minkowski spacetime.
For these spacetimes, eight components of $h_{\mu\nu}$ are eliminated,
and leaving only two physical degrees of freedom  (DOF) for the GW.
For the  fRW spacetimes, as pointed out by Ref.\cite{FordParker1977},
the conditions \eqref{coordcondition} \eqref{transvcondition} imply
the traceless condition \eqref{conditionsforGW}.
In general,  the GW conditions will change under coordinate transformations
in a given spacetime  \cite{Isaacson1968a,MTW},
see discussions in Appendix.\ref{Conditions}.
However,  under the class of synchronous-to-synchronous transformations,
  the GW   $h_{\mu\nu}$  remains invariant,
as demonstrated in Refs.\cite{WangZhang2017,WangZhang2018,WangZhang2019}.
In this paper we shall consider the spacetimes for which
the GW conditions
\eqref{coordcondition}\eqref{transvcondition}\eqref{conditionsforGW}
can be imposed consistently.

For GW, under the conditions
\eqref{coordcondition}\eqref{transvcondition}\eqref{conditionsforGW},
 using  \eqref{1storderriccitensor} \eqref{Tmunu1},
the 1st order equation \eqref{compact1st}  becomes the following
\bl
( - \Box h_{\mu\nu}  + 2\bar{R}_{\nu\rho\alpha\mu} h^{\rho\alpha}
+ \bar{R}_{\rho\mu} h^{\rho}_{~\nu}
+  \bar{R}_{\rho\nu} h^{~\rho}_{\mu}  )
= -  \frac{1}{2}  h_{\mu\nu}(\bar\rho-\bar p) \, ,
\label{eqGWfl}
\el
which is the same as  (2.8) of Ref.\cite{FordParker1977}
up to the sign convention.
Using the background  equation \eqref{ei1st},
the Ricci tensor terms are canceled by  the fluid term,
so that \eqref{eqGWfl} becomes
\bl
\Box h_{\mu\nu} -2\bar{R}_{\mu\alpha\beta\nu} h^{\alpha\beta} =0  .
\label{GWeq}
\el
This is the equation of GW and holds in the presence of fluid, as well as in the vacuum.
Moreover,  eq.\eqref{GWeq} holds even
when  the linear perturbations of the perfect fluid, $\delta\rho,\delta p,\delta u^{\mu}$, are present,
the latter are coupled only with the scalar and vector metric perturbations
\cite{Lifshitz1946,Matarrese1998,WangZhang2017,WangZhang2018,WangZhang2019,WangZhang2024}.
When the fluid is imperfect with the anisotropic stress,
say, due to free streaming neutrinos \cite{Weinberg2004,MiaoZhang2007},
 the equation \eqref{GWeq} will be modified.
When nonlinear perturbations are included,
the equation \eqref{GWeq} will also  be  modified
\cite{WangZhang2017,WangZhang2018,WangZhang2019,WangZhang2024}.
These modifications will be beyond the  scope of this paper.

The equation \eqref{eqGWfl}
also follows from the action
\bl
I_{\text{FP}}
 = \int d^4x  \frac{1}{4}\sqrt{-\gamma} \Big(h_{\mu\nu|\rho}h^{\mu\nu|\rho}
+2\bar{R}_{\mu\alpha\beta\nu}h^{\mu\nu}h^{\alpha\beta}
+  2\bar{R}_{\beta}^{~\mu} h_{\alpha\mu}h^{\alpha\beta}
+ \frac12( \bar{\rho}-\bar{p})h_{\mu\nu}h^{\mu\nu}
  \Big) .
\label{FPlagrangian}
\el
This is the Ford-Parker's action for the system of
GW and fluid,  up to the sign convention  (see  (2.9) in Ref.\cite{FordParker1977}).
By the background equation  \eqref{ei1st}
and the coordinate condition \eqref{coordcondition} for GW,
the Ricci tensor term and the fluid term cancel each other,
\bl
\label{GFcancel}
2\bar{R}_{\beta}^{~\mu} h_{\alpha\mu}h^{\alpha\beta}
+\frac12( \bar{\rho}-\bar{p})h_{\mu\nu}h^{\mu\nu}=0,
\el
so  that  \eqref{FPlagrangian} reduces to  the   action of GW
\bl
I_{gw}
  &= \int d^4x  \frac{1}{4} \sqrt{-\gamma}
  \Big(h_{\alpha\beta|\nu}h^{\alpha\beta|\nu}
+2\bar{R}_{\nu\alpha\beta\mu} h^{\mu\nu}h^{\alpha\beta}\Big) .
\label{Igw}
\el
The vanishing  variation $\delta  I_{gw} =0$ with respect to  $h_{\mu\nu}$
yields the GW equation \eqref{GWeq}.

 So far, the background equation \eqref{beisteineq}
contains only the fluid as the source.
To include the back-reaction of GW upon the  background spacetime,
one defines the conserved effective stress tensor of GW
\bl
\tau^{\mu\nu} &=-\frac{2}{\sqrt{-\gamma}}
\frac{\delta I_{gw}}{\delta\gamma_{\mu\nu}} ,
\label{uptaumunu}
\el
which consists of the second order terms like $( h_{\mu\nu})^2$.
The conservation of $\tau^{\mu\nu}$
is ensured by the GW equation \eqref{GWeq}.
Adding $\tau_{\mu\nu}$ to the source of the background equation, we arrive at
\bl
\bar G_{\mu\nu}(\gamma '_{\alpha\beta}) = \frac12 (\bar T_{\mu\nu} + \tau_{\mu\nu})
  .\label{backreacteq}
\el
This equation will determine a new background metric  $\gamma'_{\mu\nu}$
different from  $\gamma_{\mu\nu}$ that was determined by eq.\eqref{beisteineq}.
In this way,  GW generates  an impact upon the background metric.
In practice,  the magnitude of $\tau_{\mu\nu}$ is often smaller than $\bar T_{\mu\nu}$,
and its effect can be treated as a perturbation.
Therefore, the equation \eqref{backreacteq} gives a perturbation scheme of back-reaction.
As we shall see in Sect. \ref{sectionAnal},
the perturbation  scheme \eqref{backreacteq}
also arises as a result of a general scheme of the back-reaction.

\section{The Hilbert-Einstein action
and the fluid action expanded to the second order }\label{action}

It is revealing to show that
the GW action $I_{gw}$ can be also extracted from
the total action of gravity and fluid \cite{FordParker1977}
\bl
I & = I_G + I_F =\int d^4x\sqrt{-g}
\Big(- \frac{1}{16\pi G} R-\frac12\big((\rho+p)u^{\mu}u^{\nu}g_{\mu\nu}
-(\rho+3p)\big)\Big),\label{HEactionperfectfluid}
\el
where  $I_G$ is the Hilbert-Einstein  action and  $I_F$ is the fluid action.
The Einstein tensor and the stress tensor of the fluid are defined by
\bl
 G_{\mu\nu}&=-\frac{1}{\sqrt{-g}}\frac{ \delta I_{G}}{\delta g^{\mu\nu}},
\label{loweg}
 \\
 T_{\mu\nu}&=\frac{2}{\sqrt{-g}}\frac{ \delta I_{F}}{\delta g^{\mu\nu}},
\label{loweF}
\el
and  the corresponding upper indexed  tensors  are
\bl
 G^{\mu\nu}&=\frac{1}{\sqrt{-g}}\frac{ \delta I_{G}}{\delta g_{\mu\nu}},
\label{uppereg}
 \\
 T^{\mu\nu}&=-\frac{2}{\sqrt{-g}}\frac{ \delta I_{F}}{\delta g_{\mu\nu}} .
\label{upperef}
\el
In \eqref{uppereg}  a plus sign is consistent with
$\delta g ^{\mu\nu} = - g^{\mu\alpha} g^{\nu\beta}\delta g_{\alpha\beta}$
to ensure $G_{\mu\nu}=g_{\mu\alpha} g_{\nu\beta}G^{\alpha\beta}$.
Under the metric perturbation  \eqref{expandg},
the  action \eqref{HEactionperfectfluid} is expanded
up to the second order (see  Appendix \ref{AppendixA})
\bl
I&\simeq\int d^4x\sqrt{-\gamma}\Big(1 + \frac12 h
-\frac{1}{4} h_{\alpha\nu}h^{\alpha\nu} +\frac18 h^2 \Big)\nn
\\
&~~~~~ \times\Big(-\bar{R} -R^{(1)} -R^{(2)}
-\frac12\big((\bar{\rho}+\bar{p})\bar{u}^{\mu}\bar{u}^{\nu}\gamma_{\mu\nu}-(\bar\rho+3\bar p)\big)
-\frac12(\bar{\rho}+\bar{p})\bar{u}^{\mu}\bar{u}^{\nu}h_{\mu\nu}\Big),
\label{howtoexpand2nd}
\\
 &= I^{(0)}+I^{(1)}+I^{(2)},
\el
where
\bl
 I^{(0)} & =I_{G}^{(0)}+I_{F}^{(0)},
 \\
  I^{(1)} & =I_{G}^{(1)}+I_{F}^{(1)},
 \\
  I^{(2)} & =I_{G}^{(2)}+I_{F}^{(2)},
\el
the geometric actions of various order  are
\bl
I^{(0)}_G & = \int d^4x\sqrt{-\gamma} \Big(-\bar R \Big) , \label{I0G}
\\
I^{(1)}_G & = \int d^4x \sqrt{-\gamma}
\Big( -R^{(1)} - \frac12 h  \bar R \Big)\nn
\\
&= \int d^4x\sqrt{-\gamma}
\Big(-(h_{\alpha\sigma}^{~~~|\sigma|\alpha}
-h_{|\alpha}^{~~|\alpha}
-h^{\beta\sigma}\bar{R}_{\beta\sigma})
-\frac12 h\bar{R} \Big),
\label{I1stnofld1}
\\
 I^{(2)}_G  &=\int d^4x\sqrt{-\gamma}\Big( - R^{(2)} -\frac12 h  R^{(1)}
+\frac14 h_{\alpha\nu}h^{\alpha\nu} \bar{R}
-\frac18h^2 \bar{R} \Big)\nn
\\
&=  \int d^4x \sqrt{-\gamma}
\Big( -  \frac12 h^{\alpha\beta}_{~~~|\beta} h_{\alpha\nu}^{~~~|\nu}
         +  \frac12 h^{\alpha\beta}_{~~~|\beta}h_{|\alpha}
- \frac14 h^{|\alpha}h_{|\alpha}
+\frac12 h h^{\beta\sigma}\bar{R}_{\beta\sigma}
     -\frac1{8}h^2\bar{R}
     \nn \\
&~~~~~~~~~~~~~~~~~~  +  \frac14 \big(   h_{\alpha\beta|\nu}h^{\alpha\beta|\nu}
+ 2 \bar{R}_{\nu\rho\alpha\mu} h^{\mu\nu}h^{\rho\alpha}
- 2 \bar{R}_{\sigma\alpha} h^{\nu\alpha}  h^{\sigma}_{~\nu}
+ h_{\alpha\nu}h^{\alpha\nu}\bar{R} \,  \big)  \Big),
 \label{nofldI2nd}
\el
and  the  fluid actions of various order   are
\bl
 I^{(0)}_F & = \int d^4x\sqrt{-\gamma}
\Big( -\frac12 \big((\bar{\rho}+\bar{p}) \bar{u}^{\mu}\bar{u}^{\nu}\gamma_{\mu\nu}
-(\bar\rho+3\bar p)\big) \Big),\label{I0thnofld}
\\
I^{(1)}_F & = \int d^4x \sqrt{-\gamma}
\Big( - \frac14 h[(\bar{\rho}+\bar{p})
   \bar{u}^{\mu}\bar{u}^{\nu}\gamma_{\mu\nu}-(\bar\rho+3\bar p)]
   -\frac12 (\bar{\rho}+\bar{p})\bar{u}^{\mu}\bar{u}^{\nu}h_{\mu\nu} \Big) ,
\label{Fl1st1}
\\
 I^{(2)}_F  &=\int d^4x \sqrt{-\gamma}
\Big(-\frac14h(\bar{\rho}+\bar{p})\bar{u}^{\mu}\bar{u}^{\nu}h_{\mu\nu}
  -\frac{1}{16}h^2[(\bar{\rho}+\bar{p})
\bar{u}^{\mu}\bar{u}^{\nu}\gamma_{\mu\nu}
-(\bar\rho+3\bar p)]
\nn
\\
&~~~~~~~~~~~~~~~~~~
  +\frac18h_{\alpha\beta}h^{\alpha\beta}[(\bar{\rho}+\bar{p})
      \bar{u}^{\mu}\bar{u}^{\nu}\gamma_{\mu\nu}-(\bar\rho+3\bar p)]
\Big). \label{I2nd}
\el
The background metric  $\gamma_{\mu\nu}$
and the perturbation $h_{\mu\nu}$ are regarded as independent variables.
In general,   $h$ and $h^{\mu\nu}_{~~|\nu}$
 contain the scalar and vector metric perturbations.
The  variation of the 0th order action
with respect to the background metric  $\gamma_{\mu\nu}$
leads to  the 0th  order Einstein equation \eqref{beisteineq},
\bl
\frac{1}{\sqrt{-\gamma}}
  \frac{\delta  }{\delta \gamma^{\mu\nu}}(I_G^{(0)}  +I^{(0)}_F )
  =0 ,
\label{I0theqfl}
\el
where
\bl
\bar{ G}_{\mu\nu} & \equiv - \frac{1}{\sqrt{-\gamma}}
      \frac{\delta I^{(0)}_G}{\delta \gamma^{\mu\nu}} ,
\label{downtwoindexG}
\\
\bar T_{\mu\nu} & \equiv   \frac{2}{\sqrt{-\gamma}}
           \frac{\delta I^{(0)}_F}{\delta \gamma^{\mu\nu}}.
\label{fluiddown2G}
\el

So far,
the GW conditions \eqref{coordcondition}  \eqref{transvcondition} \eqref{conditionsforGW}
have not been used in the above expressions \eqref{I0G}---\eqref{I2nd}.
Varying the 1st order actions  \eqref{I1stnofld1} and \eqref{Fl1st1}
with respect to the background metric $\gamma_{\mu\nu}$
and then imposing the TT condition  $h^{\mu\nu}_{~~|\nu}=h=0$, we obtain
\bl
\delta I^{(1)}_G  & = \int d^4x \sqrt{-\gamma}\Big(
\bar{R}_{\rho\sigma} h^{\sigma}_{\lambda}
-\frac12  h_{\lambda\rho} \bar{R}
-\bar{R}_{\rho\nu\beta\lambda}  h^{\nu\beta}
+\frac12\Box h_{\rho\lambda}
 \Big)
  \delta \gamma^{\lambda\rho},
\label{d1e}
\\
\delta I^{(1)}_F & = \int d^4x \sqrt{-\gamma}
    \frac{1}{2}  \bar{p}\,  h_{\lambda\rho} \delta \gamma^{\lambda\rho}  .
 \label{542}
\el
In deriving \eqref{d1e}, we have taken  $\delta h_{\mu\nu}=0$
as $h_{\mu\nu}$ is independent of $\gamma_{\mu\nu}$,
and used the following formulae
\[
\delta R_{\mu\nu}=(\delta\Gamma^{\rho}_{\mu\nu})_{|\rho}
-(\delta\Gamma^{\rho}_{\mu\rho})_{|\nu},
\]
and
\bl
(\delta\Gamma^{\rho}_{\beta\sigma})_{|\rho}
& =  \frac12 \gamma^{\rho\theta}(\delta\gamma_{\beta\theta|\sigma|\rho}
+\delta\gamma_{\sigma\theta|\beta|\rho}
-\delta\gamma_{\beta\sigma|\theta|\rho}) ,
\nn
\\
(\delta\Gamma^{\rho}_{\beta\rho})_{|\sigma}
&= \frac12 \gamma^{\rho\theta}(\delta\gamma_{\beta\theta|\rho|\sigma}
+\delta\gamma_{\rho\theta|\beta|\sigma}
-\delta\gamma_{\beta\rho|\theta|\sigma}) ,
\nn
\\
h_{\mu\nu|\alpha|\beta} - h_{\mu\nu|\beta|\alpha}
& = -R_{\mu\rho\alpha\beta} h^{\rho}_{~\nu}
       -R_{\nu \rho\alpha\beta} h^{~~\rho}_{\mu}.
 \label{commRR}
\el
Then setting
\bl
\frac{1}{\sqrt{-\gamma}}
  \frac{\delta  }{\delta \gamma^{\mu\nu}}(I_G^{(1)}  +I^{(1)}_F )
  =0 ,
  \label{deltaI1GF}
\el
gives   the following
\bl
\Big(
\frac12 \bar{R}_{\rho\sigma} h^{\sigma}_{\lambda}
+ \frac12 \bar{R}_{\lambda \sigma} h^{\sigma}_{\rho}
-\frac12 \bar{R} h_{\rho\lambda}
-\bar{R}_{\rho\nu\beta\lambda}  h^{\nu\beta}
+\frac12\Box h_{\rho\lambda}
 \Big)
=  \frac12 (  -\bar{p} ) \,  h_{\lambda\rho}   ,
\label{G1munueq}
\el
which, by use of the cancelation equation \eqref{GFcancel},
 leads to the GW equation \eqref{GWeq}.
If the GW conditions \eqref{coordcondition}
 \eqref{transvcondition} \eqref{conditionsforGW}
are imposed first,
the 1st order actions will be vanishing, $I^{(1)}_{G}=I^{(1)}_{F}=0$.
(The variation of the 1st order actions \eqref{I1stnofld1} \eqref{Fl1st1}
with respect to $h_{\mu\nu}$
also leads to the background equation \eqref{beisteineq}.)

The second order actions \eqref{nofldI2nd} \eqref{I2nd}
are the central part of our concern.
To facilitate the calculation of the GW stress tensor
and the comparison with Ford and Parker's result \cite{FordParker1977},
we split \eqref{nofldI2nd} and  \eqref{I2nd} into the following parts
\bl
I^{(2)}_G & = I^{(2)}_{g1} +I^{(2)}_{g2} +I_{gw} ,
 \label{totG123I}
 \\
 I^{(2)}_F &= I^{(2)}_{f1}+I^{(2)}_{f2} ,
 \label{I2F}
\el
where
the geometric part
\bl
& I^{(2)}_{g1}= \int d^4x\sqrt{-\gamma}
 \Big( -  \frac12 h^{\alpha\beta}_{~~~|\beta} h_{\alpha\mu}^{~~~|\mu}
   +  \frac12 h^{\alpha\beta}_{~~~|\beta}h_{|\alpha}
  - \frac14 h^{|\alpha}h_{|\alpha}
  +\frac12h  h^{\beta\sigma}\bar{R}_{\beta\sigma}
  -\frac1{8}h^2\bar{R}\Big) ,
\label{121I}
   \\
& I^{(2)}_{g2}= \int d^4x\sqrt{-\gamma} \frac14
   \Big(- 2 \bar{R}_{\mu\alpha} h^{\mu}_{~\beta}  h^{\alpha\beta}
     + h_{\alpha\beta} h^{\alpha\beta} \bar{R} \Big),
  \label{123I}
\el
and $I_{gw}$ is the GW action given by \eqref{Igw},
and the fluid part
\bl
I^{(2)}_{f1}= & \int d^4x \sqrt{-\gamma}
\Big( -\frac14h(\bar{\rho}+\bar{p})\bar{u}^{\mu}\bar{u}^{\nu}h_{\mu\nu}
-\frac{1}{16}h^2 ( (\bar{\rho}+\bar{p})
\bar{u}^{\mu}\bar{u}^{\nu}\gamma_{\mu\nu}-(\bar\rho+3\bar p) )
  \Big),
\label{I2F1} \\
I^{(2)}_{f2} = & \int d^4x \sqrt{-\gamma}
\frac18 h_{\alpha\beta}h^{\alpha\beta}
  \Big((\bar{\rho}+\bar{p})
      \bar{u}^{\mu}\bar{u}^{\nu}\gamma_{\mu\nu}-(\bar\rho+3\bar p) \Big)
\nn
\\
= & \int d^4x \sqrt{-\gamma}
\frac14   h_{\alpha\beta}h^{\alpha\beta}  ( - \bar p )
  . \label{I2F3}
\el
Although the second order geometric action  $I^{(2)}_{G}$ is unique,
it can appear in different forms.
Integrating the first term in \eqref{121I} by parts twice
and using the formula \eqref{formula1}
leads to the following  relation,
\bl
h^{\alpha\beta}_{~~~|\beta} h_{\alpha\mu}^{~~~|\mu}
 =  h^{\alpha\beta}_{~~|\mu} h^{\mu}_{~\alpha|\beta}
+\bar{R}_{\alpha \mu\nu\beta} h^{\alpha\beta} h^{\mu\nu}
+\bar{R}_{ \mu\beta}h^{\alpha\beta}  h^{\mu}_{~\alpha} ,
 \label{1stterm}
\el
which is valid up to some dropped surface terms.
Then  $I^{(2)}_{G}$ can be also grouped as the following
form adopted by MacCallum and Taub  \cite{MacCallum1973}
\bl
I^{(2)}_{G} & = J_1+J_2 ,
\label{IG2J1J2}
\el
with
\bl
J_1 & \equiv \int d^4x\sqrt{-\gamma} \Big[
     -\bar{R}_{\sigma\alpha}h^{\nu\alpha}h^{\sigma}_{~\nu}
   +\frac14 h_{\alpha\nu}h^{\alpha\nu}\bar{R}
   +\frac12 h   h^{\beta\sigma}\bar{R}_{\beta\sigma}
    -\frac1{8}h^2\bar{R}    \Big] ,
\label{J1}
\\
J_2 &  \equiv
\int d^4x \sqrt{-\gamma}
 \Big[ \frac14 h_{\alpha\beta|\nu} h^{\alpha\beta|\nu}
   -\frac12 h^{\alpha\beta}_{~~|\nu} h^{\nu}_{~\alpha|\beta}
    +  \frac12 h^{\alpha\beta}_{~~~|\beta} h_{|\alpha}
     -\frac14  h^{|\alpha}h_{|\alpha}    \Big] .
\label{J2}
\el
$J_1$ and $J_2$ are also adopted in  Refs.\cite{stein2011,Deepen2022}.
We notice that,   in Ref.\cite{Butcher2009},
the action $S_{2G}$ after integrating by parts
 is equivalent to $J_2$,
but $S_{2H}$ does not have the term $\bar{R}_{\sigma\alpha}h^{\nu\alpha}h^{\sigma}_{~\nu}$
and  is not equal to $J_1$ of \eqref{J1}.

The variation of the action  $J_2$ of \eqref{J2} with respect to $h_{\mu\nu}$
leads to an equation as the following
\bl
-\frac12 \Box  h^{\alpha\beta }
  + \frac12 ( h^{\nu\alpha|\beta}\, _{|\nu} +  h^{\nu\beta|\alpha}\, _{|\nu} )
  -  \frac14 ( h^{|\alpha|\beta} +  h^{|\beta|\alpha} )
  - \frac12 \gamma^{\alpha\beta} h^{\mu\nu}\, _{|\nu|\mu}
  + \frac12 \gamma^{\alpha \beta}  h^{|\nu}_{~~|\nu}  =0,
\label{118}
\el
which can be rewritten   as
\bl
-\frac12 \Box \tilde{h}_{\alpha\beta}
+ \frac12 ( \tilde{h}_{\alpha}^{~~\rho}\, _{|\beta|\rho}
   + \tilde{h}_{\beta}^{~~\rho}\, _{|\alpha|\rho} )
-\frac12 \gamma_{\alpha\beta} \tilde{h}^{\lambda \rho}\, _{|\lambda |\rho} =0 ,
\label{MacCallum-Taubkeq}
\el
with $\tilde{h}_{\alpha\beta}\equiv h_{\alpha\beta}-\frac12 h \gamma_{\alpha\beta}$.
It should be pointed out that eq.\eqref{118},
as well as eq.\eqref{MacCallum-Taubkeq},
is not the correct equation of GW in the presence of matter,
even though it often appeared in literature
\cite{Isaacson1968a,MacCallum1973,stein2011,Butcher2009}.
(We check that $G_{abcd} h^{cd}$ in Ref.\cite{Butcher2009}
is not equal to the first order
perturbed  Einstein tensor, $G_{abcd} h^{cd} \ne G^{(1)}_{ab}$,
  in the presence of matter,
on contrary to the claim below eq.(13) in Ref.\cite{Butcher2009}.)

Now impose the TT gauge condition.
Many terms become  vanishing,   and the actions are simplified.
The 2nd order actions \eqref{121I}   \eqref{I2F1} are  vanishing,
\bl \label{I1is0}
I^{(2)}_{g1} & = I^{(2)}_{f1}=0 ,
\el
and so are their variations with respect to the background metric,
\bl
\delta I^{(2)}_{g1} &  = \delta I^{(2)}_{f1}=0 .
\label{deltaI1is0}
\el
By use of eq.\eqref{GFcancel},
the 2nd order actions \eqref{123I} \eqref{I2F3}  cancel each other,
\bl
I^{(2)}_{g2} + I^{(2)}_{f2}
&= \int d^4x\sqrt{-\gamma}
        \frac{1}{4}\Big( - 2 \bar{R}_{\mu\alpha} h^{\mu}_{~\beta}  h^{\alpha\beta}
 + h_{\alpha\beta} h^{\alpha\beta} \bar{R}
 -  h_{\alpha\beta} h^{\alpha\beta} \bar{p}  \Big)
 =0  .
\label{2ndactionGW}
\el
Thus, the second order action in the TT gauge  reduces to
\bl
I^{(2)}= I_{gw} .
\label{I2equaltoIG2}
\el
This confirms the  GW action \eqref{Igw} in  the last section.
The important point  is that
 $I^{(2)}_{g2}$  will not affect the GW equation \eqref{GWeq},
nor contribute to the GW stress tensor \eqref{uptaumunu},
because it has been canceled out
by the fluid  $I^{(2)}_{f2}$ through eq.\eqref{2ndactionGW}.
This cancelation should be taken into account
for the back-reaction,
as we shall see in Sect. \ref{effectivest} and Sect. \ref{sectionAnal}.

In the TT gauge, the actions
\eqref{J1} and \eqref{J2}  reduce  to the following
\bl
J_1 & =  \int d^4x\sqrt{-\gamma} \Big(
     -\bar{R}_{\sigma\alpha}h^{\nu\alpha}h^{\sigma}_{~\nu}
   +\frac14 h_{\alpha\nu}h^{\alpha\nu}\bar{R} \Big) ,
\label{J1GW}
\\
J_2 & =   \int d^4x\sqrt{-\gamma}
 \Big( \frac14 h_{\alpha\beta|\nu} h^{\alpha\beta|\nu}
   -\frac12 h^{\alpha\beta}_{~~|\nu} h^{\nu}_{~\alpha|\beta}  \Big) ,
\label{J2GW}
\el
and   the relation \eqref{1stterm} reduces to
\bl
 h^{\alpha\beta}_{~~|\mu} h^{\mu}_{~\alpha|\beta}
+\bar{R}_{\alpha \mu\nu\beta} h^{\alpha\beta} h^{\mu\nu}
+\bar{R}_{ \mu\beta}h^{\alpha\beta}  h^{\mu}_{~\alpha}   =0 .
 \label{1stterm0}
\el
So  $J_2 $ can be also written as
\bl
 J_2 =  I_{gw}+ \int d^4x\sqrt{-\gamma}
 \Big( \frac{1}{2} \bar{R}_{ \mu\beta}h^{\alpha\beta} h^{\mu}_{~\alpha} \Big)
   , \label{I2equaltoIG2-2}
\el
containing an extra   $\bar{R}_{ \mu\nu}$ term than  $I_{gw}$.
Eq.\eqref{MacCallum-Taubkeq} in the TT gauge reduces to
\bl
- \Box  h^{\alpha\beta }
  + ( h^{\nu\alpha|\beta}\, _{|\nu} +  h^{\nu\beta|\alpha}\, _{|\nu} ) =0 ,
\label{MeacCallumTaubeq}
\el
which, by use of the formula \eqref{commRR},  can be rewritten as
(see (2.1a) in Ref.\cite{Isaacson1968b})
\bl
- \Box  h^{\alpha\beta }
    +  2  \bar{R}^{\alpha}\, _{\mu\nu}\, ^{\beta} h^{\mu\nu}
    +   (  \bar{R}^{\alpha\nu}  h^{\beta}_{~\nu}
    +\bar{R}^{\beta\nu}  h^{\alpha}_{~\nu} ) =0 .
\label{FierzPaulieq1}
\el
Obviously, eq.\eqref{FierzPaulieq1} differs  from the correct GW equation \eqref{GWeq},
 due to the  two $\bar{R}^{ \mu\nu}$ terms
 that came from the  $\bar{R}_{ \mu\nu}$ term in $J_2$ of \eqref{I2equaltoIG2-2}.
This  $\bar{R}_{ \mu\nu}$ term in  $J_2$,  together with the whole  $J_1$,
should be canceled out by the fluid term,
as we have shown in  \eqref{2ndactionGW}.
So the division of the second order action into $J_1+ J_2$ is not pertinent.

In fact,
$J_2$  is the curved-spacetime generalization of the Fierz-Pauli action
\cite{FierzPauli1939,Padmanabhan2008,Butcher2009}.
As is known, the Fierz-Pauli action
for a massless spin-2 field in the flat spacetime
leads to some inconsistency (see p180 -- p186 in Ref.\cite{MTW}).
Here we see that  $J_2$ in curved spacetimes
does not lead to the equation of GW.
Isaacson first used $J_2$ to give the associated equation \eqref{MacCallum-Taubkeq},
but rejected it in searching for an effective stress tensor of GW.
(See (5.7) in Ref.\cite{Isaacson1968a}.)

\section{The conserved  $\tau^{\mu\nu}$
and the nonconserved  $T^{\mu\nu}_\text{MT}$}\label{sect4FP}

For a general curved spacetime  we calculate
the variation $\delta I_{gw}$ with respect to $\gamma_{\alpha\beta}$
(see \eqref{C20IG2} in Appendix \ref{Appvaaction}).
By the definition \eqref{uptaumunu},
we obtain the conserved effective stress tensor of GW
as the following
\bl
\tau^{\alpha\beta} =&\frac{1}{2}
\Big(  h_{\mu\nu}^{~~~|\alpha} h^{\mu\nu|\beta}
-\frac12\gamma^{\alpha\beta} ( h_{\mu\nu|\gamma}h^{\mu\nu|\gamma}
+2\bar{R}_{\mu\rho\theta\nu} h^{\mu\nu}h^{\rho\theta} )
\nn
\\
&   + (h^{\mu\alpha}_{~~|\rho}h^{\rho\beta}_{~~|\mu}
    +  h^{\mu\beta}_{~~|\rho}h^{\rho\alpha}_{~~|\mu} )
 +  ( \bar{R}^{\nu~~\alpha}_{~\rho\sigma} h^{\beta}_{~\nu}h^{\rho\sigma}
       +\bar{R}^{\nu~~\beta}_{~\rho\sigma} h^{\alpha}_{~\nu}h^{\rho\sigma} )
\nn
\\
&  -( h^{\beta}_{~\nu |\mu} h^{\mu\nu |\alpha}
 + h^{\alpha}_{~\nu|\mu  } h^{\mu\nu |\beta} )
 -2 h^{\alpha\beta}\, _{|\sigma|\rho}h^{\rho\sigma}
 +(h^{\sigma\nu }h^{\alpha~|\beta}_{~\nu~~|\sigma}
+ h^{\sigma\nu }h^{\beta~|\alpha}_{~\nu~~|\sigma})
 \Big) .
\label{covt2}
\el
This expression is
a generalization of Ford-Parker's result in fRW spacetimes \cite{FordParker1977},
and is valid for the class of curved spacetimes
 in which the GW conditions \eqref{coordcondition}\eqref{transvcondition}\eqref{conditionsforGW}
can be imposed consistently.
Note that  the expression  \eqref{covt2} can be further simplified in applications.
Writing $h_{\mu\nu}(x) = \sum_{\bf k} h_{\mu\nu}(k) e^{i{\bf k}\cdot {\bf x}}$,
by $\int e^{i{(\bf k- k')}\cdot {\bf x}}  d^3 x =0$ for ${\bf k} \ne {\bf k}'$,
only the Fourier modes with the same $\bf k$ will remain
in  $h_{\mu\nu}(x) h_{\alpha \beta}(x)$ in the action $ I_{gw}$.
This applies to \eqref{covt2} too.
We notice that the stress tensor (3.36) in Ref.\cite{Giovannini2019prd} (with some typos)
and   (3.114) in Ref.\cite{Giovannini2020} differ from our result  \eqref{covt2}.

The lower-indexed conserved  GW stress tensor is defined  by
\bl
\tau_{\alpha\beta} & =\frac{2}{\sqrt{-\gamma}}
\frac{\delta I_{gw} }{\delta\gamma^{\alpha\beta}} ,
\label{deflowtau}
\el
so that  $\tau_{\alpha\beta}= \gamma_{\alpha\mu}\gamma_{\beta\nu} \tau^{\mu\nu}$.
The plus sign of   \eqref{deflowtau} is consistent with
the minus sign of \eqref{uptaumunu},
because $\delta\gamma_{\alpha\beta}=
-\gamma_{\lambda\alpha}\gamma_{\rho\beta}\delta\gamma^{\rho\lambda}$.
Similarly,  the mixed indexed stress tensor is given by
$\tau^{\alpha}_{~\beta}= \gamma_{\beta\sigma} \tau^{\alpha\sigma}
 =  \gamma^{\alpha\sigma} \tau_{\sigma \beta}$.
So,  $\tau_{\alpha\beta}$,  $\tau^{\alpha\beta}$,
and $\tau^{\alpha}_{~\beta}$ represent
the same tensor defined on the background spacetime.
In contrast, the second order Einstein tensor $G^{(2)}_{\mu\nu}$
is actually not a true tensor,
as we shall see in Sect.\ref{effectivest}.

Now we  prove that
the GW  stress tensor \eqref{covt2} is conserved.
Taking the covariant four-divergence upon \eqref{covt2} gives
\bl
 \tau^{\alpha\beta}\, _{|\beta}=&\frac{1}{2}\
\Big(
 h_{\mu\nu}^{~~~|\alpha}\, _{|\beta} h^{\mu\nu|\beta}
 +  h_{\mu\nu}^{~~~|\alpha}\Box h^{\mu\nu}
\nn
\\
&  + h^{\mu\alpha}_{~~|\rho|\beta}h^{\rho\beta}_{~~|\mu}
+ h^{\mu\alpha}_{~~|\rho} h^{\rho\beta}_{~~|\mu|\beta}
+ h^{\mu\beta}_{~~|\rho|\beta}h^{\rho\alpha}_{~~|\mu}
+ h^{\mu\beta}_{~~|\rho}h^{\rho\alpha}_{~~|\mu|\beta}
\nn
\\
& - h^{\beta}_{~\nu|\mu|\beta }  h^{\mu\nu |\alpha}
- h^{\beta}_{~\nu |\mu }  h^{\mu\nu|\alpha}\, _{|\beta}
- h^{\alpha}_{~\nu |\mu|\beta}   h^{\mu\nu|\beta}
- h^{\alpha}_{~\nu|\mu} \Box h^{\mu\nu}
\nn
\\
&
+  h^{\sigma\nu }\, _{|\beta}h^{\alpha~|\beta}_{~\nu~|\sigma}
+ h^{\sigma\nu }(h^{\alpha~|\beta}_{~\nu~|\sigma})_{|\beta}
+ h^{\sigma\nu }\, _{|\beta} h^{\beta~|\alpha}_{~\nu~|\sigma}
+ h^{\sigma\nu }(h^{\beta~|\alpha}_{~\nu~|\sigma})_{|\beta}
\nn
\\
& -2h^{\alpha\beta}_{~~|\sigma|\rho|\beta}h^{\rho\sigma}
  -2h^{\alpha\beta}_{~~|\sigma|\rho}h^{\rho\sigma}_{~~|\beta}
\nn
\\
&+  [ (\bar{R}^{\nu~~\alpha}_{~\rho\sigma})_{|\beta}
h^{\beta}_{~\nu}h^{\rho\sigma}
+\bar{R}^{\nu~~\alpha}_{~\rho\sigma}
h^{\beta}_{~\nu|\beta}h^{\rho\sigma}
+\bar{R}^{\nu~~\alpha}_{~\rho\sigma}
h^{\beta}_{~\nu}h^{\rho\sigma}_{~~|\beta} ]
\nn
\\
&+  [ (\bar{R}^{\nu~~\beta}_{~\rho\sigma})_{|\beta}
h^{\alpha}_{~\nu}h^{\rho\sigma}
+\bar{R}^{\nu~~\beta}_{~\rho\sigma}
h^{\alpha}_{~\nu|\beta}h^{\rho\sigma}
+\bar{R}^{\nu~~\beta}_{~\rho\sigma}
h^{\alpha}_{~\nu}h^{\rho\sigma}_{~~|\beta} ]
\nn
\\
&-\frac12\gamma^{\alpha\beta}[2h_{\mu\nu|\gamma|\beta}h^{\mu\nu|\gamma}
+2\bar{R}_{\nu\rho\theta\mu|\beta} h^{\mu\nu}h^{\rho\theta}
+4\bar{R}_{\nu\rho\theta\mu} h^{\mu\nu}_{~~|\beta}h^{\rho\theta}  ]\Big) .
\label{covaianttmunuGW4div}
\el
The  right hand side of \eqref{covaianttmunuGW4div} is a complicated expression
and can be further simplified as the following.
By using  the GW equation \eqref{GWeq}
and the following formulae (see  (43.21) in Ref.\cite{Vfork1964})
\bl
U_{\mu\nu|\alpha|\beta}-
U_{\mu\nu|\beta|\alpha}
&=-R_{\mu\rho\alpha\beta} U^{\rho}_{~\nu}
-R_{\nu \rho\alpha\beta} U^{~~\rho}_{\mu}
 ,   \label{formula1}
\\
U_{\mu\nu\lambda |\alpha|\beta}-
U_{\mu\nu\lambda|\beta|\alpha}
&=-R_{\mu\rho\alpha\beta} U^{\rho}_{~\nu\lambda}
-R_{\nu\rho\alpha\beta} U_{\mu ~~\lambda}^{~\rho}
-R_{\lambda \rho\alpha\beta} U_{\mu\nu}^{~~\rho},\label{formula2}
\el
with $U_{\mu\nu}$, $U_{\mu\nu\lambda}$  being some  tensors,
and Bianchi identities \cite{Vfork1964}
\bl
\bar R^{\alpha}_{~\rho\beta\mu}
+ \bar R^{\alpha}_{~\mu\rho\beta}
+ \bar R^{\alpha}_{~\beta\mu\rho}&=0,\label{formula4}
\\
\bar R_{\alpha\beta\sigma\rho|\theta}
+\bar R_{\alpha\beta\theta\sigma|\rho}
+\bar R_{\alpha\beta\rho\theta|\sigma}&=0,\label{formula5}
\el
and \cite{BarthChristensen1983}
\bl
 \bar R_{\alpha\rho\beta\gamma}\, ^{|\gamma}
&= \bar R_{\alpha\beta|\rho}
  - \bar R_{\rho\beta|\alpha},\label{formula3}
\el
after some algebraic calculations,
\eqref{covaianttmunuGW4div} is rewritten as the following
\bl
 \tau^{\alpha\beta} \, _{|\beta}
&=\frac{1}{2} \Big( ( \bar{R}^{~\alpha}_{\rho}  h^{\rho}_{~\nu|\sigma}
+ \bar{R}_{\rho\sigma}  h^{\alpha}_{~\nu}\, ^{|\rho}
 -2 \bar{R}_{\sigma\rho}  h^{\rho \alpha}_{~~|\nu} )  h^{\sigma\nu}
\nn
\\
& ~~~~ + ( \bar{R}^{~\alpha}_{\beta}\, _{|\sigma} h^{\beta}_{~\rho}
   -\bar{R}^{\nu}_{~\sigma|\rho}  h^{\alpha}_{~\nu}
   - \bar{R}^{~~~|\nu}_{\rho\sigma}  h^{\alpha}_{~\nu} ) h^{\rho\sigma}  \Big)
  . \label{csvFP}
\el
In  presence of fluid ($\bar R_{\mu\nu} \ne 0$),
by using the background equation \eqref{ei1st}
and the GW conditions \eqref{coordcondition} \eqref{transvcondition} \eqref{conditionsforGW},
the right hand side of \eqref{csvFP} is zero,
\bl
 \tau^{\alpha\beta}\, _{|\beta}=0  . \label{FordPconservedpaper}
\el
So $\tau^{\alpha\beta}$ is covariantly conserved,
and is referred to as the conserved GW stress tensor.
In  absence of matter  ($\bar R_{\mu\nu} = 0$)
 eq.\eqref{FordPconservedpaper} holds too.

MacCallum and Taub \cite{MacCallum1973} proposed an effective stress tensor
for the back-reaction by varying $J_2$ with respect to $\gamma_{\alpha\beta}$,
but did not give its explicit expression.
By calculation (see Appendix  \ref{j2append}),
we derive  the MacCallum-Taub effective stress tensor
 in TT gauge as the following
\bl
T_{\text{MT}}^{\alpha\beta} & \equiv - \frac{2}{\sqrt{-\gamma}}
\frac{\delta J_2 }{\delta\gamma_{\alpha\beta}}
\nn
\\
& =    \frac{1}{2}h_{\mu \nu}\,^{ |\alpha} h^{\mu \nu \mid \beta}
-\gamma^{\alpha \beta}(\frac{1}{4} h_{\mu \nu \mid \rho} h^{\mu \nu \mid \rho}
-\frac{1}{2} h^{\mu \nu}\,_{\mid \rho} h_{~\mu \mid \nu}^\rho)
\nn \\
& ~~~ +h^{\beta \rho \mid \nu} h_{~\nu \mid \rho}^\alpha
+h^{\mu \alpha \mid \nu} h_{~\mu \mid \nu}^\beta
-(h^{\alpha \rho \mid \nu} h_{\nu \rho}\,^{\mid \beta}
+h^{\beta \rho \mid \nu} h_{\nu \rho}\,^{\mid \alpha})
\nn \\
& ~~~ +\bar{R}^{\alpha}_{~\rho\gamma\nu}h^{\rho\gamma}h^{\beta\nu}
+\bar{R}_{\rho\nu}h^{\rho\alpha}h^{\beta\nu}
+\bar{R}^{\beta}_{~\rho\gamma\nu}h^{\rho\gamma}h^{\alpha\nu}
+\bar{R}_{\rho\nu}h^{\rho\beta}h^{\alpha\nu}
 -h^{\beta \alpha \mid \nu}{ }_{\mid \mu} h_{~\nu}^\mu
    \, .
\label{MTstress}
\el
This is a  tensor on the background spacetime, like $\tau^{\mu\nu}$.
The expression \eqref{MTstress}
  differs from (20b) of Ref.\cite{stein2011} in the TT gauge.
As we have checked, the four-divergence of \eqref{MTstress} is nonzero,
\bl
T^{\mu\nu}_\text{MT}\,_{|\nu} \ne 0 ,
\label{MTnonconserv}
\el
ie, the MacCallum-Taub stress tensor is not conserved, unlike $\tau^{\mu\nu}$.
Obviously, when  the  $\bar{R}_{ \mu\nu}$ term in $J_2$
is canceled out by the fluid term,
$T_{\text{MT}}^{\mu\nu}$ will reduce to
the conserved $\tau^{\mu\nu}$.

\section{The second order Einstein tensor  $G^{(2)}_{\mu\nu}$ }\label{effectivest}

The second order perturbed Einstein tensor  $G^{(2)}_{\mu\nu}$
as the effective stress tensor of GW
was initially discussed by Brill and Hartle \cite{BrillHartle1964}
to study the gravitational geons,
and further developed by Isaacson \cite{Isaacson1968a,Isaacson1968b,MTW}
for the vacuum case,
and  then used in the context of the expanding Universe
\cite{AbramoPRD1997,AbramoPRD1999,BrandTaka2018}.

Under the metric perturbation \eqref{expandg},
the Einstein tensor and the fluid stress tensor
can be expanded in powers of $h_{\alpha\beta}$ as the following
\bl\label{expeff}
G_{\mu\nu} & = \bar G_{\mu\nu}(\gamma_{\alpha\beta})
    + G^{(1)}_{ \mu\nu}(\gamma_{\alpha\beta},h_{\alpha\beta})
    +G^{(2)}_{\mu\nu}(\gamma_{\alpha\beta},h_{\alpha\beta})  + O(h^3) ,
    \\
T_{\mu\nu} & = \bar T_{\mu\nu}(\gamma_{\alpha\beta})
    + T^{(1)}_{ \mu\nu}(\gamma_{\alpha\beta},h_{\alpha\beta})
    + T ^{(2)}_{\mu\nu}(\gamma_{\alpha\beta},h_{\alpha\beta})  + O(h^3) ,
\el
where $G^{(1)}_{ \mu\nu}$ and $G^{(2)}_{ \mu\nu}$ are respectively
the first and second order perturbed Einstein tensors,
and    $T^{(1)}_{ \mu\nu}$ and $T^{(2)}_{ \mu\nu}$ are
the perturbed stress tensors of fluid.
Then the Einstein equation $G_{\mu\nu} = \frac12 T_{ \mu\nu}$
can be decomposed into  two equations
\bl
\bar G_{\mu\nu} & = \frac12 \bar T_{\mu\nu}
- G^{(2)}_{\mu\nu} + \frac12 T^{(2)}_{ \mu\nu} ,  \label{2ndexpeff}
\\
G^{(1)}_{ \mu\nu} & = \frac12 T^{(1)}_{ \mu\nu}  ,  \label{linexpeff}
\el
and the higher order $ O(h^3) $ terms are neglected.
Plugging the expression $G^{(1)}_{ \mu\nu}$ of \eqref{G1}
and  $T^{(1)}_{\mu\nu}$ of \eqref{Tmunu1}
into the 1st order equation \eqref{linexpeff} gives the GW equation \eqref{GWeq}.
Note that the 2nd order quantities,
$-G^{(2)}_{\mu\nu}$ and $\frac12 T^{(2)}_{ \mu\nu}$,
are kept into the 0th-order equation \eqref{2ndexpeff},
and both serve effectively as part of the source.
Thus, \eqref{2ndexpeff} is regarded as the back-reaction equation,
and $ G^{(2)}_{\mu\nu}$ (up to a factor $-2$)
is  called the effective stress tensor of GW
\cite{BrillHartle1964,Isaacson1968a,Isaacson1968b,AbramoPRD1997}
\bl
T^{\text{eff} }_{\mu\nu} \equiv      - 2 G^{(2)}_{\mu\nu}
   =-2 \big(R^{(2)} _{\mu\nu} -\frac12 h_{\mu\nu}   R^{(1)}
   -\frac12\gamma_{\mu\nu}  R^{(2)}\big)
   . \label{effdef}
\el
(Isaacson  \cite{Isaacson1968b} used  the simplified definition
$T^{\text{eff}}_{\mu\nu} \equiv  -2 \big(R^{(2)}_{\mu\nu}
 -\frac{1}{2} \gamma_{\mu\nu} \gamma^{\alpha\beta} R^{(2)}_{\alpha\beta} \big)$
 for the vacuum case with $\bar{R}_{\mu\nu}=0$ and  ${R}^{(1)}_{\mu\nu}=0$.)
In principle, the equations \eqref{2ndexpeff} and \eqref{linexpeff} should be solved
self-consistently as a set \cite{BrillHartle1964,Isaacson1968a,Isaacson1968b},
and this task will be complicated.
We refer to the set \eqref{2ndexpeff} and \eqref{linexpeff}
  as the general scheme of back-reaction.
In practice,   nevertheless, $G^{(2)}_{\mu\nu}$ and $T^{(2)}_{\mu\nu}$
are generally much lower than $\bar T_{ \mu\nu}$ in magnitude
and can be treated as a perturbation.
Consequently this general scheme will reduce to
the perturbation scheme \eqref{backreacteq},
when a nonconserved part of $G^{(2)}_{\mu\nu}$ will be canceled by the fluid term,
as we shall show in Sect.\ref{sectionAnal}.

Initially,  Refs. \cite{BrillHartle1964,Isaacson1968a,Isaacson1968b}
studied the vacuum case,
where eq.\eqref{2ndexpeff} and eq.\eqref{linexpeff} took the following simple form
\bl
\bar G_{\mu\nu} & = - G^{(2)}_{\mu\nu} ,  \label{vacuum2nd}
\\
G^{(1)}_{ \mu\nu} & = 0  . \label{vacuumlin}
\el
By the perturbation treatment,
the nonconserved part of $G^{(2)}_{\mu\nu}$  in absence of fluid
will be vanishing,
and  eq.\eqref{vacuum2nd} will reduce to
\bl
\bar G_{\mu\nu}  = \tau_{\mu\nu}   ,
\label{tauvac2nd}
\el
(see  Sect.\ref{sectionAnal}.)
Eq.\eqref{tauvac2nd} can describe   an expanding Universe
driven only by the relic GW background.

Ref.\cite{Butcher2009} considered   the following scheme
\bl
\bar G_{\mu\nu} & =   \frac12 \bar T_{\mu\nu} ,
\nn
\\
G^{(1)}_{ \mu\nu} & =   - G^{(2)}_{\mu\nu} ,
\el
where $- G^{(2)}_{\mu\nu}$ is taken as a source for
the first order  Einstein equation,
not for the background Einstein equation.
This scheme actually amounts to studying
the second order metric perturbations
\cite{Matarrese1998,WangZhang2017,WangZhang2018,WangZhang2019}.
The model discussed in Ref.\cite{AbottDeser1982}
also belongs to this scheme.
In this paper we shall not consider this scheme,
as it is not about the back-reaction upon the back-ground spacetime.

In the following we shall examine some difficulties of
the 2nd order Einstein tensor as the effective tress tensor.
Plugging the expressions  of $R^{(2)}_{\mu\nu}$,  $R^{(1)}$,  $R^{(2)}$,
(see  \eqref{Rmunu2}   \eqref{R1st}  \eqref{R2st})
into the definition \eqref{effdef},
we obtain the 2nd order  Einstein tensor
\bl
  G^{(2)} _{\mu\nu}=&\frac{1}{2}\Big(
\frac{1}{2}h_{\alpha\beta|\mu}h^{\alpha\beta}_{~~|\nu}
+h^{\alpha\beta}(h_{\alpha\beta|\mu|\nu}+h_{\mu\nu|\alpha|\beta}
-h_{\alpha\mu|\nu|\beta}-h_{\alpha\nu|\mu|\beta})
\nn
\\
& +h_{\nu}^{~\alpha|\beta}(h_{\alpha\mu|\beta}-h_{\beta\mu|\alpha})
-(h^{\alpha\beta}_{~~~|\beta}-\frac12 h^{|\alpha})
(h_{\alpha\mu|\nu}+h_{\alpha\nu|\mu}-h_{\mu\nu|\alpha})\Big)
\nn
\\
&-\frac12 h_{\mu\nu}\Big( h_{\alpha\beta}^{~~~|\alpha|\beta}
-h^{|\alpha}_{~~|\alpha}
-h^{\alpha\beta}\bar{R}_{\alpha\beta}\Big)
\nn
\\
&-\frac12 \gamma_{\mu\nu} \Big(\frac{1}{2}\big(
\frac{3}{2}h_{\alpha\beta|\sigma}h^{\alpha\beta|\sigma}
+h^{\alpha\beta}\Box h_{\alpha\beta}
-h^{\sigma\alpha|\beta} h_{\sigma\beta|\alpha} \big)
\nn
\\
& +\frac12h^{\alpha \beta}\Box h_{\alpha \beta}
-\bar{R}_{\alpha\rho\sigma\beta} h^{\alpha\beta}h^{\rho\sigma}\Big)
\nn
\\
& - \frac12 \gamma_{\mu\nu} \frac12
\Big( h^{\alpha\beta}( 2 h_{~ |\alpha \beta}
-4\gamma^{\sigma\rho}h_{\alpha\sigma |\rho\beta})
-(h^{\alpha\beta}_{~~~|\beta}-\frac12 h^{|\alpha})
(2h_{\alpha\rho}^{~~~|\rho}
-h_{|\alpha})\Big) .
 \label{2ndEinsteintensor}
\el
In the TT gauge,
by use of the background equation and the GW equation,
 \eqref{2ndEinsteintensor} becomes
\bl
  G^{(2)} _{\mu\nu} &= \frac12 \Big(
\frac{1}{2}h_{\alpha\beta|\mu}h^{\alpha\beta}_{~~|\nu}
+h^{\alpha\beta}(h_{\alpha\beta|\mu|\nu}+h_{\mu\nu|\alpha|\beta}
-h_{\alpha\mu|\nu|\beta}-h_{\alpha\nu|\mu|\beta})
\nn
\\
&~~~ +h_{\nu}^{~\alpha|\beta}(h_{\alpha\mu|\beta}-h_{\beta\mu|\alpha})
  -  \gamma_{\mu\nu} \frac{1}{2}  \big(
\frac{3}{2}h_{\alpha\beta|\lambda}h^{\alpha\beta|\lambda}
+ h^{\alpha\beta} \Box h_{\alpha\beta}
-h^{\lambda\alpha|\beta}h_{\lambda\beta|\alpha}\big) \Big) .
\label{2Gmunulow}
\el

Given  $G_{\mu\nu}$ of \eqref{expeff},
the upper indexed Einstein tenor  $G^{\mu\nu}$ is given by
\bl
G^{\mu\nu}= g^{\mu\alpha} g^{\nu\beta} G_{\alpha\beta}
= \bar{G}^{\mu\nu} + G^{(1)\, \mu\nu} +G^{(2)\, \mu\nu}
   + O(h^3).
\el
From this we get the upper indexed perturbed
$G^{(1)\mu\nu}$  and $G^{(2)\mu\nu}$  as the following
\bl
G^{(1)\mu\nu} &=    \gamma^{\mu\alpha} \gamma^{\nu\beta} G^{(1)}_{\alpha\beta}
 - \Big( h^{\mu\alpha}\gamma^{\nu\beta} \bar G_{\alpha\beta}
      + h^{\nu\alpha}\gamma^{\mu\beta} \bar G_{\alpha\beta} \Big),
\\
G^{(2)\mu\nu} & =  \gamma^{\mu\alpha} \gamma^{\nu\beta} G^{(2)}_{\alpha\beta}
   - \Big( h^{\mu\alpha}\gamma^{\nu\beta}  G^{(1)}_{\alpha\beta}
  +  h^{\nu\alpha}\gamma^{\mu\beta}   G^{(1)}_{\alpha\beta}
  \nn \\
& ~~~~  -h^{\mu}_{~\sigma}h^{\sigma\alpha}\gamma^{\nu\beta}\bar{G}_{\alpha\beta}
  -h^{\nu}_{~\sigma}h^{\sigma\beta}\gamma^{\mu\alpha}\bar{G}_{\alpha\beta}
  -h^{\mu\alpha}h^{\nu\beta}\bar{G}_{\alpha\beta}\Big) .
\label{upperG2}
\el
It is observed that
\bl
G^{(1)\mu\nu}\ne \gamma^{\mu\alpha} \gamma^{\nu\beta} G^{(1)}_{\alpha\beta} ,
\\
G^{(2) \mu\nu} \neq \gamma^{\mu\alpha}\gamma^{\nu\beta} G^{(2)}_{\alpha\beta}  ,
\label{upperlowerne}
\el
so,  $G^{(2)}_{\mu\nu}$  and $G^{(2) \mu\nu}$ are not
the associated covariant and contravariant tensors defined on the background spacetime,
neither are $G^{(1)}_{\mu\nu}$ and $G^{(1) \mu\nu}$.
In the TT gauge,
by use of the background equation   and  the GW equation,
the upper indexed effective stress tensor is given by
\bl
T^{\text{eff}\, \mu\nu}  & \equiv  -2 G^{(2)\mu\nu}
\nn
\\
&=  - \Big(    \frac{1}{2} h_{\alpha\beta}\, ^{|\mu} h^{\alpha\beta|\nu}
   +h^{\alpha\beta}\big( h_{\alpha\beta}\, ^{|\mu|\nu} +h^{\mu\nu}\, _{|\alpha|\beta}
   -h^{~\mu|\nu}_{\alpha~~|\beta}-h^{~\nu|\mu}_{\alpha~~|\beta}\big)
 \nn
   \\
& ~~   +h^{\nu\alpha|\beta} \big(h^{~\mu}_{\alpha~|\beta} -h^{~\mu}_{\beta~|\alpha}\big)
\nn
\\
& ~~  -\frac{1}{2}\gamma^{\mu\nu}\big(
   \frac{3}{2}h_{\alpha\beta|\rho}h^{\alpha\beta|\rho}
   +h^{\alpha\beta}\Box h_{\alpha\beta}
  -h^{\rho\alpha|\beta}h_{\beta\rho|\alpha}\big)
   \nn \\
&~~ +\bar{R}_{\sigma}^{~\mu} h^{\alpha\nu}h^{\sigma}_{~\alpha}
   +\bar{R}_{\sigma}^{~\nu} h^{\alpha\mu}h^{\sigma}_{~\alpha}
     - h^{\mu\alpha} h^{\nu}_{\alpha}     \bar{R}  \Big)  .
   \label{fromeinstein}
\el
Similarly,    the  mixed-indexed Einstein tensor is defined as
\bl
G^{\mu }_{ ~~\nu} & =  g^{\mu\alpha}   G_{\alpha\nu} ,
 \label{rsloDefGmunu2}
\el
From this we get
\bl
G^{(1) \mu }_{~~~~\nu}
& = \gamma^{\mu\alpha}  G^{(1)}_{\alpha\nu} -h^{\mu\alpha} \bar G_{\alpha\nu}  ,
\label{raisloG1}
\\
G^{(2) \mu }_{~~~~\nu}
& = \gamma^{\mu\alpha}   G^{(2)}_{\alpha\nu}
-h^{\mu\alpha}   G^{(1)}_{\alpha\nu}
+h^{\mu}_{~\beta}h^{\beta\alpha} \bar G_{\alpha\nu} ,
\label{raisloG2}
\el
which tells that generally
$G^{(1) \mu }_{~~~~\nu}  \ne \gamma^{\mu\alpha} G^{(1)}_{\alpha\nu}$,
and $G^{(2) \mu }_{~~~~\nu}  \ne \gamma^{\mu\alpha} G^{(2)}_{\alpha\nu}$.
The mixed indexed  effective stress tensor is given  by
\bl
T^{\text{eff}\,  \mu}\, _{\nu} & \equiv  -2 G^{(2)\mu}\, _{\nu}
\nn
\\
= &-\Big(\frac{1}{2}h_{\alpha\beta|\mu}h^{\alpha\beta}_{~~|\nu}
+h^{\alpha\beta} (h_{\alpha\beta|\mu|\nu}
+h^{\mu}_{~\nu|\alpha|\beta}
-h^\mu_{~\alpha|\nu|\beta}
-h_{\alpha\nu|\mu|\beta})
\nn
\\
& ~~~ +h_{\nu}^{~\alpha|\beta} (h^{\mu}_{~\alpha|\beta}-h^{\mu}_{~\beta|\alpha})
\nn
\\
&~~~  -   \frac{1}{2} \gamma^{\mu}_{~\nu}  \big(
\frac{3}{2}h_{\alpha\beta|\lambda}h^{\alpha\beta|\lambda}
+ h^{\alpha\beta} \Box h_{\alpha\beta}
-h^{\lambda\alpha|\beta}h_{\beta\lambda|\alpha}\big)
\nn
\\
& ~~~  -\bar{R}_{\rho\alpha} h^{\rho}_{~\nu}h^{\mu\alpha}
   + \bar{R}_{\rho\nu} h^{\mu\alpha}h^{\rho}_{~\alpha}  \Big) .
\label{uplowTeff}
\el
Since the three expressions
$T^{\text{eff}}_{\mu\nu}$, $T^{\text{eff}\, \mu}\, _{\nu}$ and $T^{\text{eff}\, \mu\nu}$
are generally not related to each other by $\gamma_{\mu\nu}$,
the associated effective stress tensor is not uniquely defined.
This is a difficulty.

Another difficulty of the effective stress tensor
that it is not covariantly conserved
with respect to the   background metric.
Taking the covariant four divergence at both sides of  \eqref{fromeinstein},
 using the formulae \eqref{formula1} --- \eqref{formula3}, we have
\bl
(T^{\text{eff}\, \mu\nu })_{|\nu}=&-h^{\alpha\beta}
   \bar{R}_{\rho~}^{~\mu} h_{\alpha\beta} ^{~~|\rho}
-\bar{R}_{\rho\alpha|\beta}h^{\alpha\beta}h^{\rho\mu}
  -h^{\mu\alpha}\bar{R}_{\rho}^{~\nu}h^{\rho}_{~\alpha|\nu } \nn
+h^{\alpha\beta}\bar{R}_{\rho\beta} h_{\alpha}^{~\rho|\mu}
\nn
\\
  &- h_{~\alpha|\beta}^{\mu}
  \bar{R}^{~\alpha}_{\rho}h^{\rho\beta}
  -h^{\mu\alpha}_{~~|\nu}\bar{R}_{\rho}^{~\nu}h^{\rho}_{~\alpha }
  +h^{\mu\beta}_{~~|\nu}h_{\beta}^{~\nu}\bar{R}
  +h^{\mu\beta}h_{\beta}^{~\nu}\bar{R}_{|\nu}
   -h^{\mu\alpha}\bar{R}_{\rho~|\nu}^{~\nu}h^{\rho}_{~\alpha } .
   \label{tmunufourdivv}
\el
The right hand side is nonzero
in presence of fluid  with $\bar R_{\mu\nu} \ne 0$.
So,  $T^{\text{eff}\, \mu\nu }$ is generally not conserved
in the presence of matter.
We shall also show this explicitly for fRW spacetimes in Sect.\ref{sectionfordpark}.

The expression of $T^{\text{eff} }_{\mu\nu}$ is quite complicated.
Brill and Hartle applied an approximation  on it \cite{BrillHartle1964,Isaacson1968b}.
This Brill-Hartle (BH) approximation
assumes  that wavelengths of GW be much smaller
than the  scale of the background spacetime
and that the amplitude of GW be small.
Under the BH approximation,
$T^{\text{eff} }_{\mu\nu}$ of \eqref{effdef}  will reduce to the following
  \cite{BrillHartle1964,Isaacson1968a,Isaacson1968b}
\bl
T^{\text{BH}}_{\mu\nu} =
  - \frac{1}{32\pi G} h^{\alpha\beta}\, _{|\mu} h_{\alpha\beta |\nu} .
  \label{RHeff}
\el
As we find, after the BH approximation
the conserved $\tau_{\mu\nu}$ of \eqref{covt2}
will also reduce to the expression  \eqref{RHeff}.
Ref.\cite{SuZhang2012} applied the BH approximation
to the stress pseudo-tensor of GW defined by Landau and Lifshitz \cite{LandauLifshitz1975},
also arrived  at the expression \eqref{RHeff}.
Although the expression \eqref{RHeff} is often used in literature
\cite{Weinberg1972,Flanagan2005,maggiore2018,WangZhangChen2016,ZhangWang2018JCAP},
it is not exactly conserved,
but will be conserved under the BH approximation \cite{MTW}.
In cosmology where wavelengths of GW  are often comparable to
the scale of the background spacetime,
the BH approximation is not applicable.
In this paper we treat $h_{\mu\nu}$ as a fundamental field,
and do not apply  the BH approximation.

\section{Expressing  $G^{(2)}_{\mu\nu}$ in terms of the perturbed actions }\label{sectionAnal}

As   seen in Sect. \ref{action},
by expanding the action up to the second order,
$I_{gw}$ gives the conserved stress tensor $\tau_{\mu\nu}$,
while all the other geometric  actions are canceled out by the fluid actions.
We now express  $G^{(2)}_{\mu\nu}$  in terms of the perturbed actions,
reveal that
the difference between $G^{(2)}_{\mu\nu}$  and $\tau_{\mu\nu}$
is a nonconserved part
which will be canceled out by the corresponding fluid term.

The full Einstein tensor $G_{\mu\nu}$
 has been defined by \eqref{loweg}.
Expanding both sides of  \eqref{loweg}  up to the second order of perturbation,
one has
\bl
\bar{ G}_{\mu\nu}+  G^{(1)}_{\mu\nu}+  G^{(2)}_{\mu\nu}
& \simeq  - \frac{1}{\sqrt{-\gamma}}\Big(1-\frac12h +  \frac14 h^{\sigma\rho}h_{\sigma\rho}
   + \frac18 h^2   \Big)
\Big(\frac{\delta I^{(0)}_G}{\delta \gamma^{\alpha\beta}}
+\frac{\delta I^{(1)}_G}{\delta \gamma^{\alpha\beta}}
+\frac{\delta I^{(2)}_G}{\delta \gamma^{\alpha\beta}} \Big)\nn
\\
&~~~~~~~~~~~~~~~~~ \times \Big(\delta^{\alpha}_{~\mu}\delta^{\beta}_{~\nu}
+\delta^{\alpha}_{~\mu}h_{\nu}^{~\beta}
+\delta^{\alpha}_{~\nu}h_{\mu}^{~\beta}
+h_{\mu}^{~\alpha}h_{\nu}^{~\beta}\Big)
. \label{beforevaritation1}
\el
where
$\frac{\delta\gamma^{\alpha\beta}}{\delta g^{\mu\nu} }
     \simeq  (\delta^{\alpha}_{~\mu}\delta^{\beta}_{~\nu}
+\delta^{\alpha}_{~\mu}h_{\nu}^{~\beta}
+\delta^{\alpha}_{~\nu}h_{\mu}^{~\beta}
+h_{\mu}^{~\alpha}h_{\nu}^{~\beta})$ has been used (see \eqref{gmunutosecondorder}),
and  the actions $I^{(0)}_G, I^{(1)}_G, I^{(2)}_G$
are given in \eqref{I0G} \eqref{I1stnofld1} \eqref{totG123I}, respectively.
For GW in the TT gauge,  $ I^{(2)}_{g1}=0$ and
$I^{(2)}_G= I^{(2)}_{g2} +I_{gw} $,
so that  eq.\eqref{beforevaritation1} reduces to
\bl
\bar{ G}_{\mu\nu}+  G^{(1)}_{\mu\nu}+  G^{(2)}_{\mu\nu}
& \simeq - \frac{1}{\sqrt{-\gamma}}\Big(1 +  \frac14 h^{\sigma\rho}h_{\sigma\rho}  \Big)
\Big(\frac{\delta I^{(0)}_G}{\delta \gamma^{\alpha\beta}}
+\frac{\delta I^{(1)}_G}{\delta \gamma^{\alpha\beta}}
+\frac{\delta I^{(2)}_{g2}}{\delta \gamma^{\alpha\beta}}
+\frac{\delta I_{gw}}{\delta \gamma^{\alpha\beta}}
 \Big)\nn
\\
&~~~~~~~~~~~~~~~~~ \times \Big(\delta^{\alpha}_{~\mu}\delta^{\beta}_{~\nu}
+\delta^{\alpha}_{~\mu}h_{\nu}^{~\beta}
+\delta^{\alpha}_{~\nu}h_{\mu}^{~\beta}
+h_{\mu}^{~\alpha}h_{\nu}^{~\beta}\Big)
. \label{beforevaritation}
\el
We read off the 0th order Einstein tensor $\bar G_{\mu\nu}$
as given in \eqref{downtwoindexG},
and the first order Einstein tensor given by
\bl
G_{\mu\nu}^{(1)}
&=- \frac{1}{\sqrt{-\gamma}}\frac{\delta I^{(0)}_G}{\delta \gamma^{\alpha\beta}}
    (\delta^{\alpha}_{~\mu}h_{\nu}^{~\beta}+\delta^{\alpha}_{~\nu}h_{\mu}^{~\beta})
- \frac{1}{\sqrt{-\gamma}}\frac{\delta I^{(1)}_G}{\delta \gamma^{\mu\nu}}
 , \label{downtwoindexG0rder1}
\el
which, by use of  \eqref{downtwoindexG} \eqref{d1e},  is explicitly written
as the following
\bl
 G_{\mu\nu}^{(1)} & =
\bar G_{\mu\alpha}h^{\alpha}_{~\nu}
     +\bar{G}_{\nu\alpha}h^{\alpha}_{~\mu}
- \big(\frac12\bar{R}_{\mu\sigma}h_{\nu}^{~\sigma}
+\frac12\bar{R}_{\nu\sigma}h_{\mu}^{~\sigma}
-\frac12 h_{\mu\nu}\bar{R}
-\bar{R}_{\mu\rho\sigma\nu} h^{\rho\sigma}
+\frac12\Box h_{\mu\nu}
 \big)  ,
    \label{Gmunu1}
\el
agreeing with the expression \eqref{G1}.
We read off the second order  Einstein tensor from \eqref{beforevaritation}
as the following
\bl
 G_{\mu\nu}^{(2)}
& = -  \frac14   h^{\sigma\rho}h_{\sigma\rho}
    \frac{1}{\sqrt{-\gamma}}\frac{\delta I^{(0)}_G}{\delta \gamma^{\mu\nu}}
- \frac{1}{\sqrt{-\gamma}}\frac{\delta I^{(0)}_G}{\delta \gamma^{\alpha\beta}}
    h_{\mu}^{~\alpha}h_{\nu}^{~\beta}
-\frac{1}{\sqrt{-\gamma}}\frac{\delta I^{(1)}_G}{\delta \gamma^{\alpha\beta}}
(\delta^{\alpha}_{~\mu}h_{\nu}^{~\beta}+\delta^{\alpha}_{~\nu}h_{\mu}^{~\beta})
\nn \\
& ~~~~  -\frac{1}{\sqrt{-\gamma}}\frac{\delta I^{(2)}_{g2}}{\delta \gamma^{\mu\nu}}
 -\frac{1}{\sqrt{-\gamma}}\frac{\delta I_{gw} }{\delta \gamma^{\mu\nu}}
.
 \label{barGmunu2dow}
\el
This expression is equivalent to the definition \eqref{2Gmunu},
 nevertheless,  reveals a finer composition.
We double check \eqref{barGmunu2dow} as the following.
Summing up \eqref{C20IG2} and \eqref{D13}, we have
 \bl
 \frac{1}{\sqrt{-\gamma}}
 \frac{\delta ( I^{(2)}_{g2} +  I_{gw} ) }{\delta \gamma^{\mu\nu}}
 = & -  \frac{1}{2} \Big(
    \frac{1}{2}h_{\sigma\rho|\mu}h^{\sigma\rho}_{~~|\nu}
    +h^{\sigma\rho}\big(h_{\sigma\rho|\mu|\nu}+h_{\sigma\rho|\mu|\nu}
    -h_{\sigma\mu|\nu|\rho}-h_{\sigma\nu|\mu|\rho}\big)
    \nn
    \\
 &  +h_{\nu}^{~\sigma|\rho}\big(h_{\sigma\mu|\rho}
 -h_{\rho\mu|\sigma}\big)- \gamma_{\mu\nu} \frac{1}{2} \big(
    \frac{3}{2}h_{\sigma\rho|\theta}h^{\sigma\rho|\theta}
    +h^{\sigma\rho}\Box h_{\sigma\rho}
    -h^{\theta\sigma|\rho}h_{\rho\theta|\sigma}\big)\nn
 \\
 & + \frac14\gamma_{\mu\nu}h_{\sigma\rho}h^{\sigma\rho}\bar{R}
 -\frac{1}{2} h_{\sigma\rho}h^{\sigma\rho}\bar{R}_{\mu\nu}
 - h_{\mu}^{~\alpha}  h_{\alpha \nu} \bar{R}
 +  \bar{R}_{\sigma\mu} h^{\rho}_{~\nu}h^{\sigma}_{~\rho}
 +\bar{R}_{\sigma\nu} h^{\rho}_{~\mu}h^{\sigma}_{~\rho}   \Big) .
    \label{deltaaphabeyaI2}
 \el
Substituting  \eqref{downtwoindexG} \eqref{d1e}   \eqref{deltaaphabeyaI2}
into  \eqref{barGmunu2dow},  using the background equation,
we obtain  the expression \eqref{2Gmunulow} of $G_{\mu\nu}^{(2)}$
that was derived from the definition \eqref{2Gmunu}.

Now we rewrite $ G_{\mu\nu}^{(2)}$ into two parts
\bl
 G_{\mu\nu}^{(2)} = X_{\mu\nu}   - \frac12 \tau_{\mu\nu} ,
  \label{compis}
\el
where   $\tau_{\mu\nu}$ is the conserved stress tensor
defined by \eqref{deflowtau}, and
\bl
X_{\mu\nu}  \equiv &
-  \frac14   h^{\sigma\rho}h_{\sigma\rho}
    \frac{1}{\sqrt{-\gamma}}\frac{\delta I^{(0)}_G}{\delta \gamma^{\mu\nu}}
- \frac{1}{\sqrt{-\gamma}}\frac{\delta I^{(0)}_G}{\delta \gamma^{\alpha\beta}}
    h_{\mu}^{~\alpha}h_{\nu}^{~\beta}
\nn
\\
&    -\frac{1}{\sqrt{-\gamma}}\frac{\delta I^{(1)}_G}{\delta \gamma^{\alpha\beta}}
 (\delta^{\alpha}_{~\mu}h_{\nu}^{~\beta}+\delta^{\alpha}_{~\nu}h_{\mu}^{~\beta})
 -\frac{1}{\sqrt{-\gamma}}\frac{\delta I^{(2)}_{g2}}{\delta \gamma^{\mu\nu}}
  ,  \label{eqation259}
\el
which consists of a mixture of perturbed actions, and is the nonconserved part.

In presence of fluid,  $X_{\mu\nu}$
in the back-reaction equation \eqref{2ndexpeff}
is canceled out by the   fluid term,
\bl
 X_{\mu\nu} -  \frac12 T^{(2)}_{\mu\nu}   =0 .
 \label{cancel2}
\el
This is simply checked as the following.
In analogy to  \eqref{barGmunu2dow},
the second order fluid stress tensor is given by
(see  the definitions \eqref{loweg} and \eqref{loweF})
\bl
\frac12  T_{\mu\nu}^{(2)}
& =   \frac14   h^{\sigma\rho}h_{\sigma\rho}
    \frac{1}{\sqrt{-\gamma}}\frac{\delta I^{(0)}_F}{\delta \gamma^{\mu\nu}}
+  \frac{1}{\sqrt{-\gamma}}\frac{\delta I^{(0)}_F}{\delta \gamma^{\alpha\beta}}
    h_{\mu}^{~\alpha}h_{\nu}^{~\beta}
+  \frac{1}{\sqrt{-\gamma}}\frac{\delta I^{(1)}_F}{\delta \gamma^{\alpha\beta}}
(\delta^{\alpha}_{~\mu}h_{\nu}^{~\beta}+\delta^{\alpha}_{~\nu}h_{\mu}^{~\beta})
\nn
\\
& +  \frac{1}{\sqrt{-\gamma}}\frac{\delta I^{(2)}_{f2}}{\delta \gamma^{\mu\nu}} .
\label{Tmunu2w}
\el
Eq.\eqref{I0theqfl} gives the cancelation
$\delta I^{(0)}_G +  \delta I^{(0)}_F=0$,
 eq.\eqref{deltaI1GF} gives the cancelation
$\delta I^{(1)}_G + \delta I^{(1)}_F=0$,
and eq.\eqref{2ndactionGW}  gives the cancelation
$ I^{(2)}_{g2} +  I^{(2)}_{f2} =0$, so that
\bl
\delta I^{(2)}_{g2} + \delta I^{(2)}_{f2} =0 .
\label{3GFcancl}
\el
Thus the cancelation \eqref{cancel2} holds.
Alternatively, one can directly  show  that eq.\eqref{3GFcancl}
 holds under the conditions \eqref{normalization} and  \eqref{dbv}
that the velocity normalization and the background Einstein equation
hold also on  the  varied background metric.
(See  \eqref{normalization} --- \eqref{fI3v} in  Appendix \ref{G3delta}.)

In absence of fluid,  $\bar R_{\mu\nu} = \bar R= R^{(1)}=0$,
so  $I^{(0)}_{G}= I^{(1)}_{G}= I^{(2)}_{g2}=0$,
and,  by definition \eqref{eqation259}, one has
\bl
X_{\mu\nu}=0
\label{Xmunu=0} .
\el
Hence,  regardless of the fluid,  $X_{\mu\nu}$ will be removed and
only the conserved $\tau_{\mu\nu}$ remains in $G^{(2)}_{\mu\nu}$,
and,   consequently,  the back-reaction equation \eqref{2ndexpeff}
in the general scheme
reduces to the back-reaction equation \eqref{backreacteq}
in the perturbation scheme.

If the variations in \eqref{eqation259}  are performed,
$X_{\mu\nu}$ will have a lengthy  expression as the following
\bl
X_{\mu\nu} & =    \frac12  \Big[  \gamma_{\mu\nu}
\big[ - \frac12 \bar{R}_{\sigma\alpha}h^{\alpha\beta}h^{\sigma}_{~\beta}
 +  \frac12 (h^{\alpha\beta}_{~~~|\sigma}h^{\sigma}_{~\alpha|\beta}
    +h^{\alpha\beta}h^{\sigma}_{~\alpha|\beta|\sigma})
 - (h^{\alpha\beta}\Box h_{\alpha\beta}
      +h_{\alpha\beta|\sigma} h^{\alpha\beta|\sigma}) \big]
\nn
\\
&   - \big(   \frac12 h^{\alpha \beta }_{~~|\nu} h_{\mu \alpha|\beta}
            + \frac12 h^{\alpha \beta }_{~~|\mu} h_{\nu \alpha|\beta}
 + \frac12 h^{\beta\alpha } h_{\mu\beta|\nu|\alpha }
 + \frac12 h^{\beta\alpha } h_{\nu\beta|\mu|\alpha }
   - \frac14 h_{\nu \beta} \Box h^{\beta}_{~\mu}
   - \frac14 h_{\mu \beta} \Box h^{\beta}_{~\nu}
\nn
\\
& -  h_{\mu}^{~ \beta|\alpha} h_{\nu\beta|\alpha}
   - \frac12 h_{\alpha\beta}h^{\alpha\beta}_{~~~ |\mu|\nu}
   - \frac12 h_{\alpha\beta}h^{\alpha\beta}_{~~~ |\nu|\mu}
 -h_{\alpha\beta|\mu} h^{\alpha\beta}_{~~~|\nu}  \big)
\nn
\\
& + ( \bar{R}_{\alpha\beta} h^{~\alpha}_{\nu} h^{~\beta}_{\mu}
   - \frac12 \bar{R}_{\beta\mu} h^{\alpha}_{~\nu} h^{\beta}_{~\alpha}
   - \frac12 \bar{R}_{\beta\nu} h^{\alpha}_{~\mu} h^{\beta}_{~\alpha} ) \Big]
      .
\label{Gnc}
\el
This would be nontrivial to remove from $G^{(2)}_{\mu\nu}$
without the guidance of decomposition \eqref{compis}.

Similarly, the upper indexed   Einstein tensor $G^{\mu\nu}$
is defined by \eqref{uppereg}.
Expanding this  to the second order,
 we have
\bl
\bar{ G}^{\mu\nu}+  G^{(1)\mu\nu}+  G^{(2)\mu\nu}
& \simeq \frac{1}{\sqrt{-\gamma}}(1-\frac12h+\frac14h^{\alpha\nu}h_{\alpha\nu}
 + \frac18h^2 )
 \Big(\frac{\delta I^{(0)}_G}{\delta \gamma_{\mu\nu}}
+\frac{\delta I^{(1)}_G}{\delta \gamma_{\mu\nu}}
+\frac{\delta I^{(2)}_G}{\delta \gamma_{\mu\nu}} \Big) ,
\label{452}
\el
where $\frac{\delta \gamma_{\alpha\beta}}{\delta g_{\mu\nu} }
=\delta_{\alpha \, \mu} \delta_{ \beta\, \nu}$.
In the TT gauge,  from \eqref{452} we read off
\bl
\bar{ G}^{\mu\nu} & =\frac{1}{\sqrt{-\gamma}}
\frac{\delta I^{(0)}_G}{\delta \gamma_{\mu\nu}},\label{bardeaction}
\\
 G^{(1)\mu\nu} & =\frac{1}{\sqrt{-\gamma}}\frac{\delta I^{(1)}_G}
 {\delta \gamma_{\mu\nu}}
 = - \Big( \frac12 \bar{R}^\mu_{~\sigma} h^{\sigma\nu}
+ \frac12  \bar{R}^\nu_{~\sigma} h^{\sigma\mu}
-\frac12 \bar{R} h^{\mu\nu}
-\bar{R}^\mu_{~\alpha\beta}\, ^\nu  h^{\alpha\beta}
+\frac12\Box h^{\mu\nu}  \Big)
 ,\label{1stbardeaction}
\\
G^{(2)\mu\nu}
&  = \frac14 h^{\alpha\beta}h_{\alpha\beta}
\frac{1}{\sqrt{-\gamma}} \frac{\delta I^{(0)}_G}{\delta \gamma_{\mu\nu}}
+\frac{1}{\sqrt{-\gamma}}
\frac{\delta I^{(2)}_{g2} }{\delta \gamma_{\mu\nu}}
+\frac{1}{\sqrt{-\gamma}} \frac{\delta I_{gw} }{\delta \gamma_{\mu\nu}}.
\label{Gupmunu2}
\el
Comparing  \eqref{Gupmunu2} with \eqref{barGmunu2dow},
it is seen that $G^{(2)\mu\nu}$ has a composition different from
the lower indexed $G^{(2)}_{\mu\nu}$ of \eqref{barGmunu2dow}.
Plugging  \eqref{bardeaction} \eqref{deltaaphabeyaI2}
into \eqref{Gupmunu2}, we obtain the expression  \eqref{fromeinstein}
that was derived by expanding $G^{\mu\nu}$ to the second order.
Analogously  to  \eqref{compis},  we can write
\bl
 G^{(2)\mu\nu} =   X^{\mu\nu} - \frac12 \tau^{\mu\nu}  ,
 \label{upcompis}
\el
where $\tau^{\mu\nu}$ is the conserved defined by  \eqref{uptaumunu} and
\bl
X^{\mu\nu}
&    = \frac14 h^{\alpha\beta}h_{\alpha\beta}
         \frac{1}{\sqrt{-\gamma}} \frac{\delta I^{(0)}_G}{\delta \gamma_{\mu\nu}}
+\frac{1}{\sqrt{-\gamma}}  \frac{\delta I^{(2)}_{g2} }{\delta \gamma_{\mu\nu}} ,
\label{upGnc2}
\el
is the nonconserved part.
In analogy  to  \eqref{cancel2},
$X^{\mu\nu}$ is canceled by the fluid term
\bl
X^{\mu\nu} -   \frac12 T^{(2)\mu\nu} =0 ,
\el
where $\frac12 T^{(2)\mu\nu}$ is given by \eqref{upT2}.
So only the conserved $\tau^{\mu\nu}$ remains.

Comparing $X^{\mu\nu}$ of \eqref{upGnc2} with
 $X_{\mu\nu}$ of \eqref{eqation259},
we see  that
\bl
X^{\mu\nu} \ne \gamma^{\mu\alpha} \gamma^{\nu\beta} X_{\alpha\beta}   .
\el
So,  $X_{\mu\nu}$ and  $X^{\mu\nu}$ are not a tensor on
the background spacetime,  unlike $\tau_{\mu\nu}$ and $\tau^{\mu\nu}$.
This leads to
$G^{(2) \mu\nu} \neq \gamma^{\mu\alpha}\gamma^{\nu\beta} G^{(2)}_{\alpha\beta}$.

The conclusion: by its composition,
$G^{(2)}_{\mu\nu}$ contains a nonconserved and nontensorial part $X_{\mu\nu}$.
Interestingly, $X_{\mu\nu}$ is canceled out  by the fluid term
in the back-reaction equation,
and  $G^{(2)}_{\mu\nu}$  reduces to the conserved $\tau_{\mu\nu}$.
Consequently,  the   general scheme \eqref{2ndexpeff}
reduces to the perturbation scheme  \eqref{backreacteq}.

\section{The  effective stress tensors in fRW spacetimes }\label{sectionfordpark}

We shall demonstrate explicitly the three effective stress tensors in a fRW spacetime
with the  metric in the synchronous coordinate
\bl
d s^2=a(\tau)^2(d\tau^2-(\delta_{ij}+H_{ij})dx^idx^j) .
\label{RWmetric}
\el
So $\gamma_{\mu\nu}=a^{2}\eta_{\mu\nu}$,
 $h_{0\mu}  =0$, $h_{ij}=-a^2 H_{ij}$.
We consider GW only, and $H_{ij}$ represents the tensorial metric perturbation.
The indices of $H_{ij}$ will be raised by  $\delta^{ij}$,
so that  $H_{ij}$ can be treated as a scalar field \cite{FordParker1977,Giovannini2010}.
For $a'\equiv d a/d\tau \ne 0$,
the condition \eqref{transvcondition} leads to
\bl
 -\frac{a'}{a^3} H^{i}_{~i} & =0, \label{traceless}
\\
\partial_i H^{ij} & =0  , \label{transverse}
\el
which are four constraints upon $H_{ij}$, often called
the traceless and transverse constraints  in literature.
Therefore, $H_{ij}$ has two independent polarizations,
say, $h^+ =H^1_{~1}= - H^2_{~2}$ and $h^\times =H^1_{~2}$
for a GW propagating along $x^3$-direction.
The fluid  velocity  is $\bar{u}^\mu=a^{-1}(1,0,0,0)$,
so that the coordinate condition $\bar{u}^\mu  h_{\mu\nu}=0$  is satisfied.
Under the synchronous-to-synchronous coordinate transformations,
the tensorial metric perturbation  $H_{ij}$ is invariant
\cite{WangZhang2017,WangZhang2018,WangZhang2019,WangZhang2024}.
The background equation \eqref{ei1st}
gives  two Friedmann equations
\bl
3\frac{a'^2}{a^2}-3\frac{a''}{a}&=a^2\frac{1}{4}(\bar{\rho}+3\bar{p}),
\label{secondfridman}
\\
\frac{a''}{a}+\frac{a'^2}{a^2}&=a^2\frac14(\bar\rho-\bar{p})
  . \label{firstfridman}
\el

{\bf The conserved  $\tau_{\mu\nu}$ }

From the  action $I_{\text{FP}}$ of \eqref{FPlagrangian},
Ford and Parker obtained  the action of GW  in a fRW spacetime \cite{FordParker1977}
\bl
I_{gw} & =\int d^4x \sqrt{-\gamma} \frac{1}{4}
   \gamma^{\mu\nu}H_{ij,\mu}H^{ij}_{~~,\nu}
\nn
\\
& =  \int d^4x \sqrt{-\gamma} \frac{1}{4 a^2}
      \Big( H_{ij,0}H^{ij}_{~~,0}  -H_{ij,k}H^{ij}_{~~,k}  \Big)  .
\label{LagrangianinfRW}
\el
(See \eqref{h00alphacodalpha}---\eqref{Enurhoalmimunurhoal} in Appendix F.)
This action is equivalent to that of
a pair of minimally-coupling massless scalar field.
The variation of \eqref{LagrangianinfRW} with respect to $H_{ij}$
leads to the GW equation \cite{Lifshitz1946,Grishchuk1974,FordParker1977}
 \bl
 H_{ij}''+2\frac{a'}{a} H_{ij}'
-\nabla^2H_{ij}=0, \label{BighijeomofGW}
\el
with $\nabla^2\equiv \partial_k\partial^k$.
The variation of  \eqref{LagrangianinfRW} with respect to $\gamma_{\mu\nu}$
gives the Ford-Parker stress tensor of GW in fRW spacetimes \cite{FordParker1977}
\bl
\tau_{\mu\nu} & =\frac{1}{2}\Big[H_{ij,\mu}H^{ij}_{~~,\nu}
- \frac12\gamma_{\mu\nu}  \gamma^{\alpha\beta}H_{ij,\alpha}H^{ij}_{~~,\beta}
\Big] ,
\label{FordParkerGiovainni}
\el
with the  components
\bl
\tau_{00}
&=\frac{1}{4}\Big[H_{ij,0}H^{ij}_{~~,0}+ H_{ij, l}H^{ij}_{~~,l} \Big],
\label{calf00}
\\
\tau_{mn}
&=\frac{1}{2}\Big[H_{ij,m}H^{ij}_{~~,n}
+ \frac12 \delta_{mn} ( H_{ij,0}H^{ij}_{~~,0}-H_{ij,l}H^{ij}_{~~,l})
\Big] ,
\label{calfij}
\\
\tau_{0n}
&=\frac{1}{2}\Big[H_{ij,0}H^{ij}_{~~,n}
\Big] .
\label{calf0n}
\el
The  conservation is satisfied
 \bl
\tau^{\nu}_{~~\mu |\nu}
=\frac1{2a^2}
\Big\{H_{mn}''+\frac{2a'}{a}
\partial_0H_{mn}-\partial^j\partial_jH_{mn}
\Big\}\partial_{\mu}H^{mn}=0 ,
\label{campoast}
\el
using  the GW equation \eqref{BighijeomofGW}.
The conservation  can be explicitly written in the components
\bl
 \tau^{\nu}_{~~0 |\nu}
&  = \rho_{gw}' +3\frac{a'}{a} (\rho_{gw} +p_{gw})
+ \partial_{j} \tau^{j}_{~~0}    =0,
\label{cs3}
\\
\tau^{\nu}_{~~i |\nu}
&   = \partial_{0} \tau^{0}_{~~i} +4 \frac{a'}{a} \tau^{0}_{~~i}
         + \partial_{j} {\tau}^{j}_{~~i}    =0  ,
\label{pcs}
\el
where
the energy density  and pressure of GW are
\bl
\rho_{gw} & =\gamma_{00} \, \tau^{00}
=\frac1{4 a^2}\Big( H_{ij,0}H^{ij}_{~~,0} +H_{ij,l} H^{ij, \, l} \Big),
\label{ourT00mid}
\\
p_{gw}   & =-\frac{1}{3}  \gamma_{ij} \, \tau^{ij}
=\frac{1}{4a^2}
\Big( H_{ij,0}H^{ij}_{~~,0}-\frac13  H_{ij,l}  H^{ij,\, l}\Big) ,
\label{calculationoFP}
\el
$\tau^{0i}$ represents the energy flux,
and $\tau^{ij}$ with ($i\ne j$)
represents the anisotropic stress of GW  \cite{Weinberg2004}.
In general these off-diagonal components are not zero.
For the relic GW forming an isotropic background,
one can take $\tau^{0i} =0$ and $\pi^{ij}=0 $,
where
\bl
\pi^{ij} & \equiv \tau^{ij} - \frac13 \delta^{ij}   \tau^{m}_{~~ m} ,
\label{definationpi}
\el
 the GW stress tensor has a diagonal form
\bl
  \tau^{ \mu}\, _{\nu}   = \text{diag}  (\rho_{gw}, -p_{gw}, -p_{gw}, -p_{gw})    ,
   \label{tmunuhomoandiso}
\el
and  \eqref{cs3} becomes the familiar conservation equation
\bl
 \rho'_{gw}  +3\frac{a'}{a}(\rho_{gw}+p_{gw})=0  .
\label{RWconservation}
\el
The relic GW stress tensor can be used for the back-reaction
in the perturbation scheme \eqref{backreacteq}
and the special scheme \eqref{tauvac2nd}.

We can also apply the general expression  $\tau^{\mu\nu}$ of \eqref{covt2}
to the  fRW spacetimes.
Calculation gives  the following
(see \eqref{firstterm00stGiov}---\eqref{sigmamuinu0sihgmag}
in Appendix \ref{usefulquantites})
\bl
\tau^{00} &=\frac1{2a^4}\Big(\frac12H_{mn,0}H^{mn}_{~~,0}
  +\frac{1}{2}H_{mn, l}H^{mn}_{~~~, l}\Big),\label{ourT00up}
\\
\tau^{ij} &=\frac{1}{2a^4}
\Big(
\partial_iH_{mn}\partial_jH^{mn}
+\frac12\delta_{ij}\big(H_{mn,0}H^{mn}_{~~,0}-H_{mn,k}H^{mn}_{~~,k}\big)
\nn
\\
&~~~~~
- (\partial_{m}H_{i}^{~n}\partial_{j}H^{m}_{~n}
    + \partial_{m}H_{j}^{~n}\partial_{i}H^{m}_{~n} )
+   (H^{nm}\partial_m \partial_{j}H_{in}
     +H^{nm}\partial_m \partial_{i}H_{jn} )
\nn
\\
&~~~~~ +  ( \partial^m H^{n}_{~i}\partial_n H_{mj}
               +\partial^mH^{n}_{~j}\partial_n H_{mi} )
  -2 H^{m n}\partial_{n} \partial_{m} H_{ij}
\Big),\label{ourTijup}
\\
\tau^{0i} &=\frac1{2a^4}
\Big(-\partial_0H_{mn} \partial_{i}H^{mn}
    + \partial_m H_{i}^{~n} \partial_{0}H_{n}^{~m}
    -H^m_{~n}\partial_m \partial_0 H^{in} \Big) .
\label{0icomponentFPrdp}
\el
Taking the four divergence of \eqref{ourT00up} \eqref{ourTijup} \eqref{0icomponentFPrdp},
using  the GW equation \eqref{BighijeomofGW},
we find that
\bl
\tau^{\nu}_{~~ 0|\nu} &= 0,\label{conseversedinfrw1}
\\
\tau^{\nu}_{~~i|\nu }  &=  0 .
\label{conseversedinfrw}
\el
This confirms the conservation \eqref{FordPconservedpaper}
for general curved spacetimes.
The expressions  \eqref{ourTijup} \eqref{0icomponentFPrdp}
seemingly   contain several  extra terms than
the expressions  \eqref{calfij} \eqref{calf0n}.
As pointed out below \eqref{covt2}, in each product $H_{ij}(x) H_{mn}(x)$
only the   Fourier  modes with the same $\bf k$ remain.
Then, by use of the transverse condition \eqref{transverse},
 these extra terms are vanishing,
\bl
 \partial_{m}H_{i}^{~n}\partial_{j}H^{m}_{~n}
&  \propto  k^m \epsilon^{\lambda}_{m n}({\bf k}) =0 ,
 \label{ijzero}
\\
 H^{nm}\partial_m \partial_{j}H_{in}  &  \propto  k^m \epsilon^{\lambda}_{m n}({\bf k}) =0 ,
\\
\partial^m H^{n}_{~i}\partial_n H_{mj}   &  \propto  k^m \epsilon^{\lambda}_{m n}({\bf k}) =0 ,
\\
   H^{m n}\partial_{n} \partial_{m} H_{ij}
&  \propto  k^m \epsilon^{\lambda}_{m n}({\bf k}) =0 .
 \label{nmzero}
\\
 \partial_m H_{i}^{~n} \partial_{0}H_{n}^{~m}
 & \propto  k^m \epsilon^{\lambda}_{m n}({\bf k}) =0 ,
\\
 H^m_{~n}\partial_m \partial_0 H^{in}
    &  \propto  k^m \epsilon^{\lambda}_{m n}({\bf k}) =0   ,
 \label{zeroc}
\el
where $\epsilon^{\lambda}_{m n}({\bf k})$ being the polarization of GW
 (see eq.\eqref{foursonstraints}).
Therefore,    the expressions  \eqref{ourTijup} and \eqref{0icomponentFPrdp}
give  respectively \eqref{calfij} and \eqref{calf0n}.
That is,
the general expression   $\tau^{\mu\nu}$ of \eqref{covt2}
reduces to the Ford-Parker stress tensor \eqref{FordParkerGiovainni}
in the fRW spacetimes.
The transversality  formulae \eqref{ijzero} -- \eqref{zeroc} were also used
in Ford and Parker's treatment on the action $I_{\text{FP}}$ \cite{FordParker1977}.
In quantum states, or in the statistic average,
the expectation value of these extra terms are vanishing too
\cite{Giovannini2010,Giovannini2020,Giovannini2019prd}.

On the other hand,
the stress tensor (3.114) in Ref.\cite{Giovannini2020}
does not satisfy the conservation,   and, in fRW spacetimes,
does not reduce to the Ford-Parker stress tensor \eqref{FordParkerGiovainni},
as we have checked.

{\bf The nonconserved } $T_{\text{MT}}^{\mu\nu}$

In fRW spacetimes the action \eqref{I2equaltoIG2-2}  becomes
\bl
J_2 & =  \int d^4x \sqrt{-\gamma} \frac{1}{4 a^2 }
   \Big(  H_{ij,0}H^{ij}_{~~,0}  -   H_{ij,k}H^{ij}_{~~,k}
   -2  (\frac{a''}{a}+ \frac{a'^2}{a^2})  H^{ij} H_{ij}\Big)  ,
\label{J2fRW}
\el
where the term $2(\frac{a''}{a}+ \frac{a'^2}{a^2})$
comes from the Ricci tensor term.
This action  resembles that of
a pair of effectively coupling massive scalar fields,
with a coupling  $\xi \bar R = 2 \frac{a''}{a^3}$ (ie, $\xi=\frac13$)
and an effective mass  $2 \frac{a'^2}{a^4}$.
The equation \eqref{FierzPaulieq1} in fRW spacetimes becomes
\bl
\partial_0^2 H_{mn}+2\frac{a'}{a}\partial_0 H_{mn}
  -\partial_i\partial^i H_{mn}
  + 2(\frac{a''}{a}+\frac{a'^2}{a^2})H_{mn} =0 ,
\label{Isacequ}
\el
which differs  from the GW equation \eqref{BighijeomofGW}.
The MacCallum-Taub  stress tensor  \eqref{MTstress} in fRW spacetimes
has the components,
\bl
T_{\text{MT}}^{00}
&=\frac{1}{4a^4}\Big(\partial_0 H_{ij}\partial_0 H^{ij}
+H_{i j, k} H^{i j}{ }_{, k}
-2\partial_{n} H^{ik}\partial_{k} H^{ni}+4\frac{a'}{a}H_{ij}\partial_0 H^{ij}\Big) ,
\label{MT00}
\\
T_{\text{MT}}^{0i}
&=-\frac{1}{a^4}\Big(\frac{1}{2}\partial_0H_{mn}\partial_i H^{mn}
+\frac{a^{\prime}}{a} H_{m n} \partial_i H^{m n}
-\partial_m H_i{ }^k \partial_0 H_k{ }^m\Big) ,
\label{MT0i}
\\
T_{\text{MT}}^{ij}
&=\frac{1}{a^4}\Big(2 (\frac{a^{\prime 2}}{a^2}+\frac{a''}{a}) H_i^n H_{n j}
+\frac12\partial_i H_{a b} \partial_j H^{a b}
\nn
\\
& ~~~ +\delta^{ij}(\frac14H_{i j, 0} H_{, 0}^{i j}
-\frac14H_{i j, k} H^{i j}{ }_{, k}+ \frac{a'}{a}H^{im}\partial_{0}H^{mi}
+\frac12\partial_{n} H^{ik}\partial_{k} H^{ni})
\nn
\\
& ~~~
-\partial_0 H^{ki} \partial_0 H^{j}_{~k}
+\partial_{b}H^{m i}\partial^b H^{j}_{~m}
+\partial^m H_i^k \partial_k H_{m j}
\nn
\\
&  ~~~ -H^{m k} \partial_k \partial_m H_{i j}
-\partial_m H_i{ }^k \partial_j H_k^m
-\partial_m H_j{ }^k \partial_i H_k^m\Big) ,
\label{MTij}
\el
and the associated energy density and pressure   are
\bl
\rho_{\text{MT}} & = T_\text{MT}\, ^{0}_{~0}
=\frac{1}{4a^2}\Big( \partial_0 H_{ij}\partial_0 H^{ij}
+H_{i j, k} H^{i j}{ }_{, k}
-2\partial_{n} H^{ik}\partial_{k} H^{ni}
+4\frac{a'}{a}H_{ij}\partial_0 H^{ij}\Big) ,
\label{spdemt}
\\
p_{\text{MT}}&=- \frac13 T_{\text{MT}}\, ^{i}_{~i}
\nn
\\
&= -\frac13\frac{1}{4a^2}\Big( 8 (\frac{a^{\prime 2}}{a^2}
+\frac{a''}{a}) H_i^n H_{n i}
+3\partial_i H_{a b} \partial_i H^{a b}
+12 \frac{a'}{a}H^{im}\partial_{0}H^{mi}
\nn
\\
& ~~~  -\partial_0 H^{ki} \partial_0 H^{i}_{~k}
+2\partial^m H_i^k \partial_k H_{m i}\Big) .
\label{spmt}
\el
The four divergence is given by
\bl
T^{0\nu}_{\text{MT} ~ |\nu}  &=\frac{1}{a^4}\Big(\partial_0^2 H_{mn}
+2\frac{a'}{a}\partial_0 H_{mn}-\partial_i\partial^i H_{mn}
+2(\frac{a''}{a}+\frac{a'^2}{a^2})H_{mn}\Big)
\Big(\frac12\partial_0 H^{mn}+\frac{a'}{a}H^{mn}\Big)
\nn
\\
&=\frac{1}{a^4}2(\frac{a''}{a}
+\frac{a'^2}{a^2})H_{mn}\Big(\frac12\partial_0 H^{mn}
+\frac{a'}{a}H^{mn}\Big) \ne 0 ,
\label{dgewyug}
\el
which is nonzero in the presence of matter,
where   the  GW equation \eqref{BighijeomofGW} has been used.
Thus, in fRW spacetimes
the MacCallum-Taub stress tensor is  nonconserved,
confirming the general result \eqref{MTnonconserv}.

{ \bf The nonconserved $T^{\text{eff}}_{\mu\nu}$ }

As shown  in Sect.\ref{effectivest},
$T^{\text{eff}}_{\mu\nu}$, $T^{\text{eff}\, \mu}\, _{\nu}$, and $T^{\text{eff}\, \mu\nu}$
are not a  tensor, and  are not conserved  on the background spacetime.
We explicitly  demonstrate these in the fRW spacetimes.
Applying the formula \eqref{fromeinstein} to  the fRW spacetimes gives
   (see Appendix \ref{usefulquantites})
\bl
T^{\text{eff}\, 00} = & \frac{1}{a^4}\Big(
\frac14\partial_0H_{ij}\partial_{0}H^{ij}
+\frac14 \partial_m H_{ij}  \partial^m H^{ij}
\nn
\\
&  +\frac{2 a'}{a}H_{ij}\partial_0H^{ij}
- \partial_m(H_{ij}\partial^mH^{ij})
+\frac12 \partial^{j}( H^{im}\partial_{m}H_{ij})\Big),
\label{de2G00AB}
\\
T^{\text{eff}\, ij}
= & \frac{1}{a^4}\Big(\partial_0 H_{i}^{~m}\partial_{0}H_{m j}
-\frac12\partial_iH_{mn}\partial_j H^{mn}
-  H^{mn}\partial_{i} \partial_j H_{mn}
-\partial_{m}H_{i}^{~n}\partial^{m}H_{nj}
\nn
\\
&  +  H^{nm}\partial_m\partial_{i}H_{jn}
         + H^{nm}\partial_m\partial_{j}H_{in}
+\partial_l (H_{j}^{~k}\partial_kH_{i}^{~l})
-\partial_{l}(H^{ml} \partial_{m}H_{ij})
\nn
\\
&  +\delta_{ij}\big[-\frac34\partial_0H_{mn}\partial_{0}H^{mn}
+\frac34\partial_mH^{ln}\partial^mH_{ln}
-\frac12\partial_{m}(H_{n l}\partial^{l} H^{nm})\big]
\nn
\\
&  -2(\frac{a'^2}{a^2}-2\frac{a''}{a}) H^{i}_{~m}H^{mj} \Big),
\label{tijab}
\\
T^{\text{eff}\, 0i}
&=\frac{1}{a^4}\Big(
  \frac{1}{2}\partial_0 H_{mn}\partial_i H^{mn}
 + H_{mn}\partial_i \partial_0 H^{mn}
  - H^{n}_{~m} \partial_{n} \partial_{0}H^{im}\Big),
\label{ti0ab}
\el
where the GW equation \eqref{BighijeomofGW} has been used.
By use of the transverseness  formulae \eqref{ijzero} ---\eqref{zeroc},
the above expressions \eqref{de2G00AB} \eqref{tijab} \eqref{ti0ab} reduce to
\bl
T^{\text{eff}\, 00} & = \frac{1}{a^4}\Big(
\frac14\partial_0H_{ij}\partial_{0}H^{ij}
+\frac14 \partial_m H_{ij}  \partial^m H^{ij}
  +\frac{2 a'}{a}H_{ij}\partial_0H^{ij}
- \partial_m(H_{ij}\partial^mH^{ij})  \Big) ,
\label{red00}
\\
T^{\text{eff}\, ij}
&=\frac{1}{a^4}\Big(\partial_0 H_{i}^{~m}\partial_{0}H_{m j}
-\frac12\partial_iH_{mn}\partial_j H^{mn}
- H^{mn} \partial_{i} \partial_j H_{mn}
-\partial_{m}H_{i}^{~n}\partial^{m}H_{nj}
\nn
\\
&~~~  +\delta_{ij}\big[-\frac34\partial_0H_{mn}\partial_{0}H^{mn}
+\frac34\partial_mH^{ln}\partial^mH_{ln}   \big]
 -2(\frac{a'^2}{a^2}-2\frac{a''}{a})  H^{i}_{~m}H^{mj} \Big),
\label{redij}
\\
T^{\text{eff}\, 0i}
&=\frac{1}{a^4}\Big(
  \frac{1}{2}\partial_0 H_{mn}\partial_i H^{mn}
 + H_{mn}\partial_i \partial_0 H^{mn} \Big) .
\label{red0i}
\el
Applying four divergence upon \eqref{red00} \eqref{redij} \eqref{red0i} gives
\bl
T^{\text{eff}\, 0\nu}\, _{|\nu} &=\frac{2}{a^4}
(\frac{a''}{a}-2\frac{a'^2}{a^2})H_{ij}
\partial_0H^{ij}+\frac{2}{a^4}\frac{a'}{a}(2\frac{a''}{a}
      -\frac{a'^2}{a^2})    H^{mn}H_{nm} \ne 0 ,
\label{TNU0BU}
\\
T^{\text{eff}\, i \nu}\, _{|\nu} &=-\frac{2}{a^4}(\frac{a'^2}{a}
     -2\frac{a''}{a}) H^{m}_{~n}\partial_m H^{in} \ne 0 ,
  \label{TNUnBU}
\el
being nonconserved  explicitly.
As we have checked,
eqs.\eqref{TNU0BU} \eqref{TNUnBU} also follow from eq.\eqref{tmunufourdivv}
in fRW spacetimes.

The  mixed indexed  $T^{\text{eff} ~\mu}\, _{\nu}$,
by use of \eqref{tumudownu2},
has  the components  in fRW spacetimes
\bl
T^{\text{eff} \, 0}\, _{0} & =\gamma_{00}T^{\text{eff} ~ 00} , \label{Tup0down0}
\\
T^{\text{eff} \, i}\, _{0} & =\gamma_{00} T^{\text{eff} ~ 0i} , \label{Tupidown0}
\\
T^{\text{eff}\, i}\, _{j}  & =\gamma_{jk} T^{\text{eff} ~ ki}
 -\frac{2}{a^2}(\frac{a'^2}{a^2}
   -2\frac{a''}{a})  H^{i}_{~n}H^{n}_{~j} .  \label{Tupidownj}
\el
Thus, $T^{\text{eff} ~\mu}\, _{\nu} \ne \gamma_{\nu\alpha}T^{\text{eff} ~\mu \alpha}$.
So, $T^{\text{eff} ~\mu}\, _{\nu} $ is not a tensor on the fRW  spacetimes.
Refs.\cite{AbramoPRD1997,AbramoPRD1999,BrandTaka2018}
adopted $T^{\text{eff} ~\mu}\, _{\nu}$ to discuss cosmology.
We check that plugging \eqref{de2G00AB} \eqref{tijab} \eqref{ti0ab}
into \eqref{Tup0down0} \eqref{Tupidown0} \eqref{Tupidownj}
leads to the mixed indexed (8)$\sim$(11) in Ref.\cite{BrandTaka2018}.
When a spatial average  \cite{AbramoPRD1997,BrandTaka2018}
is applied on  $T^{\text{eff} ~\mu}\, _{\nu}$,
the spatial total derivative terms in \eqref{de2G00AB} \eqref{tijab}
 will be vanishing,  and  the averaged energy density and pressure are
\bl
\rho_{\text{eff}}& =\langle T^{\text{eff} \,  0}\, _{0}\rangle
 =\frac1{a^2}\Big(\frac14\langle\partial_0 H_{ij}\partial_{0} H^{ij} \rangle
+\frac14\langle\partial_m H_{ij}\partial^m H^{ij}\rangle
+2\frac{a'}{a}\langle H_{ij} \partial_0 H^{ij}\rangle\Big),
\label{t00ABgw}
\\
p_{\text{eff}}&=-\frac{1}{3}\langle T^{\text{eff} ~i}\, _{i} \rangle
=\frac{1}{3a^2}\Big(
-\frac{5}{4}\langle\partial_0 H_{i}^{~n}\partial_{0}H_{n}^{~i}\rangle
+ \frac74 \langle\partial_m H^{ij} \partial^m H_{ij} \rangle \Big) ,
\label{pressureABgw}
\el
(here \eqref{pressureABgw} differing from (19) of ref.\cite{BrandTaka2018}
that had a term $\propto  H_{ij}\partial_0 H^{ij}$),
and the four divergence is nonzero
\bl
\langle  T^{\text{eff}~ \nu}\, _{0|\nu} \rangle  =
\rho_{\text{eff}}\, ' +3 \frac{a'}{a} (\rho_{\text{eff}} +p_{\text{eff}})
  &
  = \frac{2}{a^2} \Big( (\frac{a'}{a})' -(\frac{a'}{a })^2   \Big)
     \langle H_{ij}\partial_0H^{ij}\rangle \ne 0 ,
  \label{nonconsv}
\el
ie.,  $T^{\text{eff} ~\mu}\, _{\nu}$ is  nonconserved  explicitly.
Eq.\eqref{nonconsv} here is equivalent to
eq.(88) in Ref.\cite{AbramoPRD1997}
and eq.(25) in Ref.\cite{AbramoPRD1999},
and was also discussed in Refs.\cite{Giovannini2006,SuZhang2012}.

The lower indexed $T^{\text{eff}}_{\mu \nu}$
is related to the mixed indexed
\bl
T^{\text{eff}}_{\mu \nu} = \gamma_{\mu \sigma} T^{\text{eff}\, \sigma}\, _{\nu} ,
\label{uplower}
\el
which is valid under the condition \eqref{coordcondition}.
(See \eqref{proof2g}  \eqref{RWTeff}.)
We find that  $T^{\text{eff}}_{\mu \nu}$ is not conserved either.

{\bf The energy density spectra   in de Sitter space}

We shall show that the conserved   $\tau^{\mu\nu}$
has a positive spectral energy density,
and the nonconserved $T_{\text{MT}}^{\mu\nu}$
and  $T^{\text{eff} ~\mu}\, _{\nu}$ have a negative one
at long wavelengths \cite{BrandTaka2018}.
To be  specific, in the following we shall work in
the de Sitter space with  $a=-1/(H\tau)$.
Consider the GW as a quantum field.
The  field operator $H_{ij}$ can be written as
\bl
H_{ij}(x) & =
\int \frac{d^3 k}{(2 \pi)^{3 / 2}} \sum_{\lambda =+, \times}
\epsilon^{\lambda}_{i j}({\bf k})
\left[a_{\mathbf{k}}^\lambda h^\lambda_k (\tau) e^{i \mathbf{k} \cdot \mathbf{x}}
+a_{\mathbf{k}}^{\lambda \dagger} h^{\lambda *}_k(\tau)
  e^{-i \mathbf{k} \cdot \mathbf{x}}\right],\label{Hijquantum}
\el
where $a_{\mathbf{k}}^\lambda$ and $a_{\mathbf{k}}^{\lambda \dag}$
are the annihilation and creation operators for the $\lambda$ mode,
satisfying $[a_{\mathbf{k}}^\lambda, a_{\mathbf{k}^{\prime}}^{\lambda' \dagger}]
=\delta_{\lambda\lambda'} \delta^3(\mathbf{k}-\mathbf{k}^{\prime})$,
and $\epsilon_{ij}^{\lambda}({\bf k})$ is the polarization tensor
satisfying \cite{FordParker1977,Zhang2005CQG,Zhang2006CQG,ZhangWang2018JCAP,
BaskaranGrishchuk2006}
\bl
\delta_{ij} \epsilon_{ij}^{\lambda}({\bf k})  & =0,
~~~
 k^i \epsilon_{ij}^{\lambda}({\bf k})   =0,
\label{foursonstraints}
\\
\epsilon^{\lambda\, ij}({\bf k}) \epsilon_{ij}^{\lambda'}({\bf k})
  & = 2 \delta^{\lambda\lambda'},
 ~~~
 \epsilon^{\lambda}_{ij}({\bf k})   =\epsilon^{\lambda}_{ij}(-{\bf k}) .
\label{urnorm}
\el
The conditions \eqref{foursonstraints} correspond to  \eqref{traceless} \eqref{transverse}.
It is assumed that  $h^{+}_k, h^{\times}_k$ be statistically independent
and $h^{+}_k=h^{\times}_k=h_k$.
Eq.\eqref{BighijeomofGW} gives the equation of the $k$-modes
\bl
h^{''}_k   +2\frac{a'}{a} h^{ '}_k
 +k^2 h_k  =0 ,
\label{heq}
\el
and the normalized solution of GW  in the de Sitter space is \cite{ZhangWang2018JCAP}
\bl
h_k(\tau)=\frac1a\sqrt{\frac{\pi}{2}}
\sqrt{\frac{-\tau}{2 }}e^{i \pi(\beta+1) / 2}
H_{\beta+\frac{1}{2}}^{(1)}(-k\tau).
\label{solutionofhdesitter}
\el
The  Bunch-Davies  vacuum is defined by
$a_{\mathbf{k}}|0\rangle_{\bf k}=0$.
For  the conserved $\tau^{\mu\nu}$
in a quantum state $|\phi \rangle$
the expectation value
of \eqref{ourT00mid} \eqref{calculationoFP} are
\bl
\langle \phi| \rho_{gw}  |\phi \rangle
& =\int_0^{\infty} \rho_k\frac{d k}{k}
+ \int  \frac{d k}{k} \rho_{ k} \sum_{\lambda=\pm} n_{\bf k}^{\lambda},
\\
\langle \phi| p_{gw} |\phi \rangle
& =\int p_k \frac{d k}{k}
+ \int  \frac{d k}{k} p_k \sum_{\lambda=\pm } n_{\bf k}^{\lambda},
\el
where the first terms are the vacuum part,
the second terms are
the graviton part with $n_{\bf k}^{\lambda}$ being the number distribution
of gravitons of polarization $\lambda$,
the conserved spectral energy density and pressure of GW are
\bl
\rho_{  k}  &=\frac{k^3}{2a^2\pi^2}\Big( |\partial_0 h_k|^2 +k^2|h_k|^2  \Big)
\label{FPsrhoenerg}
\\
&   =   \frac{k^4}{4\pi^2a^4}
2 (1+\frac{1}{2 k^2\tau^2}),\label{Fordparkerenergy}
\\
p_k & = \frac{k^3}{2a^2\pi^2}
\Big( |\partial_0 h_k|^2   -\frac13 k^2|h_k|^2 \Big)
\label{FPspenerg}
\\
 & =   \frac{k^4}{12 a^4 \pi^2}  2 (1 - \frac{1}{2 k^2\tau^2}) .
\label{Fordparkerpressure}
\el
The spectra \eqref{FPsrhoenerg}---\eqref{Fordparkerpressure}
are the same as  those of  a pair of
minimally-coupling massless scalar fields \cite{ZhangYeWang2020,ZhangWangYe2020}.
$\rho_{k} \ge 0$ for all $k$, and $p_{k} \simeq  -\frac13 \rho_{k}$ at small $k$.
Physically,  an energy density should be nonnegative,
while a negative pressure is allowed.
For instance, the inflaton scalar field has
a positive energy density and a negative pressure during inflation.
These behaviors are typical for a scalar field.

Similarly, for the MacCallum-Taub  $T_{\text{MT}}^{\mu\nu}$,
the expectation value of  \eqref{spdemt} \eqref{spmt}
leads to the following
\bl
\rho^{\text{MT}}_{k}&=\frac{k^3}{2\pi^2a^2}[|\partial_0 h_k|^2
+k^2 |h_k|^2 +2 \frac{a'}{a} (h_k\partial_0 h^*_k+h_k^*\partial_0 h_k)]
\label{MTenergydensity1}
\\
 & =\frac{k^4}{4\pi^2a^4} 2 (1-\frac{3}{2 k^2\tau^2})
 ,\label{MTenergydensity1d}
\\
p^{\text{MT}}_k&=\frac{k^3}{2\pi^2a^2}\Big(\frac13|\partial_0h_k|^2
-( k^2+\frac83(\frac{a^{\prime 2}}{a^2}+\frac{a''}{a})) |h_k|^2-2\frac{a'}{a}
(h_k\partial_0h_k^*+h_k^*\partial_0h_k)\Big)
  \label{MTpressure2}
\\
& =-\frac{k^3}{4\pi^2a^4}\frac{2}{3}
(1+\frac{15}{2 k^2 \tau ^2}+\frac{12}{k^4 \tau ^4})
  .  \label{MTpressure2d}
\el
Note that $\rho^{\text{MT}}_{k}<0$  at small $k$, which is unphysical.

For   $T^{\text{eff} ~\mu}\, _{\nu}$,
the  expectation value of \eqref{t00ABgw} \eqref{pressureABgw}
leads to
\bl
\rho_k^{\text{eff}} &  =\frac{k^3}{2a^2\pi^2}\Big( |\partial_0 h_k|^2 +k^2|h_k|^2
+4\frac{a'}{a}(h_k\partial_0h^{*}_{k}+h_k^*\partial_0h_{k}) \Big) \label{afterreal}
\\
& =\frac{k^4}{4\pi^2a^4} 2 (1-\frac{7}{2 k^2\tau^2}),
\label{effectiveenergy}
\\
p_k^{\text{eff}} & = \frac{k^3}{2a^2\pi^2}\Big(
-\frac53 |\partial_0 h_k|^2 + \frac73 k^2|h_k|^2   \Big)
\label{pressspectreff}
\\
&  =  \frac{k^4}{12  \pi^2 a^4}    2 (1 +  \frac{7}{2 k^2\tau^2}).
\label{effectivepressure}
\el
Again $\rho^{\text{eff}}_{k}<0$  at small $k$, which is unphysical.
We notice that eqs. (27) (28) in Ref.\cite{BrandTaka2018}
had typos of a nonreal term  $h_k\partial_0h^{*}_{k}$, and differ from
the expressions \eqref{afterreal} \eqref{pressspectreff} here.

\begin{figure}[htb]
	\centering
	\includegraphics[width = .62\linewidth]{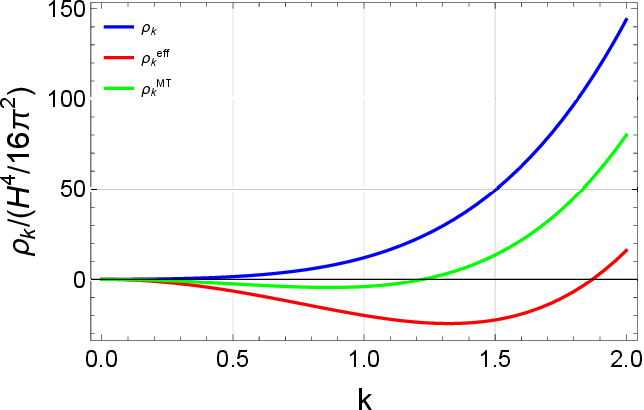}
	\caption{ The spectral energy densities.
Blue:   $\rho_{k} >0$   for all $k$.
Red:  $\rho^{\text{eff}}_{k} <0 $   at small $k$.
Green:  $\rho^{\text{MT}}_{k}<0 $   at small $k$.
A negative energy spectrum is unphysical,
so $\rho^{\text{eff}}_{k}$ and $\rho^{\text{MT}}_{k}$ are unphysical.
For illustration a time $\tau= -1$ is taken in the plot.}
	\label{fig1}
\end{figure}

\begin{figure}[htb]
	\centering
	\includegraphics[width = .6\linewidth]{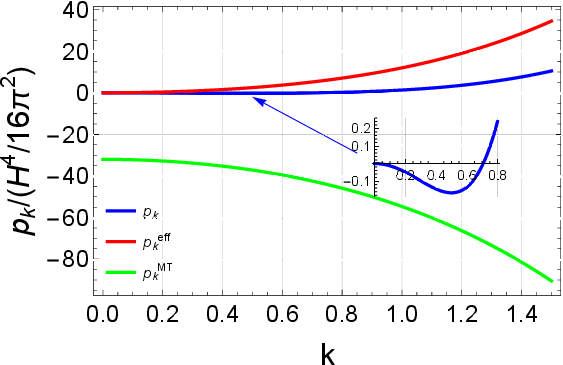}
	\caption{ The spectral pressures.
Blue:    $p_{k}<0$ for small $k$.
Red:  $p^{\text{eff}}_{k}>0$ for all $k$.
Green:  $p^{\text{MT}}_{k}<0$ for all $k$.
Physically, both positive and negative pressures are allowed.
}
	\label{fig2}
\end{figure}

For comparison,
Fig.\ref{fig1}  plots the spectral energy densities
$\rho_{k}$,  $\rho^{\text{eff}}_{k}$, and $\rho^{\text{MT}}_{k}$,
showing that $\rho_{k}$ is positive,
but $\rho^{\text{eff}}_{k}$ and $\rho^{\text{MT}}_{k}$
are negative at low $k$.
Fig.\ref{fig2}  plots the spectral pressures
 $p_k$, $p^{\text{eff}}_{k}$,  and $p^{\text{MT}}_{k}$.

Note that the conserved
$\rho_{k}\propto k^4$  and $p_{k}\propto k^4$ at high $k$,
so that the corresponding vacuum energy density and pressure
$\int_0^{\infty} \rho_k\frac{d k}{k}$ and
$\int_0^{\infty} p_k\frac{d k}{k}$
will be ultra-violet divergent.
These UV divergences can be removed by
the second order adiabatic regularization,
the regularized vacuum energy density and pressure will be vanishing,
in a similar manner to a minimal-coupling massless scalar field
 \cite{ZhangYeWang2020,ZhangWangYe2020}.
The regularization treatment in Refs. \cite{WangZhangChen2016,ZhangWang2018JCAP}
was performed on the BH averaged stress tensor \eqref{RHeff},
and is not pertinent to the conserved $\tau_{\mu\nu}$.

\section{Discussion and Conclusion}\label{Disscusson}

There were three candidates
proposed for the effective stress tensor of GW,
but the differences of these stress tensors
have not been sufficiently examined in literature.
In this paper we have carried out a comprehensive study
of these stress tensors,
traced the origin of the differences,
and found out the adequate candidate and the associated back-reaction scheme.

The first  candidate is  the second order perturbed
Einstein tensor up to a coefficient,
 $T^{\text{eff} }_{\mu\nu} \equiv -\frac{1}{8\pi G}G^{(2)}_{\mu\nu}$,
proposed by Brill and Hartle \cite{BrillHartle1964},
and developed by Isaacson \cite{Isaacson1968a,Isaacson1968b,MTW}.
This is nonconserved and has a negative spectral energy density.
The second is $T_{\text{MT}}^{\mu\nu} \equiv - \frac{2}{\sqrt{-\gamma}}
\frac{\delta J_2 }{\delta\gamma_{\mu\nu}}$ proposed by MacCallum and Taub \cite{MacCallum1973},
where $J_2$  is a second order perturbed action.
$T_{\text{MT}}^{\mu\nu}$ is nonconserved and
has a negative spectral energy density, too.
The third is $\tau_{\mu\nu}  =\frac{2}{\sqrt{-\gamma}}
\frac{\delta I_{gw} }{\delta\gamma^{\mu\nu}}$,
 proposed by Ford and Parker for fRW spacetimes  \cite{FordParker1977},
where  $I_{gw}$ is the action of GW.
$\tau_{\mu\nu}$ is covariantly conserved and has a positive spectral energy density.

It  is not simple to compare these three effective stress tensors,
because a given stress tensor in terms of $h_{\mu\nu}$,  by integrating by parts,
can appear in different forms.
Nevertheless, when expressed in terms of the perturbed actions,
the compositions and
the differences of these stress tensors become more transparent.
Firstly, we observe  that  $G^{(2)}_{\mu\nu}$ is actually not a tensor
on the background spacetime,
ie, generally $G^{(2)\mu\nu} \ne \gamma^{\mu\alpha}\gamma^{\nu\beta} G^{(2)}_{\alpha\beta}$
and  $G^{(2) \mu }_{~~~~\nu}  \ne  \gamma^{\mu\alpha} G^{(2)}_{\alpha\nu}$.
Therefore,  $G^{(2)}_{\mu\nu}$ itself is not legitimate for the back-reaction equation.
More importantly,  we find that $G^{(2)}_{\mu\nu}$ contains
a nonconserved and nontensorial  part $X_{\mu\nu}$,
which should be canceled out by the second order fluid stress tensor
in the back-reaction equation,
or vanishing in absence of matter.
After this correction,  $G^{(2)}_{\mu\nu}$ will reduce to the conserved $\tau_{\mu\nu}$,
the second order action reduces to the GW action $I_{gw}$,
and the back-reaction equation \eqref{2ndexpeff} of the general scheme
will reduce to the back-reaction equation \eqref{backreacteq}
of the perturbation scheme.
In retrospect, Brill and Hartle, and Isaacson studied
the vacuum case with  $\bar R_{\mu\nu}=0$,
but did not explicitly remove all the contributions of $\bar R_{\mu\nu}$
from   $G^{(2)}_{\mu\nu}$.
Subsequent literatures \cite{AbramoPRD1997,AbramoPRD1999,BrandTaka2018}
adopted the whole $G^{(2)}_{\mu\nu}$,
thus ended up with the unphysical consequences, such as
the nonconservation and a negative spectral energy density.

MacCallum and Taub  introduced $T_{\text{MT}}^{\mu\nu}$
without giving its expression.
We have derived  $T_{\text{MT}}^{\mu\nu}$ from  $J_2$,
calculated its four divergence, and found that it is not conserved,
and that $J_2$ does not give the correct GW equation.
In fact,  $J_2$ is the curved-spacetime generalization of
the Fierz-Pauli action of spin-2 massless field,
the latter in the flat spacetime has been known to have
some inconsistency problem \cite{FierzPauli1939,MTW}.
The difficulty with $J_2$ is
due to an $\bar R_{\mu\nu}$ term in \eqref{I2equaltoIG2-2}.
After canceling the $\bar R_{\mu\nu}$ term by the fluid term,
$J_2$ will also reduce to  $I_{gw}$,
and  $T_{\text{MT}}^{\mu\nu}$  will reduce to the conserved $\tau_{\mu\nu}$.
In literature,  the action $J_2$ was first mentioned by Isaacson,
who finally abandoned it in searching for an effective GW stress tensor.
MacCallum and Taub regrouped the second order Hilbert-Einstein action
into two parts, $J_1+J_2$, with $J_1$ consisting of the $\bar R_{\mu\nu}$ terms.
But they did not notice that $J_2$ still implicitly contains the $\bar R_{\mu\nu}$ term
via the relation \eqref{1stterm0}.
(Isaacson studied the the vacuum case with $\bar R_{\mu\nu}=0$,
so that $J_1=0$.)

Ford and Parker studied a system of GW and fluid.
The introduction of a fluid is helpful
to reveal that the nonconserved part  $X_{\mu\nu}$
should be removed from the effective stress tensor.
For a fRW spacetime, they derived $I_{gw}$ of \eqref{LagrangianinfRW}
and gave the conserved $\tau_{\mu\nu}$ of \eqref{FordParkerGiovainni}.
For a general curved spacetime,
we have derived $\tau_{\mu\nu}$ of \eqref{covt2},
obtained the covariant four-divergence of \eqref{csvFP},
and shown  its conservation.
Both the equation of GW and the conserved $\tau_{\mu\nu}$ follow  from
 the same action $I_{gw}$,
and this is  a consistent feature for a well-defined field theory.

We have demonstrated  the three candidates in a fRW spacetime.
We have checked that our general expression \eqref{covt2} of $\tau_{\mu\nu}$
reduces to
Ford-Parker's result \eqref{FordParkerGiovainni} for a fRW spacetime,
and that  $T_{\text{MT}}^{\mu\nu}$ is nonconserved,
and that $T^{\text{eff} }_{\mu\nu}$ is nonconserved and nontensorial.
Furthermore,  in de Sitter space,   we have shown that
$\tau_{\mu\nu}$ has a positive energy density spectrum,
while $T^{\text{eff} }_{\mu\nu}$ and $T_{\text{MT}}^{\mu\nu}$
have a negative one at long wavelengths.

The conclusion:   $\tau_{\mu\nu}$ is adequate as the effective stress tensor of GW
for the back-reaction in the perturbation scheme \eqref{backreacteq},
while $T^{\text{eff} }_{\mu\nu}$ and $T_{\text{MT}}^{\mu\nu}$ are unphysical
and, after canceling the respective nonconserved part,
 will reduce to $\tau_{\mu\nu}$.

Although this paper  studies the case of the fluid without perturbation,
it is expected that the conserved  $\tau_{\mu\nu}$
will be also valid when the linear  perturbations of the fluid are present.
This is because  the matter perturbations do not couple  with GW,
but  only couple  with the scalar and vector metric perturbations.

\appendix
\numberwithin{equation}{section}

\

\section{Expanding the relevant tensors up to the second order}\label{AppendixA}

In this appendix, we list the perturbed  quantities up to the second order.
The connection coefficient is defined as
\bl
\Gamma^{ \mu}_{\nu\sigma}= \frac12 g^{\mu\beta}
 ( g_{\beta\sigma,\nu} +g_{\nu\beta,\sigma}- g_{\sigma\nu,\beta}),
 \label{1stGamma}
\el
the Riemann tensor is defined by
\bl
R^{\alpha}_{~\mu\beta\nu}
=\partial_{\beta}\Gamma^{\alpha}_{\mu\nu}
-\partial_{\nu}\Gamma^{\alpha}_{\mu\beta}
+\Gamma^{\alpha}_{\lambda \beta}\Gamma^{\lambda}_{\mu\nu}
-\Gamma^{\alpha}_{\lambda\nu}\Gamma^{\lambda}_{\mu\beta} ,
\label{defriem}
\el
and  the Ricci tensor is
$R_{\mu\nu}=  R^{\alpha}_{~~\mu\alpha\nu} $.

From the metric perturbation  \eqref{expandg},
the corresponding upper indexed metric  is
\bl
g^{\mu\nu}=\gamma^{\mu\nu}-h^{\mu\nu}
  +h^{\mu}_{~\alpha}h^{\alpha\nu}
    + O(h^3) ,\label{gmunutosecondorder}
\el
where $h^{\mu\nu}=\gamma^{\mu\alpha}\gamma^{\nu\beta}h_{\alpha\beta}$,
and
\bl
\sqrt{-g} &  \simeq \sqrt{-\gamma}(1+ \frac12 h
 -\frac{1}{4} h_{\alpha\nu}h^{\alpha\nu}
 +\frac18h^2).
\label{sqrminusgtosecond}
\el
The connection coefficient   is
\bl
\Gamma^{\mu}_{\nu\rho}=\bar\Gamma^{\mu}_{\nu\rho}
              + \Gamma^{\mu(1)}_{\nu\rho}
              + \Gamma^{\mu(2)}_{\nu\rho},\label{Gammaupto2nd}
\el
where  $\bar{\Gamma}_{\nu\rho}^{\mu}$ is constructed
by the background metric, and
\bl
\Gamma^{\mu(1)}_{\nu\rho}&=\frac12\gamma^{\mu\beta}(h_{\beta\rho|\nu}
        +h_{\nu\beta|\rho}- h_{\rho\nu|\beta}),
\label{1stGamma}
\\
\Gamma_{\nu\rho}^{\mu(2)}
 &=-\frac12h^{\mu\beta}
(h_{\beta\rho|\nu}+ h_{\beta\nu|\rho}- h_{\nu\rho|\beta}).
\label{rpl}
\el
The  Ricci tensor  is
\bl
R_{\mu\nu}=\bar{R}_{\mu\nu}+R^{(1)}_{\mu\nu}+R^{(2)}_{\mu\nu}
   + O(h^3) ,\label{Riccitupto2nd}
\el
where $\bar{R}_{\mu\nu}$ is constructed by the background metric, and
\bl
R^{(1)}_{\mu\nu}&=\frac12\gamma^{\alpha\gamma}(
h_{\alpha\nu|\mu|\gamma}
+ h_{\mu\gamma|\nu|\alpha}
- h_{\mu\nu|\alpha|\gamma}
-h_{\alpha\gamma|\mu|\nu}),\label{1storderriccitensor}
\\
R^{(2)}_{\mu\nu}&=\frac{1}{2}\Big( \frac{1}{2}h_{\alpha\beta|\mu}h^{\alpha\beta}_{~~|\nu}
+h^{\alpha\beta}(h_{\alpha\beta|\mu|\nu}+h_{\mu\nu|\alpha|\beta}
-h_{\alpha\mu|\nu|\beta}-h_{\alpha\nu|\mu|\beta})\nn
\\
&~~~~+h_{\nu}^{~\alpha|\beta}(h_{\alpha\mu|\beta}
-h_{\beta\mu|\alpha})-(h^{\alpha\beta}_{~~~|\beta}-\frac12 h^{|\alpha})
(h_{\alpha\mu|\nu}
+h_{\alpha\nu|\mu}-h_{\mu\nu|\alpha})\Big).\label{Rmunu2}
\el
The Ricci scalar   is
\bl
R= g^{\mu\nu} R_{\mu\nu} =\bar{R}+R^{(1)}+R^{(2)}  + O(h^3),
\label{Riccisupto2nd}
\el
where $\bar{R}$ is  the background Ricci scalar, and
\bl
R^{(1)}
&=  \gamma^{\beta\sigma}  R^{(1)}_{\beta\sigma}
   -h^{\beta\sigma}\bar{R}_{\beta\sigma}
\nn
\\
&=h_{\alpha\sigma}^{~~~|\sigma|\alpha}
-h_{|\alpha}^{~~|\alpha}
-h^{\beta\sigma}\bar{R}_{\beta\sigma},
\label{R1st}
\\
 R^{(2)} & =  \gamma^{\mu\nu} R^{(2)}_{\mu\nu}
 -h^{\mu\nu}  R^{(1)}_{\mu\nu}
+ h^{\mu}_{~\alpha}h^{\alpha\nu} \bar{R}_{\mu\nu}
\nn \\
 & =  \frac{1}{2}\big( \frac{3}{2}h_{\alpha\beta|\nu}h^{\alpha\beta|\nu}
+h^{\alpha\beta}\Box h_{\alpha\beta}
-h^{\nu\alpha|\beta} h_{\beta\nu|\alpha} \big)
+ \big( \frac12 h^{\mu\nu}\Box h_{\mu\nu}
-\bar{R}_{\nu\rho\alpha\mu} h^{\mu\nu}h^{\rho\alpha} \big)\nn
\nn
\\
& ~~~  +\frac12\big(h^{\alpha\beta}( 2 h_{~ |\alpha \beta}
-4\gamma^{\mu\nu}h_{\alpha\mu|\nu\beta})
-(h^{\alpha\beta}_{~~~|\beta}-\frac12 h^{|\alpha})
(2h_{\alpha\nu}^{~~~|\nu}
-h_{|\alpha})\big),\label{R2st}
\el
where the following  formula
\bl
h_{\mu\nu|\alpha|\beta}- h_{\mu\nu|\beta|\alpha}
& = -R_{\mu\rho\alpha\beta} h^{\rho}_{~\nu}
-R_{\nu \rho\alpha\beta} h^{~~\rho}_{\mu}  ,
\label{commu}
\el
has been  used.
The Einstein tensor up to the second order is given by
\bl
G_{\mu\nu} = \bar G_{\mu\nu}
    + G^{(1)}_{ \mu\nu}
    +G^{(2)}_{\mu\nu}  + O(h^3) ,
\el
where
\bl
\bar G_{\mu\nu} & = \bar R_{\mu\nu} -\frac12 \gamma_{\mu\nu} \bar{R} ,
\nn \\
G^{(1)}_{\mu\nu} & = R^{(1)}_{\mu\nu} -\frac12h_{\mu\nu} \bar{R}
-\frac12\gamma_{\mu\nu} R^{(1)},
\label{1Gmunu}
 \\
G^{(2)}_{\mu\nu} & =   R^{(2)} _{\mu\nu} -\frac12 h_{\mu\nu}   R^{(1)}
   -\frac12\gamma_{\mu\nu}  R^{(2)} ,
\label{2Gmunu}
\el
the expression $G^{(2)}_{\mu\nu}$  in terms of $h_{\mu\nu}$
is given by \eqref{2ndEinsteintensor} and \eqref{2Gmunulow}.

For GW in presence of fluid,
the coordinate condition \eqref{coordcondition}
and the TT condition \eqref{transvcondition} \eqref{conditionsforGW}
are imposed,  so that $R^{(1)}=0$,
\bl
R^{(1)}_{\mu\nu} & =\frac12(\bar{R}_{\rho\mu}
h^{\rho}_{~\nu}+ \bar{R}_{\rho\nu} h^{\rho}_{~\mu}),
\\
G^{(1)}_{\mu\nu} & = \frac12(\bar{R}_{\rho\mu} h^{\rho}_{~\nu}
+ \bar{R}_{\rho\nu} h^{\rho}_{~\mu})
          -\frac12h_{\mu\nu} \bar{R}
          \nn
          \\
&  ~~~  - \frac12 (\Box h_{\mu\nu} - 2 \bar{R}_{\mu\rho\sigma\nu} h^{\rho\sigma} )           ,
\label{G1}
\\
  R^{(2)} & = \frac{1}{2}\Big[
\frac{3}{2}h_{\alpha\beta|\nu}h^{\alpha\beta|\nu}
+h^{\alpha\beta}\Box h_{\alpha\beta}
-h^{\nu\alpha|\beta}h_{\beta\nu|\alpha} \Big]
\nn
\\
& ~~~  +\Big( \frac12h^{\mu\nu}\Box h_{\mu\nu}
     -\bar{R}_{\nu\rho\alpha\mu} h^{\mu\nu}h^{\rho\alpha} \Big) ,
 \label{secondorderRicciscalar}
\el
and $G^{(2)}_{\mu\nu} $  is given by \eqref{2Gmunulow}  in terms of $h_{\mu\nu}$.

Given $G_{\alpha\beta}$,
the upper indexed tensors  are given by
\bl
G^{\mu\nu} = g^{\mu\alpha} g^{\nu\beta} G_{\alpha\beta}
=\bar G^{\mu\nu}(\gamma_{\alpha\beta})
    + G^{(1) \mu\nu}(\gamma_{\alpha\beta},h_{\alpha\beta})
    +G^{(2)  \mu\nu}(\gamma_{\alpha\beta},h_{\alpha\beta})  + O(h^3)
 ,\label{upEinst}
\el
from which  we  read off the perturbed Einstein tensors
\bl
G^{(1)\mu\nu} &=    \gamma^{\mu\alpha} \gamma^{\nu\beta} G^{(1)}_{\alpha\beta}
 - \Big( h^{\mu\alpha}\gamma^{\nu\beta} \bar G_{\alpha\beta}
      + h^{\nu\alpha}\gamma^{\mu\beta} \bar G_{\alpha\beta} \Big),
      \label{upperG1}
\\
G^{(2)\mu\nu} &=    \gamma^{\mu\alpha} \gamma^{\nu\beta} G^{(2)}_{\alpha\beta}
 - \Big( h^{\mu\alpha}\gamma^{\nu\beta}  G^{(1)}_{\alpha\beta}
+  h^{\nu\alpha}\gamma^{\mu\beta}   G^{(1)}_{\alpha\beta}
  \nn \\
& ~~~ -h^{\mu}_{~\sigma}h^{\sigma\alpha}\gamma^{\nu\beta}\bar{G}_{\alpha\beta}
  -h^{\nu}_{~\sigma}h^{\sigma\beta}\gamma^{\mu\alpha}\bar{G}_{\alpha\beta}
  -h^{\mu\alpha}h^{\nu\beta}\bar{G}_{\alpha\beta}\Big) ,
\label{upperdef2G}
\el
For GW in the fluid, plugging  \eqref{G1} into the above,
\eqref{upperdef2G} can be written as
\bl
G^{(2)\mu\nu} &=    \gamma^{\mu\alpha} \gamma^{\nu\beta} G^{(2)}_{\alpha\beta}
+ \frac12 \Big(  h^{\mu}_{~\sigma}h^{\sigma\alpha} \bar{R}_{\alpha}^{\nu }
   +  h^{\nu}_{~\sigma}h^{\sigma\alpha} \bar{R}_{\alpha}^{\mu }
    -    h^{\mu\alpha} h^{\nu}_{\alpha}    \bar R \Big) ,
\label{up2G}
\el
\eqref{upperG1} and \eqref{up2G}  tell that generally
$G^{(1)\mu\nu}\ne \gamma^{\mu\alpha} \gamma^{\nu\beta} G^{(1)}_{\alpha\beta}$
and  $G^{(2)\mu\nu}\ne \gamma^{\mu\alpha} \gamma^{\nu\beta} G^{(2)}_{\alpha\beta}$,
that is,
$G^{(1)\mu\nu}$ and $G^{(2)\mu\nu}$ are not tensors on the background spacetime.

The  mixed-indexed Einstein tensor $G^{\mu }_{ ~~\nu}$ is
defined by \eqref{rsloDefGmunu2}.
The perturbed  Einstein tensors     $G^{(1) \mu }_{~~~~\nu}$ and  $G^{(2) \mu }_{~~~~\nu}$
are given by \eqref{raisloG1} and \eqref{raisloG2}.
In general $G^{(2) \mu }_{~~~~\nu}  \ne  \gamma^{\mu\alpha}   G^{(2)}_{\alpha\nu}$.
But, for the fluid model with the coordinate condition $\bar{u}^{\mu}h_{\mu\nu}=0$,
one has
\bl
 -\bar{R}_{\rho\alpha} h^{\rho}_{~\nu}h^{\mu\alpha}
 + \bar{R}_{\rho\nu} h^{\mu\alpha}h^{\rho}_{~\alpha}  =0 ,
\label{proof2g}
\el
so that
\bl
G^{(2) \mu }_{~~~~\nu}  = \gamma^{\mu\alpha}   G^{(2)}_{\alpha\nu} .
\label{RWTeff}
\el
The  mixed-indexed  $G^{(2)\,  \mu}_{~~~~~\nu}$
can be also rewritten  as the following
\bl
G^{(2)\,  \mu}_{~~~~~\nu} & = \gamma_{\nu\rho} G^{(2) \, \mu\rho}
 - (-   h_{\nu}^{~\alpha}\gamma^{\mu\beta}G^{(1)}_{\alpha\beta}
+h_{\nu\sigma}h^{\sigma\beta}\gamma^{\mu\alpha}\bar{G}_{\alpha\beta}
+h^{\mu\alpha}h_{\nu}^{~\beta}\bar{G}_{\alpha\beta})  . \label{tumudownu2}
\el

Generally,  the fluid has fluctuations in
its energy density, pressure, and velocity.
As is known,  the GW  is decoupled from the matter fluctuations at the linear level,
\cite{WangZhang2017,WangZhang2018,WangZhang2019,WangZhang2024}.
So, in study of the GW,
the matter perturbation can be set to zero for simplicity \cite{FordParker1977}
\bl
\delta\rho=\delta p=\delta u^{\mu}=0 ,
\label{nonpertmatt}
\el
(then $\delta u_{\mu} = \bar{u}^{\beta} h_{\mu\beta} = 0$).
The first order perturbed  stress tensor of fluid
can be given by directly expanding the stress tensor $T_{\mu\nu}$
under the condition \eqref{nonpertmatt},
or by variations of fluid actions,
\bl
 T^{(1)}_{\mu\nu} &= \frac{2}{\sqrt{-\gamma}}\frac{\delta I^{(0)}_F}{
  \delta \gamma^{\alpha\beta}}
    (\delta^{\alpha}_{~\mu}h_{\nu}^{~\beta}+\delta^{\alpha}_{~\nu}h_{\mu}^{~\beta})
+ \frac{2}{\sqrt{-\gamma}}\frac{\delta I^{(1)}_F}{\delta \gamma^{\mu\nu}}
 \label{definationof235}
\\
& = -\bar{p}h_{\mu\nu}  \, ,
\label{Tmunu1}
\el
where the condition $\bar{u}^{\mu}h_{\mu\nu}=0$ has been used.
The second order fluid stress tensor follows  either from expanding $T_{\mu\nu} $
of \eqref{strfld} up to the second order,
or by  the variation of fluid actions
\bl
T_{\mu \nu}^{(2)}
& = \frac{1}{4} h^{\sigma \rho} h_{\sigma \rho} \frac{2}{\sqrt{-\gamma}}
\frac{\delta I_F^{(0)}}{\delta \gamma^{\mu \nu}}+\frac{2}{\sqrt{-\gamma}}
\frac{\delta I_F^{(0)}}{\delta \gamma^{\alpha \beta}} h_\mu{ }^\alpha h_\nu{ }^\beta
\nn
\\
& ~~~  + \frac{2}{\sqrt{-\gamma}} \frac{\delta I_F^{(1)}}{\delta \gamma^{\alpha \beta}}
\left(\delta^\alpha{ }_\mu h_\nu{ }^\beta+\delta^\alpha{ }_\nu h_\mu{ }^\beta\right)
+\frac{2}{\sqrt{-\gamma}} \frac{\delta I_F^{(2)}}{\delta \gamma^{\mu \nu}}
\label{T2expression}
\\
& =0 .
 \label{2ndTmunu}
\el
Similarly, the upper indexed fluid stress tensor is
$T^{\mu\nu} =(\rho+p) u^{\mu} u^{\nu} -p g^{\mu\nu}$,
and the perturbed ones are
\bl
 T^{(1) \mu\nu} & =   \bar{p} h^{\mu\nu}  ,
   \\
 T^{(2) \mu\nu} & =  -\bar{p}  h^{\mu}_{~\alpha}h^{\alpha\nu}   .
 \label{upT2}
\el

\section{ The conditions of GW and gauge invariance    }\label{Conditions}

In this Appendix, we examine whether
the GW conditions \eqref{coordcondition} \eqref{transvcondition} \eqref{conditionsforGW}
can be imposed consistently,
and the possible gauge invariance of  GW.

In   a general curved spacetime filled with the fluid,
the GW conditions \eqref{coordcondition} \eqref{transvcondition} \eqref{conditionsforGW}
may not be imposed consistently.
For the fluid velocity  $u_{\mu}$,  its covariant derivative
can be generally decomposed as   \cite{HawkingEllis,Giovannini2020}
\bl
u_{\mu |\nu}= \dot u_\mu u_\nu + \omega_{\mu\nu}+\sigma_{\mu\nu}
     +\frac13 \theta P_{\mu\nu} \, ,
\label{decUbar}
\el
where $\dot u_\mu \equiv  u_{\mu|\nu} u^\nu $,
the volume expansion $\theta \equiv u^\mu\,_{|\mu}$,
the antisymmetric vorticity
$\omega_{\mu\nu}= u_{[\alpha|\beta]} P^{\alpha}_{\mu} P^{\beta}_{\nu}$,
the traceless shear
$\sigma_{\mu\nu}= u_{(\alpha|\beta)}P^{\alpha}_{\mu} P^{\beta}_{\nu}
-\frac13 \theta P_{\mu\nu}$,
and the projection $P_{\mu\nu}=\gamma_{\mu\nu}-u_\mu u_\nu$.

The coordinate condition $\bar{u}^{\mu}h_{\mu\nu}=0$ will impose
four constraints upon $h_{\mu\nu}$,
and the transverse condition $h^{\mu\nu}_{~~|\nu}=0$ generally
would also  impose  four constraints upon  $h_{\mu\nu}$,
so two physical DOF for the GW remain,
and one is not allowed to impose any additional conditions upon $h_{\mu\nu}$.
Taking four divergence of $\bar{u}^{\mu}h_{\mu\nu}=0$ leads to
\bl
\bar{u}_\mu h_{~~ |\nu}^{\mu \nu}
  =-\sigma_{\mu \nu} h^{\mu \nu}  -\frac{1}{3} \theta h ,
\label{interstep}
\el
which is a constraint upon the vector $h_{~~ |\nu}^{\mu \nu}$.

For the case $\theta\neq 0$,
in a general curved spacetime, when a shearless velocity field $u^\mu$
can be defined ($\sigma_{\mu\nu}= 0$, including $u^\mu\,_{|\nu}=0$),
one can simultaneously impose
the conditions \eqref{coordcondition}\eqref{transvcondition}\eqref{conditionsforGW}, and
leave two independent components of $h_{\mu\nu}$
to two polarizations of GW, see also \cite{FordParker1977}.
For instance, in a fRW spacetime ($\theta  \ne 0$),
a vector field $\bar{u}^{\mu}=\frac{1}{\sqrt{\gamma_{00}}} (1,0,0,0)
= \frac{1}{a} (1,0,0,0)$  can be introduced,
so that  $\sigma_{\mu\nu}=0$.
As we have checked, in a fRW spacetime
the traceless condition $h=0$ is implied by
the coordinate condition  \eqref{coordcondition}
and the transverse  condition \eqref{transvcondition},
so that the constraint \eqref{interstep} is satisfied.

For the case $\theta= 0$,  \eqref{interstep}
reduces to $\bar{u}_\mu h_{~~ |\nu}^{\mu \nu}=-\sigma_{\mu \nu} h^{\mu \nu}$.
In general,  the shear $\sigma_{\mu\nu}$ can be either zero or non-zero.
We first consider the shearless situation   $\sigma_{\mu\nu}=0$.
As a result, the vector $h_{~~ |\nu}^{\mu \nu}$ has only three independent components
and, consequently, $h_{~~ |\nu}^{\mu \nu}=0$ give only three independent conditions.
So we need to impose the traceless  condition  $h=0$ separately.
Thus, the conditions \eqref{coordcondition}\eqref{transvcondition}\eqref{conditionsforGW}
constitute only eight independent conditions and two physical DOF for GW remain.
For instance, in the Schwarzschild spacetime
or the Minkowiski spacetime ($\theta= 0$), one can define a vector field
$\bar{u}^{\mu}= \frac{1}{\sqrt{\gamma_{00}}}(1,0,0,0)$ for which  $\sigma_{\mu\nu}=0$.
As is checked, the 0-component of transverse condition \eqref{transvcondition} is an identity (0=0),
\eqref{transvcondition} only provides three independent conditions,
thus, the traceless condition \eqref{conditionsforGW} should be imposed separately.

For  another situation with shear $\sigma_{\mu\nu} \ne 0$
(regardless of $\theta=0$ or $\theta \ne 0$),
one has $h_{~~ |\nu}^{\mu \nu} \ne 0$ by \eqref{interstep},
so that  we are not allowed to impose $h_{~~ |\nu}^{\mu \nu}=0$,
and the specification of DOF of GW will be more complicated,
and we will not discuss this in the present work.

Under a  coordinate transformation
\bl
x^\mu \rightarrow x^\mu +\xi^\mu,
\label{gaugetran}
\el
with $\xi^\mu$ being small,
the metric perturbations change as the following
\bl
h_{\mu\nu} & \rightarrow h_{\mu\nu} \blue{-}(\xi_{\mu|\nu} + \xi_{\nu|\mu}),
\label{B3}
\\
h  &  \rightarrow h \blue{-}2 \xi^{\nu}\, _{|\nu} ,
\label{B4}
\\
h^{\mu\nu}\, _{|\nu} & \rightarrow h^{\mu\nu}_{~~~|\nu} \blue{-}
(\xi^{\mu|\nu}_{~~~|\nu}  + \xi^{\nu|\mu}_{~~~|\nu}),
\label{B5}
\el
so,  GW  and the GW conditions
are generally subject to change under coordinate transformations.
By the formula
$\xi^{\nu|\mu}_{~~~|\nu}-\xi^{\nu}_{~|\nu}\, ^{|\mu}=- \bar R^{\mu}_{~\nu }\xi^\nu$
 (ie, (43.17) in Ref.\cite{Vfork1964}),
it is seen from \eqref{B4} \eqref{B5} that
 the GW equation subject to the TT gauge conditions
are invariant if $\xi^\mu$ satisfies
$\xi^\mu_{~|\mu}=0$ and $\Box \xi^\mu -\bar R_{\mu\nu}\xi^\nu=0$.
(See Refs.\cite{Isaacson1968a,MTW} for further discussions.)
The simplest case is that there exists a Killing vector $\xi^\mu$ satisfying
\bl\label{SpecialTransformation}
\xi_{\mu|\nu}+\xi_{\nu|\mu}=0 .
\el
Then the  GW is invariant under the associated transformation,
$ h_{\mu\nu} \rightarrow h_{\mu\nu}$,
so are the GW conditions,
\eqref{coordcondition}\eqref{transvcondition}\eqref{conditionsforGW}.
For a general curved spacetime, however,
the Killing vector may not exist,
and the description of GW will be more complicated.

More relevant for cosmology is the  fRW spacetimes,
in which Killing vectors exist.
As is known  \cite{WangZhang2017,WangZhang2018,WangZhang2019},
under the class of synchronous-synchronous transformations,
the scalar and vector metric perturbations undergo changes  generally,
but the tensorial metric perturbation is invariant,
and so will be the stress tensor of GW.

\section{The variation $\delta I_{gw}$ under  $\delta\gamma_{\mu\nu}$ }\label{Appvaaction}

We present the details of our calculation of the variation  $\delta I_{gw}$
with respect to the background metric $\gamma_{\mu\nu}$.
Although the calculation itself is straightforward,
the algebraic manipulation is lengthy.
Moreover, our resulting  expression $\tau_{\mu\nu}$ of \eqref{covt2}
differs from (3.36) in Ref.\cite{Giovannini2019prd} (with some obvious typos)
and (3.114) in Ref.\cite{Giovannini2020},
and yet the calculational details were not  reported in literature,
as far as we see.
In calculation, to be specific,
we take $h_{\mu\nu}$ as being independent of $\gamma_{\mu\nu}$,
so that  $ \delta h_{\mu\nu} =0$ under $\delta\gamma_{\mu\nu}$,
but
$\delta h^{\mu\nu} = \delta (\gamma^{\mu\alpha} \gamma^{\nu\beta} h_{\alpha\beta} )\ne 0$.
From its definition  \eqref{Igw},  the GW action varies as
\bl
\delta I_{gw} &=\frac{1}{4}\int d^4x\delta(\sqrt{-\gamma})
\Big(h_{\alpha\beta|\nu}h^{\alpha\beta|\nu}
+2\bar{R}_{\mu\rho\alpha\nu} h^{\mu\nu}h^{\rho\alpha}\Big)
\nn
\\
& ~~~~ +\frac{1}{4}\int d^4x\sqrt{-\gamma}
\delta(h_{\alpha\beta|\nu}h^{\alpha\beta|\nu})
\nn
\\
& ~~~~ +\frac{1}{2}\int d^4x\sqrt{-\gamma}
\delta(\bar{R}_{\nu\rho\alpha\mu} h^{\mu\nu}h^{\rho\alpha}).
\label{C1IG2}
\el
The first integration of \eqref{C1IG2}  is simply given by
\bl
\frac{1}{4}\int d^4x\sqrt{-\gamma}
\gamma^{\alpha\beta} \frac{1}{2}  \Big(h_{\mu\nu|\gamma}h^{\mu\nu|\gamma}
+2\bar{R}_{\mu\rho\theta\nu} h^{\mu\nu}h^{\rho\theta}\Big)
\delta \gamma_{\alpha\beta} ,
\label{firstonI2}
\el
where $\delta(\sqrt{-\gamma})=\frac12\sqrt{-\gamma}\gamma^{\alpha\beta}
\delta \gamma_{\alpha\beta}$ has been used.
The second integration of \eqref{C1IG2} is  given by  three terms
\bl
& \frac{1}{4}\int d^4x\sqrt{-\gamma}
\delta(\gamma^{\alpha\beta}h_{\mu\nu|\alpha}h^{\mu\nu}_{~~~|\beta})
\nn
\\
& =  \frac{1}{4}\int d^4x\sqrt{-\gamma}\Big(
\delta \gamma^{\alpha\beta}(h_{\mu\nu|\alpha}h^{\mu\nu}_{~~~|\beta})
+\gamma^{\alpha\beta}\delta(h_{\mu\nu|\alpha})h^{\mu\nu}_{~~~|\beta}
+\gamma^{\alpha\beta}h_{\mu\nu|\alpha}\delta(h^{\mu\nu}_{~~~|\beta})\Big)
 . \label{ndline}
\el
The first term of \eqref{ndline} is given by
\bl
\frac{1}{4}\int d^4x \sqrt{-\gamma}
\delta \gamma^{\alpha\beta} (h_{\mu\nu|\alpha}h^{\mu\nu}_{~~~|\beta})
=  \frac{1}{4}\int d^4x\sqrt{-\gamma}
\Big( - h_{\mu\nu}^{~~~|\alpha}h^{\mu\nu|\beta}   \Big)
   \delta \gamma_{\alpha\beta} ,
   \label{fterm3}
\el
The second term of \eqref{ndline} is found to be
\bl
&  \frac{1}{4}\int d^4x\sqrt{-\gamma}
 \gamma^{\alpha\beta} \delta(h_{\mu\nu|\alpha}) h^{\mu\nu}_{~~~|\beta}
\nn
\\
= &  \frac{1}{4}\int d^4x\sqrt{-\gamma} ( h^{\beta}_{~\nu }\, ^{|\theta} h^{\alpha\nu}_{~~~|\theta}
 + h^{\beta}_{~\nu }\Box h^{\alpha\nu}
+ h^{\beta}_{~\nu }\, _{|\mu}h^{\mu\nu|\alpha}
+h^{\beta}_{~\nu } h^{\mu\nu|\alpha}\, _{|\mu}
- h_{\sigma\nu }h^{\alpha\nu|\beta|\sigma})\delta\gamma_{\alpha\beta} .
\label{term193}
\el
In deriving \eqref{term193},  we have used the following detailed steps
\bl
\delta\partial_{\alpha} h_{\mu\nu} & =\partial_{\alpha} \delta h_{\mu\nu}=0 ,
\\
\delta(h_{\mu\nu|\alpha})
  &  = \delta(\partial_{\alpha} h_{\mu\nu}
-\bar{\Gamma}_{\mu\alpha}^{\sigma} h_{\sigma \nu }
-\bar{\Gamma}_{\nu\alpha}^{\sigma}h_{\sigma\mu })
\nn
\\
& =   -\delta \bar{\Gamma}_{\mu\alpha}^{\sigma} h_{\sigma \nu }
-\delta \bar{\Gamma}_{\nu\alpha}^{\sigma}h_{\sigma\mu }  ,
\\
\delta\Gamma_{\mu\nu}^{\lambda}
& = \frac{1}{2}\gamma^{\lambda\rho}
(\delta \gamma_{\rho\nu|\mu} + \delta \gamma_{\mu\rho|\nu}
- \delta \gamma_{\nu\mu|\rho}) ,
\label{deltaGamma}
\\
\gamma^{\alpha\beta} \delta(h_{\mu\nu|\alpha}) h^{\mu\nu}_{~~~|\beta}
&  =-  \gamma^{\alpha\beta} \gamma^{\sigma\rho}
(\delta \gamma_{\mu\rho|\alpha}+ \delta \gamma_{\alpha\rho|\mu}
- \delta \gamma_{\mu\alpha|\rho}) h_{\sigma\nu }h^{\mu\nu}_{~~~|\beta} ,
\label{GdeltaG}
\\
\delta \gamma_{\rho\nu|\mu} & \equiv ( \delta \gamma_{\rho\nu} )_{|\mu} ,
\el
and the three terms of \eqref{GdeltaG} have been integrated by parts
with the total differential terms being dropped.
For example,  for the first term in \eqref{GdeltaG}, we do calculation
\bl
\gamma^{\alpha\beta}\gamma^{\sigma\rho} \delta \gamma_{\mu\rho|\alpha}
          h_{\sigma\nu }h^{\mu\nu}_{~~~|\beta}
&  = ( \gamma^{\alpha\beta}\gamma^{\sigma\rho} \delta \gamma_{\mu\rho}
          h_{\sigma\nu }h^{\mu\nu}_{~~~|\beta})_{|\alpha}
  -  \delta \gamma_{\mu\rho}(  \gamma^{\alpha\beta}\gamma^{\sigma\rho}
         h_{\sigma\nu }h^{\mu\nu}_{~~~|\beta})_{|\alpha}
\nn
\\
&    = \frac{1}{\sqrt{-\gamma}}( \sqrt{-\gamma}
        \gamma^{\alpha\beta}\gamma^{\sigma\rho} \delta \gamma_{\mu\rho}
          h_{\sigma\nu }h^{\mu\nu}_{~~~|\beta})_{,\, \alpha}
  -  \delta \gamma_{\mu\rho}(  \gamma^{\alpha\beta}\gamma^{\sigma\rho}
         h_{\sigma\nu }h^{\mu\nu}_{~~~|\beta})_{|\alpha}
\\
&   = - \delta \gamma_{\mu\rho}  \gamma^{\alpha\beta}\gamma^{\sigma\rho}
          (h_{\sigma\nu }h^{\mu\nu}_{~~~|\beta})_{|\alpha}
\nn
\\
&  = - \delta \gamma_{\mu\rho}  \gamma^{\alpha\beta}\gamma^{\sigma\rho}
      ( h_{\sigma\nu |\alpha }h^{\mu\nu}_{~~~|\beta}
       + h_{\sigma\nu }h^{\mu\nu}_{~~~|\beta|\alpha}  ) ,
\el
where we have  dropped the total differential term as a surface integration.
And other two terms in \eqref{GdeltaG} are treated analogously.

The third term of \eqref{ndline} is   similarly calculated,
\bl
&  \frac{1}{4}\int d^4x\sqrt{-\gamma}
 \gamma^{\alpha\beta} h_{\mu\nu|\alpha} \delta (h^{\mu\nu}_{~~~|\beta})
\nn
\\
= &  \frac{1}{4}\int d^4x\sqrt{-\gamma}
(  - h^{\alpha\nu|\lambda}h^{\beta}_{~\nu|\lambda}
+h^{\mu\alpha}\Box h_{\mu}^{~\beta}
   + h^{\beta}_{~\nu|\mu} h^{\mu\nu |\alpha}
      +  h^{\beta}_{~\nu} h^{\mu\nu|\alpha} \, _{|\mu}
       - h_{\rho\nu} h^{\beta\nu|\alpha|\rho} )  \delta\gamma_{\alpha\beta}  .
\label{secondvariation}
\el
Plugging  \eqref{fterm3} \eqref{term193} \eqref{secondvariation}
into \eqref{ndline} gives
the  second integration of \eqref{C1IG2}
\bl
&  \frac{1}{4}\int d^4x\sqrt{-\gamma}\Big(
-h_{\mu\nu}^{~~~|\alpha}h^{\mu\nu|\beta}
+ 2h^{\mu\alpha}\Box h_{\mu}^{~\beta}
+2 h^{\beta}_{~\nu|\mu }h^{\mu\nu|\alpha}
+2h^{\beta}_{~\nu }  h^{\mu\nu|\alpha}\, _{|\mu}
-2h_{\sigma\nu }h^{\alpha\nu|\beta|\sigma}
  \Big) \delta\gamma_{\alpha\beta}.
\label{secI2}
\el
The third integration of \eqref{C1IG2} is
\bl
&  \frac{1}{2}\int d^4x\sqrt{-\gamma}
\delta(\bar{R}_{\nu\rho\alpha\mu} h^{\mu\nu}h^{\rho\alpha})
  \nn \\
& ~~~ = \frac{1}{2}\int d^4x\sqrt{-\gamma}
\Big(-   2 h^{\alpha\mu}\Box h^{\beta}_{~\mu}
  +h^{\alpha\beta}_{~~|\sigma|\rho}h^{\rho\sigma}
-  h^{\mu\alpha}_{~~|\rho}h^{\rho\beta}_{~~|\mu}
-   h^{\mu\alpha}h^{\rho\beta}_{~~|\mu|\rho}
+  \bar{R}^{\alpha}_{~\rho\sigma\mu} h^{\mu\beta}h^{\rho\sigma}  \Big)
   \delta\gamma_{\alpha\beta}   ,
\label{firline1}
\el
where   the following formula  \cite{BarthChristensen1983} has been used
\bl
\delta R_{\alpha\beta\gamma\sigma} & =\frac12(
\delta\gamma_{\alpha\sigma|\beta|\gamma}
- \delta\gamma_{\sigma\beta|\alpha|\gamma}
-\delta\gamma_{\alpha\gamma|\beta|\sigma}
+ \delta\gamma_{\gamma\beta|\alpha|\sigma}
\nn
\\
& ~~~~  +\delta\gamma_{\beta\alpha|\sigma|\gamma}
-\delta\gamma_{\beta\alpha|\gamma|\sigma})
+\delta\gamma_{\mu\alpha} \bar{R}^{\mu}_{~\beta\gamma\sigma} ,
\label{Christensen}
\el
with $\delta\gamma_{\alpha\sigma|\beta|\gamma}
           \equiv  (\delta\gamma_{\alpha\sigma})_{|\beta|\gamma}$,
and these six terms of \eqref{Christensen}
have been integrated by parts twice
with the total differential terms being dropped.
Substituting \eqref{firstonI2} \eqref{secI2} \eqref{firline1}
into \eqref{C1IG2},
 and  using the GW equation \eqref{GWeq},
 we obtain
\bl
\delta I_{gw}
& =  \frac{1}{4}\int d^4x \sqrt{-\gamma}
\Big( -h_{\mu\nu}^{~~~|\alpha}h^{\mu\nu|\beta}
+ \frac12\gamma^{\alpha\beta}(h_{\mu\nu|\gamma}h^{\mu\nu|\gamma}
    + 2\bar{R}_{\nu\rho\theta\mu} h^{\mu\nu}h^{\rho\theta})
     -2h^{\mu\alpha}_{~~|\rho}h^{\rho\beta}_{~~|\mu}
\nn
\\
&~~~~~~~~~~~~
   - 2 \bar{R}^{\alpha}_{~\rho\sigma\mu} h^{\mu\beta}h^{\rho\sigma}
 +2h^{\beta}_{~\nu|\mu }h^{\mu\nu|\alpha}
 + 2 h^{\alpha\beta}_{~~|\sigma|\rho}h^{\rho\sigma}
 -2 h_{\sigma\nu }h^{\alpha\nu|\beta|\sigma}
 \Big)
      \delta \gamma_{\alpha\beta} .
\label{C20IG2}
\el
This  gives  the  conserved effective stress tensor
$\tau^{\alpha\beta}$ of \eqref{covt2}.

\section{The variation $\delta J_2$  under $\delta\gamma_{\mu\nu}$}\label{j2append}

We present the details of our calculation of the variation  $\delta I_2$
with respect to the background metric.
Our resulting expression $T^{\mu\nu}_\text{MT}$ of
\eqref{MTstress} differs from (20b) of Ref.\cite{stein2011},
which did not give the calculational detail.
From the definition \eqref{J2GW}, the action  $  J_2$ varies
with respect to the background metric
\bl
\delta  J_2
&    = \int d^4x (\delta  \sqrt{-\gamma})
    \Big( \frac14 h_{\alpha\beta|\nu} h^{\alpha\beta|\nu}
   -\frac12 h^{\alpha\beta}_{~~|\nu} h^{\nu}_{~\alpha|\beta} \Big)
\nn
\\
& ~~~ + \int d^4x\sqrt{-\gamma}
  \Big(  \frac14 \delta (h_{\alpha\beta|\nu} h^{\alpha\beta|\nu} ) \Big)
\nn
\\
& ~~~  + \int d^4x\sqrt{-\gamma} \Big(
  -\frac12 \delta( h^{\alpha\beta}_{~~|\nu} h^{\nu}_{~\alpha|\beta} ) \Big) .
\label{FierzPaulivary}
\el
The first term of \eqref{FierzPaulivary}  is simple and given by
\bl
  \int d^4x \sqrt{-\gamma}\gamma^{\mu\nu} \frac12
    \Big( \frac14 h_{\alpha\beta|\rho} h^{\alpha\beta|\rho}
   -\frac12 h^{\alpha\beta}_{~~|\rho} h^{\rho}_{~\alpha|\beta} \Big)
   \delta \gamma_{\mu\nu} .
\label{FierzPauli1}
\el
The second term  of \eqref{FierzPaulivary} is the same as \eqref{ndline}
and  the variation has been given by \eqref{secI2}.
The third term  of \eqref{FierzPaulivary} is calculated as the following
\bl
& \int d^4x\sqrt{-\gamma} \Big(
  -\frac12 \delta( h^{\alpha\beta}_{~~|\nu} h^{\nu}_{~\alpha|\beta} ) \Big)
\nn
\\
& =  \int d^4x\sqrt{-\gamma} (-\frac12) \Big( - h^{\beta \rho |\nu} h^{\alpha}_{~\nu|\rho}
   - 2 h^{\alpha\rho|\nu} h_{\nu\rho} \,^{|\beta} \Big) \delta\gamma_{\alpha\beta}
\nn
\\
& ~~  + \int d^4x \sqrt{-\gamma} (- \frac12)
    \Big(  2 h^{\alpha\gamma|\nu} h^{\beta}_{~\nu|\gamma}
    +  h^{\mu\gamma|\beta} h^{\alpha}_{~\mu|\gamma}
    +  h^{\mu\alpha|\nu} h^{\beta}_{~\mu|\nu}
\nn
\\
& ~~  ~~   +2 h^{\alpha\gamma|\nu}\, _{|\gamma}  h^{\beta}_{~\nu}
      -  h^{\beta\alpha|\nu}\, _{|\mu}   h^{\mu}_{~\nu}
       +  h^{\mu\gamma|\beta}\, _{|\gamma}  h^{\alpha}_{~\mu}
        +  \Box h^{\mu\alpha } h^{\beta}_{~\mu}
        -  h^{\mu\alpha|\beta}\, _{|\nu}  h^{\nu}_{~\mu}
          \Big) \delta \gamma_{\alpha\beta}  ,
\label{FierPauli3-2}
\el
where the formula \eqref{deltaGamma} has been used,
some terms have been integrated by parts
with the total differential terms being dropped.
Summing up \eqref{FierzPauli1}  \eqref{secI2} \eqref{FierPauli3-2},
 and using $h^{\alpha \gamma \mid \nu}{ }_{\mid \gamma} h_{~\nu}^\beta
=\bar{R}^{\alpha}_{~\rho\gamma\nu}h^{\rho\gamma}h^{\beta\nu}
+\bar{R}_{\rho\nu}h^{\rho\alpha}h^{\beta\nu}$,  we have
\bl
\delta  J_2  &  =
\int d^4x \sqrt{-\gamma}   \gamma^{\mu\nu} \frac12
    \Big( \frac14 h_{\alpha\beta|\rho} h^{\alpha\beta|\rho}
   -\frac12 h^{\alpha\beta}_{~~|\rho} h^{\rho}_{~\alpha|\beta} \Big)
   \delta \gamma_{\mu\nu}
\nn
\\
&  + \int d^4x\sqrt{-\gamma}  \frac{1}{4} \Big(
  -h_{\mu\nu}^{~~~|\alpha}h^{\mu\nu|\beta}  \Big) \delta\gamma_{\alpha\beta}
\nn
\\
& + \int d^4x\sqrt{-\gamma} (-\frac12)
  \Big(    h^{\beta\rho|\nu} h^{\alpha}_{~\nu|\rho}
   - 2 h^{\alpha\rho|\nu} h_{\nu\rho} \,^{|\beta}
  +    h^{\mu\alpha|\nu} h^{\beta}_{~\mu|\nu} \Big) \delta\gamma_{\alpha\beta}
\nn
\\
&   + \int d^4x\sqrt{-\gamma} (- \frac12)
    \Big(   2 (\bar{R}^{\alpha}_{~\rho\gamma\nu}h^{\rho\gamma}h^{\beta\nu}
+\bar{R}_{\rho\nu}h^{\rho\alpha}h^{\beta\nu} )
     -  h^{\beta\alpha|\nu}\, _{|\mu}   h^{\mu}_{~\nu}
             \Big) \delta \gamma_{\alpha\beta} .
\label{calMacCallum-Taub}
\el
This gives the MacCallum-Taub stress tensor
$T_{\text{MT} }^{\alpha\beta}$ of \eqref{MTstress}.

\section{The variation $\delta I_{g2}^{(2)}$
and its cancelation by $\delta I^{(2)}_{f2}$ }\label{G3delta}

As shown in \eqref{2ndactionGW},
$I^{(2)}_{g2}$ is canceled out by the fluid,
$I^{(2)}_{g2}+ I^{(2)}_{f2}=0$.
Then, it follows that $\delta I^{(2)}_{g2}+\delta I^{(2)}_{f2}=0$.
This cancelation occurs under the condition that
the velocity normalization and the background Einstein equation
hold also on  the varied background metric \cite{Weinberg1972}
\bl
 \delta ( \bar{u}^{\mu}\bar{u}^{\nu} \gamma_{\mu\nu}-1 ) &  =0 ,
 \label{normalization}
\\
\delta \big[
\bar{R}_{\mu\nu} - \frac12 \big(\bar{T}_{\mu\nu}-\frac12\gamma_{\mu\nu}\bar{T} \big)
 \big]  & = 0 ,\label{dbv}
\el
which amount to the following
\bl
\bar{u}^{\mu}\bar{u}^{\nu} \delta  \gamma_{\mu\nu} & = 0,
\\
\delta \bar{R}_{\mu\nu} & = -\frac14  (\bar \rho - \bar p)\delta   \gamma_{\mu\nu}  ,
\label{deltabarR}
\el
 respectively.

The variation of  $I^{(2)}_{g2} $ of \eqref{123I}
under   $\delta \gamma_{\mu\nu}$ is (also with $ \delta h_{\mu\nu} =0$)
\bl
\delta I^{(2)}_{g2}
&=\int d^4x(\delta\sqrt{-\gamma})\frac{1}{4}\Big(
- 2\bar{R}_{\alpha\beta} h^{\nu\alpha}h^{\beta}_{~\nu}
+ h_{\alpha\nu}h^{\alpha\nu} \bar{R}  \Big)
\nn
\\
&  ~~~~ +\int d^4x\sqrt{-\gamma}\frac{1}{4}
\Big(- 2\delta(\bar{R}_{\sigma\alpha}h^{\nu\alpha}h^{\sigma}_{~\nu})\Big)
\nn \\
& ~~~~ +\int d^4x\sqrt{-\gamma} \frac{1}{4}
        \delta \Big(h_{\alpha\nu}h^{\alpha\nu}\bar{R} \Big) .
 \label{i23geo}
\el
Using the conditions \eqref{dbv} \eqref{normalization},
as well as the background equation \eqref{bkR} \eqref{GFcancel},
we find
\bl
\delta I^{(2)}_{g2} & =
\frac{1}{4}\int d^4x \sqrt{-\gamma}
   \frac12 \, \bar p \, h^{\alpha\beta}h_{\alpha\beta}
  \gamma^{\mu\nu} \delta\gamma_{\mu\nu}
  \nn
  \\
& ~~~  + \int d^4x\sqrt{-\gamma}\frac{1}{4}
\Big( - (\bar\rho-\bar{p})h^{\nu\alpha}h^{\sigma}_{~\nu}  \Big)
  \delta \gamma_{\sigma\alpha}
  \nn
  \\
& ~~~ + \int d^4x\sqrt{-\gamma} \frac{1}{4}
 \Big((\bar{\rho} - 3\bar{p}) h_{\alpha}^{~\rho} h^{\alpha\sigma}
     \Big) \delta\gamma_{\sigma\rho},
\nn
\\
& = \int d^4x\sqrt{-\gamma}\frac{1}{4}
  \Big( \frac12 \, \bar p \, h^{\alpha\beta}h_{\alpha\beta}
    \gamma^{\mu\nu}
    - 2\bar{p}    h^{\alpha\mu}h^{\nu}_{~\alpha}
        \Big) \delta\gamma_{\mu\nu} .
    \label{G32}
\el
The variation of the fluid action \eqref{I2F3} is given by
\bl
\delta    I^{(2)}_{f2} = \int d^4x  \sqrt{-\gamma}
 \frac{1}{4} \Big( - \frac12 \bar{p}  \gamma^{\mu\nu}h_{\alpha\beta}  h^{\alpha\beta}\,
     +2 \bar{p} h_{\alpha}^{~\mu}    h^{\alpha\nu}  \Big)
          \delta\gamma_{\mu\nu}  .
\label{fI3v}
\el
Clearly,  the sum of  \eqref{G32} and   \eqref{fI3v} is zero,
$\delta  I^{(2)}_{g2} + \delta  I^{(2)}_{f2}=0$.

In actuality, the expression $G^{(2)}_{\mu\nu}$ contains
the varied $\delta I_{g2}^{(2)}$ as a main piece of the nonconserved part.
We can also explicitly write $\delta I^{(2)}_{g2}$ in terms of $h_{\mu\nu}$.
The first term of \eqref{i23geo} is simple and given by
\bl
\frac12 \int d^4x   \sqrt{-\gamma}
\Big[ \gamma^{\alpha\beta}  \frac14 (-
2\bar{R}_{\sigma\mu}h^{\mu\nu}h^{\sigma}_{~\nu}
+ h_{\mu\nu}h^{\mu\nu} \bar{R} ) \Big] \delta \gamma_{\alpha\beta}  .
\label{D2}
\el
The second term of \eqref{i23geo} is calculated
\bl
&-\frac12\int d^4x\sqrt{-\gamma}\Big(
h^{\nu\alpha|\rho}_{~~~~|\alpha}h^{\lambda}_{~\nu}
+h^{\nu\alpha|\rho}h^{\lambda}_{~\nu|\alpha}
+h^{\nu\alpha}h^{\lambda~|\rho}_{~\nu~~|\alpha}
-h^{\rho}_{~\nu}\Box h^{\nu\lambda}
-h^{\nu\lambda|\eta}h^{\rho}_{~\nu|\eta}
\nn
\\
&~~~~~
-
\frac12\gamma^{\lambda\rho}( h^{\nu\mu}_{~~~|\sigma}h^{\sigma}_{~\nu|\mu}
+h^{\nu\mu}h^{\sigma}_{~\nu|\mu|\sigma})
- \bar{R}_{\sigma\alpha}h^{\rho\alpha}h^{\sigma\lambda}
-\bar{R}_{\sigma}^{~\lambda} h^{\nu\rho}h^{\sigma}_{~\nu}
-\bar{R}^{\lambda}_{~\alpha} h^{\nu\alpha}h^{\rho}_{~\nu}
    \Big) \delta\gamma_{\lambda\rho},
\label{444}
\el
where  the following formulae
\bl
\delta \bar R_{\sigma\alpha} &  =
(\delta \bar \Gamma^{\rho}_{\sigma\alpha})_{|\rho}-
(\delta \bar \Gamma^{\rho}_{\sigma\rho})_{|\alpha} ,
\nn
\\
(\delta \bar \Gamma^{\rho}_{\sigma\alpha})_{|\rho}
& = \frac12   \gamma^{\rho\eta}(\delta\gamma_{\sigma\eta|\alpha|\rho}
+ \delta\gamma_{\alpha\eta|\sigma|\rho}
-\delta\gamma_{\alpha\sigma|\eta|\rho})  ,
\el
and  $h^{\mu\nu}_{~~~|\nu}=0$ have been used,
and some terms have been integrated by parts
with total integration terms being dropped.
The third term of \eqref{i23geo} is similarly calculated
\bl
& -\frac{1}{2}\int d^4x\sqrt{-\gamma}
\Big(\bar{R}h^{\lambda\nu}h^{\rho}_{~\nu}
+\frac{1}{2} h_{\alpha\nu}h^{\alpha\nu}\bar{R}^{\lambda\rho}
-h_{\mu\nu}h^{\mu\nu|\lambda|\rho}
- h_{\mu\nu}^{~~~|\lambda}h^{\mu\nu|\rho}
\nn
\\
& ~~~~~ +\gamma^{\rho\lambda} (h^{\mu\nu}\Box h_{\mu\nu}+h_{\mu\nu|\alpha}h^{\mu\nu|\alpha})
\Big)\delta\gamma_{\lambda\rho} .
  \label{321}
\el
Summing up \eqref{D2} \eqref{444} \eqref{321},
 and using the formula \eqref{formula1}, we obtain
\bl
\delta I^{(2)}_{g2} = & - \frac12 \int d^4x \sqrt{-\gamma}
 \Big[  \gamma_{\alpha\beta}  \big[ \frac14 (-
2\bar{R}_{\sigma\mu}h^{\mu\nu}h^{\sigma}_{~\nu}
+ h_{\mu\nu}h^{\mu\nu} \bar{R} )
 +  \frac12 (h^{\nu\mu}_{~~~|\sigma}h^{\sigma}_{~\nu|\mu}
    +h^{\nu\mu}h^{\sigma}_{~\nu|\mu|\sigma})
    \nn
 \\
&  - (h^{\mu\nu}\Box h_{\mu\nu}
      +h_{\mu\nu|\sigma}h^{\mu\nu|\sigma}) \big]
\nn
\\
& - \big(   h^{\nu\mu}_{~~|\beta} h_{\alpha\nu|\mu}
+ h^{\nu\mu} h_{\alpha\nu|\beta|\mu}
 - \frac12 h_{\beta \nu} \Box h^{\nu}_{~\alpha}
 -h_{\alpha}^{~ \nu|\mu} h_{\beta\nu|\mu}
  -h_{\mu\nu}h^{\mu\nu}_{~~~ |\alpha|\beta}
-h_{\mu\nu|\alpha} h^{\mu\nu}_{~~~|\beta}  \big)
\nn
\\
& - (  h_{\alpha}^{~\nu}h_{\beta\nu} \bar{R}
       +\frac{1}{2} h_{\mu\nu}h^{\mu\nu}\bar{R}_{\alpha\beta}
-\bar{R}_{\mu\nu} h^{~\mu}_{\beta} h^{~\nu}_{\alpha}
-\bar{R}_{\mu\alpha} h^{\nu}_{~\beta} h^{\mu}_{~\nu} ) \Big]
     \delta\gamma^{\alpha\beta} .
  \label{D13}
\el
This will be canceled out by the fluid term $\delta I^{(2)}_{f2}$ of \eqref{fI3v},
or  be set to zero in absence of fluid.
We have double checked that
substituting  \eqref{D13} into  \eqref{barGmunu2dow}
gives the expression $G_{\mu\nu}^{(2)}$ of \eqref{2Gmunulow}.

\section{Quantities in a flat RW spacetime}\label{usefulquantites}

In a fRW spacetime \eqref{RWmetric},
the 0th order non-vanishing components of the geometric quantities are
\bl
\bar{\Gamma}_{00}^{0}=\frac{a'}{a},
~~~\bar{\Gamma}_{ij}^{0}=\frac{a'}{a}\delta_{ij},
~~~\Bar{\Gamma}_{0j}^{i}=\frac{a'}{a}\delta^i_{j},
\label{nonvanishaffcontion}
\el
\bl
\bar{R}_{i0j0}& =a^2(\frac{a''}{a}-\frac{a'^2}{a^2})
\delta_{ij},~~~\bar{R}_{j 0 k i}=0,
\\
\bar{R}_{m i n j}& =-a^2\frac{a'^2}{a^2}(\delta_{mn}\delta_{ij}
-\delta_{mj}\delta_{in}),
\label{0thoftheremainn}
\el
\bl
\bar{R}_{00} &  =-3(\frac{a''}{a}-\frac{a'^2}{a^2}),
~~~\bar{R}_{ij}=(\frac{a''}{a}+\frac{a'^2}{a^2})\delta_{ij},
\\
\bar R & = - \frac{6}{a^2} \frac{a''}{a} ,
\label{0thofthericci}
\el
\bl
\bar G _{ 00} &
= 3 \frac{a'^2}{a^2} ,~~~
\bar G _{ij}
= \delta_{\ij}( -2  \frac{a''}{a} + \frac{a'^2}{a^2} ) .
\el
The components of the covariant derivative of $h^{\mu\nu}$ are
\bl
h^{00|\alpha}&=0,\label{h00alphacodalpha}
\\
h^{0i|0}&=0,\label{hoicdi}
\\
h^{0i|j}
&=-\frac{a'}{a^3}h^{ij}=\frac{a'}{a}\frac{1}{a^4}H^{ij},\label{hijcdi}
\\
h^{ij|0}&   =-\frac{1}{a^4}\partial_0H^{ij},\label{hijderate0}
\\
h^{ij|m}&=-a^{-2}\partial_{m}h^{ij}
=\frac{1}{a^4}\partial_{m}H^{ij},
\label{hijderatek}
\el
where $h^{ij}=-a^{-2} H^{ij}$  has been used.
From the above we get
\bl
h_{\alpha\beta|\nu}h^{\alpha\beta|\nu}
&=\frac{1}{a^2}(2\frac{a'^2}{a^2}H_{ij}H^{ij}
  +H_{ij,0}H^{ij}_{~~,0}  -H_{ij,k}H^{ij}_{~~,k})
  ,\label{abnabn}
\\
2\bar{R}_{\nu\rho\alpha\mu} h^{\mu\nu}h^{\rho\alpha}
&= -2 \frac{a'^2}{a^4}H_{ij} H^{ij}  ,
\label{Enurhoalmimunurhoal}
\el
which gives the GW action \eqref{LagrangianinfRW} in fRW spacetimes.

The 00 component of the conserved $\tau_{\mu\nu}$ of \eqref{covt2}
in fRW spacetimes involves the following terms
\bl
h_{\alpha\beta|0}h^{\alpha\beta}_{~~|0} &=\partial_0H_{ij}\partial_{0}H^{ij},
\label{firstterm00stGiov}
\\
h^{\mu}_{~0|\rho}h^{\rho}_{~0|\mu} &=  \frac{a'^2}{a^2}H_{ij }H^{ij},
\label{fifth00ourstmunu}
\\
(h_{0\nu })_{|\mu}(h^{\mu\nu})_{|0}&=-\frac{a'}{a}H_{ij}\partial_0H^{ij},
\label{covaianttmunuGWofus2nd}
\\
h_{00|\sigma|\rho}h^{\rho\sigma}&=
2\frac{a'^2}{a^2}H_{ij}H^{ij},\label{eighthth00ourstmunu}
\\\
h^{\sigma\nu }h_{0\nu|0|\sigma}&=
\frac{a'^2}{a^2}H_{ij }H^{ij}
-\frac{a'}{a} H_{ij }\partial_{0}H^{ij},
\label{forth00ourstmunu}
\el
and the $ij$ components  involve
\bl
h_{\alpha\beta|i}h^{\alpha\beta}_{~~|j}&=
-2\frac{a'^2}{a^2}H^{n}_{~i}H_{nj}
+\partial_iH_{ab}\partial_jH^{ab},\label{firsttermijstGiov}
\\
h^{\mu}_{~i|\rho}h^{\rho}_{~j|\mu}&=
\frac{a'}{a}H_{ik}\partial_0H^{k}_{~j}
+\frac{a'}{a}\partial_0H_{ki} H_{j}^{~k}
+\partial^m H^{k}_{~i}\partial_k H_{mj},\label{covaianttmunuGWofusij8th}
\\
(h_{i\nu })_{|\mu}(h^{\mu\nu})_{|j}&=-\frac{a'^2}{a^2}H_{im}H^{m}_{~j}
+  \frac{a'}{a}H^{k}_{~j}\partial_0H_{ik}
+  \partial_{m}H_{i}^{~k}\partial_{j}H^{m}_{~k},\label{covaianttmunuGWofusij2ndterm}
\\
\bar{R}^{\nu}_{~\rho\sigma i} h_{j\nu}h^{\rho\sigma}
  &= \frac{a'^2}{a^2} H_{jk}H_{i}^{~k},\label{covaianttmunuGWofusij12th}
\\
h_{ij|\sigma|\rho}h^{\rho\sigma}&=H^{mk}\partial_{k}\partial_{m}H_{ij}
+\frac{a'^2}{a^2}H_{ik}H_{j}^{~k}
+\frac{a'^2}{a^2}H_{jk} H_{ i}^{~k},
\label{covaianttmunuGWofusij14th}
\\
h^{\sigma\nu }h_{i\nu|j|\sigma}&=
H^{nm}\partial_m \partial_{j}H_{in}
+\frac{a'^2}{a^2}H_{ni}H^{n}_{~j}-\frac{a'}{a}H^{n}_{~j}\partial_0H_{i n} ,
 \label{covaianttmunuGWofusijsixth}
\\
h^{\mu}_{~i}h^{\rho}_{~j|\mu|\rho}&=
2\frac{a'^2}{a^2}H_{j}^{~k}H_{ki}+\frac{a''}{a}H_{j}^{~k}H_{ki},
\label{covaianttmunuGWofusij10th}
\el
and the $0i$ components  involve
\bl
h_{\mu\nu|0}h^{\mu\nu}_{~~|i}&=
(\partial_0H_{mn})(\partial_{i}H^{mn}),\label{P0HPMH}
\\
(h^{\mu 0})_{| \rho}( h^{\rho i})_{|\mu}&= \frac{1}{a^4}\frac{a'}{a} H^{jm}
\partial_{j}H_{m}^{~i},\label{secondtermta0i}
\\
(h_{0\nu})_{|\mu}(h^{\mu\nu})_{|i}&=-\frac{a'}{a}H_{mn}
\partial_i H^{mn},\label{apbya0HPMH}
\\
(h_{i\nu})_{|\mu}(h^{\mu\nu})_{|0}& =
    \partial_{m}H_{i}^{~k}\partial_0H_{k}^{~m},\label{i0third}
\\
\bar{R}^{\nu~~0}_{~\rho\sigma}h^{i}_{~\nu}h^{\rho\sigma}&=0,\label{intermsofR}
\\
(h^{0i})_{|\sigma|\rho}h^{\sigma\rho} &=
  2\frac{a'}{a}\frac{1}{a^4}(\partial_{m} H_n^{~i} )H_{m}^{~n},  \label{0isigmjatho}
\\
h^{\sigma\nu} h_{0\nu |i|\sigma}&=-\frac{a'}{a}(H^{j}_{~k}\partial_jH^{k}_{~i}
+H_{mk}\partial_{i}H^{mk}),\label{sigmanuousigmanu}
\\
h^{\sigma\nu} h_{i\nu|0|\sigma}  &=
    H^{j}_{~k} \partial_{j}\partial_{0}H^{ik}
    -\frac{a'}{a}  H^{j}_{~k}\partial_{j}H^{ik} .\label{sigmamuinu0sihgmag}
\el

The effective $T^{\text{eff} }_{\mu\nu}$
in the fRW spacetimes  involves the following additional terms
\bl
h^{\alpha\beta}h_{\alpha\beta|00}&=
H_{ij}(\partial_0^2H^{ij}
-\frac{a'}{a}\partial_0H^{ij}),\label{00componentG1st}
\\
h_{\alpha\beta}\Box h^{\alpha\beta}
&=h^{\alpha\beta}2\bar{R}_{\beta\mu\nu\alpha}h^{\mu\nu}=-2\frac{a'}{a^4}H^{ij}H_{ij},
 \label{00componentG4th}
\\
h^{\alpha\beta}h_{\alpha\beta|i|j}&=
H^{mn}\partial_j\partial_{i}H_{mn}
+2\frac{a'^2}{a^2}H_{j}^{~n} H_{ni}
-\delta^i_j\frac{a'}{a}H^{mn} \partial_{0}H_{mn},\label{Gij1st}
\\
h_{j\alpha|\beta}h_{i}^{~\alpha|\beta}&
=-\frac{a'^2}{a^2}H_{i}^{~n}H_{nj}
-  \partial_0 H_{i}^{~n} \partial_{0}H_{nj}
+ \partial_{m}H_{i}^{~n}\partial^{m}H_{nj},\label{hhijfin}
\\
h^{\alpha\beta}h_{\alpha\beta|0|i}&=H_{mk}\partial_i \partial_0 H^{mk}
-\frac{a'}{a}H_{mk}\partial_i H^{mk},\label{habhab0i}
\\
h^{\alpha\beta}h_{\alpha\beta|i|0}&=H_{mk}\partial_i \partial_0 H^{mk}
-\frac{a'}{a}H_{mk}\partial_i H^{mk},\label{habhabi0}
\\
h^{\alpha\beta}h_{0i|\alpha|\beta}&= -\frac{a'}{a} H_{m}^{~k}(\partial_k H_{i}^{~m}
+\partial_m H_{ki}) . \label{secondt}
\el

{\bf Acknowledgment} : Y. Zhang is supported by NSFC Grant No. 12261131497.

\end{CJK}
\end{document}